%% file: main.tex
\documentclass[a4paper,11pt]{article}
\usepackage[left=25mm,right=25mm,top=20mm,bottom=20mm]{geometry}
\usepackage{xurl}

\usepackage[colorlinks,
linkcolor=blue,
anchorcolor=blue,
citecolor=blue,
urlcolor=blue]{hyperref}
\usepackage{lineno}
\usepackage{subcaption}
\usepackage{graphicx}
\usepackage[symbol]{footmisc}
\usepackage{amsmath}
\usepackage{booktabs}
\usepackage{multirow}
\usepackage{tabularx}
\usepackage{soul,xcolor}
\usepackage{subcaption}
\usepackage[utf8]{inputenc}
\usepackage{authblk}
\usepackage{ulem}
\DeclareMathOperator{\erf}{erf}

\begin{document}

\title{Reactor Antineutrino Detection with JUNO-TAO: Commissioning Performance and First Results}

\input{Authors_2026_08_31_0918_TAO_final1001}

\maketitle


\abstract{
The Taishan Antineutrino Observatory (TAO) is a satellite experiment of JUNO, designed to measure the reactor antineutrino energy spectrum with high precision at a baseline of 43.2~m from a reactor core using a 2.8-ton spherical liquid-scintillator detector, 1.8~m in diameter, operated at \(-50\,^{\circ}\mathrm{C}\) and instrumented with 4,024 silicon photomultiplier tiles. The TAO detector installation was completed in 2025, and physics data taking began on 10 February 2026. This paper reports the detector design, calibration strategy, and performance based on commissioning data and the first physics runs. Stable cryogenic operation has been achieved, and about 90\% of the readout channels are retained for physics analysis after channel-quality selections. The detector light yield is measured to be about 4000~photoelectrons/MeV after correction for dark noise and optical crosstalk, using neutron capture on hydrogen in inverse beta decay (IBD) events. The energy resolution is about 2.8\% for \(^{68}\)Ge (two 0.511-MeV \(\gamma\) rays) at the detector center and about 3.7\% for \(^{137}\)Cs (one 0.662-MeV \(\gamma\) ray) at \(R=193\)~mm. This energy-resolution performance is among the best reported for reactor-antineutrino detectors and enables TAO to measure the reactor antineutrino spectrum with high precision. The detector energy response model has been established and characterized using calibration and physics data. Within the fiducial volume of \(R < 800\)~mm, the vertex reconstruction bias remains within \(\pm 30~\mathrm{mm}\), and the resolution is approximately \(70~\mathrm{mm}\) over most of the fiducial volume. The combined muon-tagging efficiency of the veto system is about 97\%. Using 31.3 days of reactor-on data and 29.4 days of reactor-off data, a significant excess of IBD candidates is observed in the reactor-on sample relative to the reactor-off sample, with candidate rates of 2900$\pm$10 events/day and 419$\pm$4 events/day, respectively. The measured spectrum is consistent with the model prediction within the current uncertainties, demonstrating that TAO has reached a high performance level in reactor antineutrino measurement.
}



\setstcolor{red}
\maketitle
\flushbottom

\section{Introduction}
\label{sec:intro}
\input{sec-intro.tex}

\section{The TAO Detector}
\label{sec:detector}
\input{sec-detector-v2}

\section{Data Acquisition, Control, and Processing Systems}
\label{sec:data_proc}
\input{sec-data_processing-v3}

\section{CD Performance Results} 
\label{sec:cd_performance}
\input{sec-performance-v2}

\section{Vertex and Energy Reconstruction in CD}
\label{sec:CCRec}
\input{sec-energy_cal_rec-v3}

\section{Energy Response in CD}
\label{sec:e_response}
\input{sec-energy_response-v2}

\section{Muon Detection in TAO}
\label{sec:muon_det}
\input{sec-muon_detection-v2}

\section{IBD Analysis}
\label{sec:ibd}
\input{sec-IBD-analysis-v2}

\section{Conclusions}
\label{sec:summary}
\input{sec-summary-v2}

\section*{Acknowledgement}
\clearpage
\bibliographystyle{JHEP}
\bibliography{biblio.bib}

\end{document}

%% file: Authors_2026_08_31_0918_TAO_final1001.tex
\author[34]{Thomas Adam}
\author[31]{Timo Ahola}
\author[11]{Fengpeng An}
\author[60]{Costas Andreopoulos}
\author[43]{Giuseppe Andronico}
\author[55]{Nikolay Anfimov}
\author[45]{Vito Antonelli}
\author[55]{Tatiana Antoshkina}
\author[34]{Jo\~{a}o Pedro Athayde Marcondes de Andr\'{e}}
\author[55]{Nikita Balashov}

% ...
\affil[11]{Sun Yat-sen University, Guangzhou, China}
% ...
\affil[31]{University of Jyvaskyla, Department of Physics, Jyvaskyla, Finland}
% ...
\affil[34]{IPHC, Universit\'{e} de Strasbourg, CNRS/IN2P3, F-67037 Strasbourg, France}
% ...
\affil[43]{INFN Catania and Dipartimento di Fisica e Astronomia dell Universit\`{a} di Catania, Catania, Italy}
% ...
\affil[45]{INFN Sezione di Milano and Dipartimento di Fisica dell Universit\`{a} di Milano, Milano, Italy}
% ...
\affil[55]{Joint Institute for Nuclear Research, Dubna, Russia}
% ...
\affil[60]{The University of Liverpool, Department of Physics, Oliver Lodge Laboratory, Oxford Str., Liverpool L69 7ZE, UK, United Kingdom}
% ...

%% file: sec-intro.tex
The Jiangmen Underground Neutrino Observatory (JUNO) has been designed to determine the neutrino mass ordering and to precisely measure neutrino oscillation parameters~\cite{JUNO:2015sjr,JUNO:2015zny,JUNO:2021vlw}. JUNO completed detector construction and began physics data taking on 26 August 2025. The detector performance has been shown to satisfy the design requirements~\cite{JUNO:2025fpc}, and the first measurement of reactor antineutrino oscillations, based on 59.1 days of data, was reported in Ref.~\cite{JUNO:2025gmd}.

The Taishan Antineutrino Observatory (TAO) is a satellite experiment of JUNO. Located 43.2\,m from a reactor core at the Taishan Nuclear Power Plant, TAO has been designed to measure the reactor antineutrino energy spectrum with a target energy resolution of about 2\% at 1\,MeV using a 2.8-ton gadolinium-doped Liquid-Scintillator (GdLS) target instrumented with high-performance silicon-based photosensors. The precise spectral measurement provided by TAO will serve as a nearly model-independent reference for JUNO and is expected to improve JUNO's sensitivity to the neutrino mass ordering.

In addition to its role in supporting JUNO, TAO has an independent physics program. Its high-resolution measurement of the reactor antineutrino spectrum will enable studies of fine spectral structures, provide important input for testing reactor antineutrino flux models, including those based on the summation method, and help benchmark the nuclear databases used in such calculations. TAO is also expected to contribute to investigations of the reactor antineutrino anomaly and to searches for oscillation signatures associated with light sterile neutrinos at short baselines. An overview of the TAO physics program is given in Ref.~\cite{JUNO:2020ijm}. The potential of TAO to explore the electromagnetic properties of reactor neutrinos is discussed in Ref.~\cite{Cao:2026nfm}.

Following the completion of testing of the 1:1 prototype detector~\cite{Chen2026TAOPrototype}, construction of the final TAO detector started in November 2024 at the Taishan Nuclear Power Plant. Detector installation was completed in September 2025, followed by a commissioning phase that concluded in early February 2026. Physics data taking began on 10 February 2026. This paper presents the performance of the TAO detector based on commissioning and early-physics data, together with a first measurement of the prompt energy spectrum from inverse beta decay (IBD) interactions of reactor antineutrinos with protons in the target.

The paper is structured as follows. Section~\ref{sec:detector} introduces the design of the central detector, together with the veto and calibration systems. Section~\ref{sec:data_proc} describes the data-acquisition workflow and data-quality monitoring. Sections~\ref{sec:cd_performance}, \ref{sec:CCRec}, and \ref{sec:e_response} present the main results for the central detector, including performance assessment, event reconstruction, and energy response. Section~\ref{sec:muon_det} discusses the veto-system performance, and Section~\ref{sec:ibd} finally covers the analysis of the first reactor antineutrino data.

%% file: sec-detector-v2.tex
\begin{figure}[!htbp]
    \centering
    \includegraphics[width=0.9\linewidth]{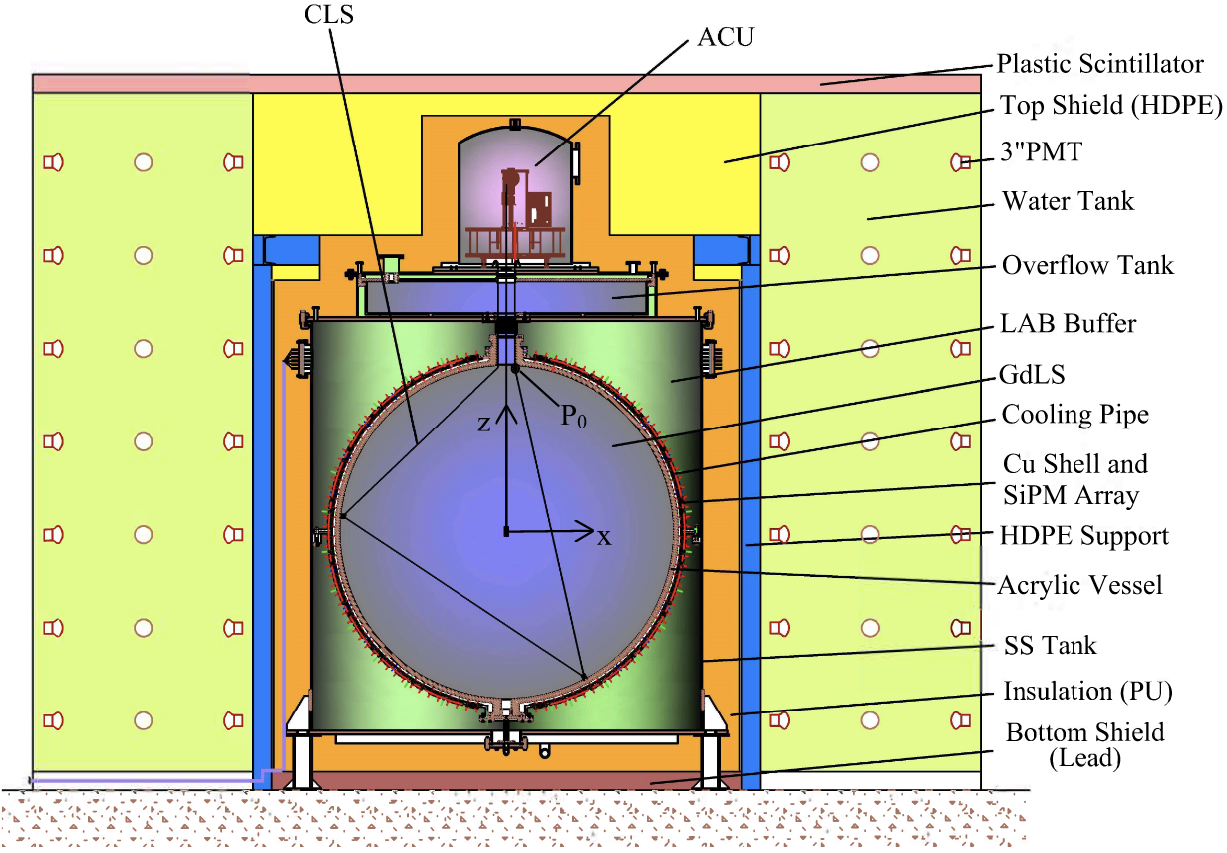}
    \caption{Cross-sectional view of the TAO detector. The Central Detector contains 2.8\,tons of Gd-loaded liquid scintillator and a $\sim$10\,m$^2$ array of silicon photomultiplier tiles mounted on two hemispherical copper shells. The detector inside the stainless steel tank is designed to operate at approximately $-50^{\circ}$C. The Central Detector is surrounded by a water-Cherenkov detector, with a plastic-scintillator tracker positioned on top.}
    \label{fig:tao_det_design}
\end{figure}

The schematic layout of the TAO detector is shown in Fig.~\ref{fig:tao_det_design}. The detector is made of three subsystems: the Central Detector (CD), the Water Tanks (WTs), and the Top Veto Tracker (TVT). The CD is a 2.8-tonne GdLS target designed to detect reactor electron antineutrinos via IBD. It is operated at a stable temperature of approximately \(-50\,^{\circ}\mathrm{C}\) with a dedicated cryogenic system and it is instrumented with an embedded calibration system for precise monitoring of the detector response. The CD is surrounded by three isolated WTs that function as a water-Cherenkov veto and provide passive shielding against external natural radioactivity and ambient neutrons in the experimental hall. The TVT, composed of plastic-scintillator (PS) strips, is installed above the CD and WTs to provide additional top-side coverage for cosmic-muon identification.

\subsection{Central Detector}

The CD adopts a two-layer spherical-cylindrical hybrid design. The inner target volume contains 2.8~tonnes of GdLS housed in a spherical acrylic vessel with a diameter of 1.8~m. This target is surrounded by an outer buffer layer consisting of $\sim$3.3~tonnes of linear alkylbenzene (LAB), contained in a cylindrical stainless-steel tank (SST) 2.1~m in diameter and 2.2~m in height. Scintillation light is detected by 4,024 silicon photomultiplier (SiPM) tiles, each measuring \(50.7\times50.7~\mathrm{mm}^2\) and made of 32 individual SiPM elements. The tiles are mounted on the inner surfaces of two hemispherical copper shells surrounding the acrylic vessel, yielding a total SiPM area of approximately \(10~\mathrm{m}^2\). The SiPMs are positioned 18~mm from the outer surface of the acrylic vessel, arranged in a way to achieve a photosensitive coverage of 83.3\%. Combined with the high photon-detection efficiency of the SiPMs, this configuration provides high light-collection efficiency and is essential for achieving the target energy resolution.

To suppress SiPM dark noise, the entire CD is maintained at approximately \(-50\,^{\circ}\mathrm{C}\) by an external cryogenic system. Chilled silicone oil circulates through cooling pipes thermally coupled to the copper shells and the exterior of the SST. The GdLS formulation is optimized on the basis of the Daya Bay recipe~\cite{DayaBay:2012aa} to preserve optical transparency at cryogenic temperatures down to \(-50\,^{\circ}\mathrm{C}\) or below~\cite{Xie:2020bqa}. The final composition consists of LAB doped with 0.13\% gadolinium, 3~g/L 2,5-diphenyloxazole (PPO), 2~mg/L 1,4-bis(2-methylstyryl)benzene (bis-MSB), 0.5\% Dipropylene Glycol n-Butyl Ether (DPnB) and 150~mg/kg butylated hydroxytoluene (BHT). Stable operation at low temperature requires a low water concentration in the LAB buffer, and a condition of 12~ppm has been actively maintained during filling and operation.

An overflow tank is installed on top of the SST to accommodate thermal expansion of the liquids. The SST is equipped with 28 flanges for routing signal and power cables, accessing cooling pipes, supplying nitrogen purge gas, and connecting internal sensors such as temperature, pressure, and liquid-level monitors. The entire CD assembly is wrapped in melamine insulation to reduce heat leakage. 

\subsection{Top Veto Tracker}

The TVT consists of 160 PS modules: 106 modules with dimensions of \(2000 (L) \times 200 (W) \times 20 (H)~\mathrm{mm}^3\) and 54 modules with dimensions of \(1500 (L) \times 200 (W) \times 20(H)~\mathrm{mm}^3\). Based on Geant4 \cite{GEANT4:2002zbu} simulation studies, the modules are arranged in four staggered horizontal layers to enable muon track reconstruction and to minimize dead regions. The four layers together cover the projected footprint of the CD and WTs with a dimension of \(4000 (L) \times 4000 (W) ~\mathrm{mm}^2\), with a vertical gap of 2.7~cm between neighboring layers.

Each PS module is instrumented with eight 1.5-mm-diameter wavelength-shifting fibers coupled to four SiPMs at both ends~\cite{TVT-2023-1,TVT-2023,TVT-2025}. The modules are fabricated from extruded polystyrene-based scintillator doped with p-terphenyl and POPOP. The SiPM signals are processed by custom-designed printed circuit boards that provide summed analog outputs of the four SiPMs at each end. 

Owing to the four-layer redundancy, the TVT maintains high muon-tagging efficiency despite small inactive regions in individual layers. Simulations show that, even for a single-strip detection efficiency of 97\% at a threshold of 10 photoelectrons (PE), the probability for a muon to trigger at least three of the four layers exceeds 99\%~\cite{TVT-2023}. Background studies including natural radioactivity from the \(^{238}\)U, \(^{232}\)Th, and \(^{40}\)K decay chains indicate that a three-out-of-four coincidence requirement effectively suppresses ambient backgrounds while preserving high muon-detection efficiency~\cite{TVT-2023}.

\subsection{Water Tank}

The WT system is designed to satisfy the spatial constraints of the experimental hall, which provides an available volume of \(5.1~\mathrm{m} \times 5.1~\mathrm{m} \times 3.7~\mathrm{m}\) (height). It consists of three individual WTs arranged in a dodecagonal configuration, with a thickness of 1.2~m and a total mass of approximately 46~tons of ultrapure water. To maximize the light-collection efficiency, the inner surfaces of the tanks are lined with reflective Tyvek film. The 3-inch PMTs are arranged in an approximately uniform rectangular grid with a density of about one PMT per \(0.5~\mathrm{m}^2\), corresponding to a photocathode coverage of 0.8\%. The WT system employs 300 3-inch PMTs to provide sufficient redundancy for stable operation~\cite{JUNO-SPMT-mass}. In addition, each water tank is equipped with a pressure sensor for water-level monitoring and a PT100 sensor for water-temperature monitoring.

A 1~tonne/hour pure-water system is installed to provide ultrapure water for detector cleaning during installation and for filling the water tanks. By default, no online water circulation is applied during operation, as the water quality can be maintained at an adequate level over timescales of several months~\cite{WT-prototype-2026}. Nevertheless, a manual filling and circulation pipe system has been designed and installed.

Additional neutron and gamma shielding is provided by HDPE layers installed at the top and bottom of the CD. These layers are segmented into modular pieces to match the detector geometry and facilitate assembly. A 60-cm-thick HDPE layer moderates fast neutrons from above, while a dedicated 10-cm-thick HDPE layer at the bottom, located between the CD and the 10-cm-thick lead shielding on the floor, attenuates neutrons produced in the lead itself~\cite{TAO-VETO-neutron-Li_2022,li2022ambientneutronmeasurementtaishan}. 

\subsection{Calibration System}

The calibration system has been designed to monitor the detector stability and to characterize the detector energy response, including position-dependent non-uniformity, energy non-linearity, and energy resolution. As described in Ref.~\cite{Xu:2022mdi}, the TAO calibration strategy employs two complementary hardware systems: the automated calibration unit (ACU) and the cable loop system (CLS).

The ACU is mounted on top of the detector and deploys calibration sources along the central vertical (\(z\)) axis. The system reuses hardware from the Daya Bay ACU, but was significantly revised for TAO, including a new deployment axis for the UV light injection system, the integration of the CLS, and modifications introduced to ensure compatibility with the low-temperature operating environment of TAO. The ACU has been designed to deploy a \(^{68}\)Ge source and a combined gamma source containing \(^{137}\)Cs, \(^{54}\)Mn, \(^{40}\)K, and \(^{60}\)Co, which are used to calibrate the energy scale and constrain the detector non-linearity. The ACU also includes a UV light injection system for monitoring the stability of the SiPM gain, photon-detection efficiency, and timing response.

The CLS is used to calibrate the position dependence of the detector response beyond the central axis. In this system, a \(^{137}\)Cs source attached to a stainless-steel wire is moved along a closed loop passing through off-axis positions in the LS. The wire trajectory is defined by dedicated anchors in the acrylic vessel, allowing measurements at predefined calibration points and enabling precise mapping of the detector spatial response.

The calibration system has been implemented in TAO, with one modification during the commissioning stage. Although both the ACU and CLS are installed, a technical issue in the ACU combined-source deployment mechanism limited operation during the first data-taking period to deployment of the \(^{68}\)Ge source only. Consequently, the initial calibration used the \(^{68}\)Ge source and the \(^{137}\)Cs source deployed by the CLS for the energy scale and spatial non-uniformity calibration. The energy non-linearity was further constrained using natural radioactive backgrounds and the 2.22-MeV gamma ray from neutron capture in IBD events. A future maintenance intervention is planned to restore the ACU's full functionality.

%% file: sec-data_processing-v3.tex
\begin{figure}[!htbp]
    \centering
    \includegraphics[width=0.9\linewidth]{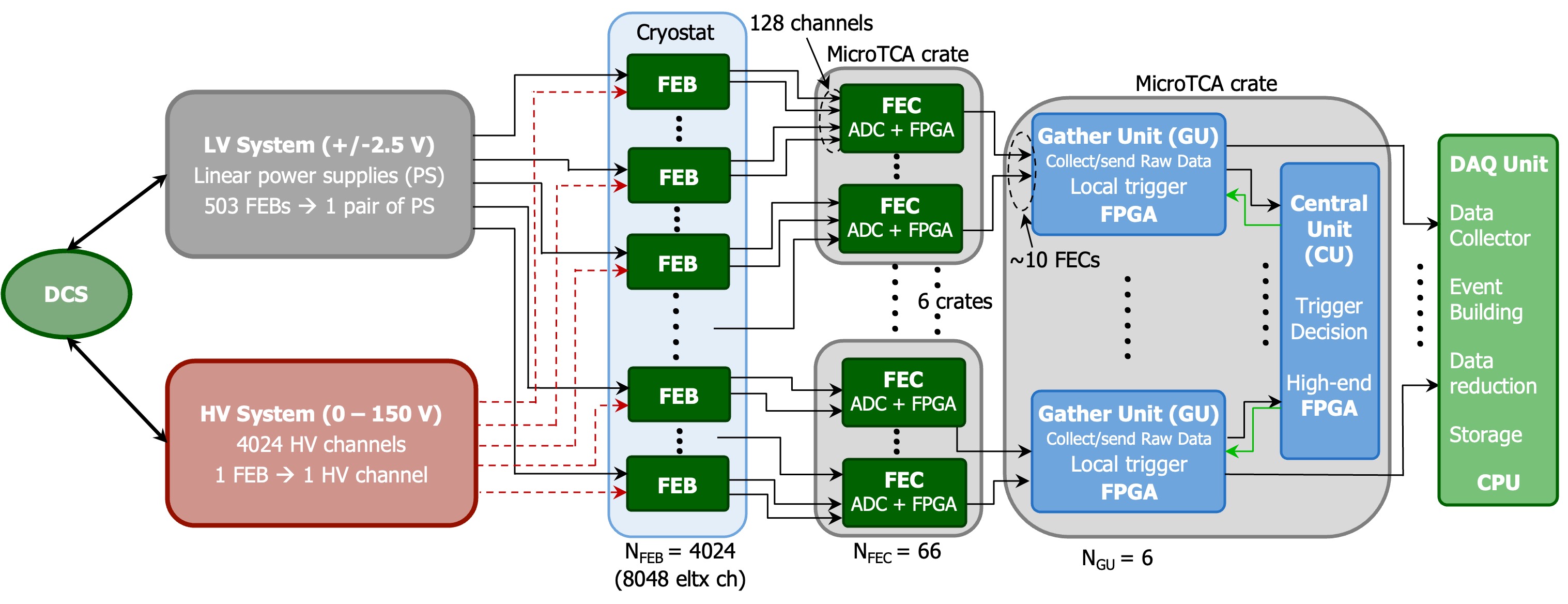}
    \caption{Architecture of the online data-processing system of CD.}
    \label{fig:tao_online_systems}
\end{figure}

As shown in Fig.~\ref{fig:tao_online_systems}, the CD online data-processing architecture is made of three main subsystems: the readout electronics, the trigger system, and the data acquisition (DAQ) system.

\subsection{Electronics}

The central detector (CD) contains 4,024 SiPM tiles, corresponding to 8,048 readout channels in total. The front-end boards (FEBs)~\cite{Venettacci:2024hwy} are mounted on the outer surfaces of the copper shells and are designed to operate at \(-50\,^{\circ}\mathrm{C}\). The analog signals are transmitted to the front-end controllers (FECs) located outside the detector, where they are digitized by analog-to-digital converter (ADC) cards with a programmable output data rate of 125~MSPS. The ADC cards are mounted on the FECs and subsequently processed by field-programmable gate arrays (FPGA) on the same boards. A total of 66 FECs are installed in six MicroTCA crates~\cite{Zhang:2023rqc}. Waveform reconstruction is performed in the field-programmable gate arrays of the FECs using the continuous over-threshold integration (COTI) algorithm~\cite{JUNO:2025fpc}, and the reconstructed quantities are transmitted to the trigger system. A small fraction of waveform data can also be sent directly to the DAQ for debugging.

A White Rabbit (WR)-based timing and synchronization system~\cite{Wlostowski2011WhiteRabbit} has been implemented for the TAO readout electronics. At present, the ADC chips' channel-to-channel timing offsets are corrected in the firmware using offline time calibration results. The SiPM bias voltages are supplied by an independent high-voltage (HV) system~\cite{Selyunin:2023ahs} with one channel per SiPM tile, while the FEBs are powered by low-voltage linear power supplies.

The Top Veto Tracker electronics use the same overall architecture and hardware platform as the CD, with dedicated FEBs and low-voltage power supplies adapted to the TVT readout. The TVT contains 320 readout channels, and the reconstructed pulse parameters are transmitted to the TVT trigger system.

The Water Tank electronics reuse the small-PMT readout system developed for JUNO~\cite{SPMT-diru-20220310,catiroc-Conforti_2021,SPMT-JUNO-XU2026171301}. The signals from the 300 PMTs installed in the three WTs are transmitted to front-end electronics housed in stainless-steel enclosures (UWBs) located in the experimental hall~\cite{JUNO:2026nzc}. Charge and timing information are digitized, processed, and transmitted to the DAQ through the associated global control units (GCUs), which also provide slow control, power regulation, clock distribution, and synchronization. 4 GCUs are used in the WT electronics system and they share the same WR timing system as the CD and TVT.

\subsection{Trigger and DAQ}

The CD trigger system employs a two-tier architecture consisting of six gather units (GUs) and one central unit (CU)~\cite{Wang:2026rsi}. Each GU buffers raw data from around 10 FECs and performs real-time analysis of the timestamps of the hits on the SiPMs. The timing information of all the GUs are sent to the CU, which makes a global trigger decision based on the multiplicity of fired channels in a fixed time window. Once the trigger condition is satisfied, the corresponding data within a configured readout time window, which is slightly larger than the trigger time window, are extracted and sent to the DAQ.

For physics data taking, the CD trigger window is set to 320~ns and the multiplicity threshold to 2000 fired channels, corresponding to an energy of about 0.7~MeV. For calibration runs, the threshold is reduced to 1400 fired channels, corresponding to about 0.5~MeV, while the trigger window remains unchanged. The readout window is set to 1.0~\(\mu\)s for physics runs and 1.44~\(\mu\)s for calibration runs. An external trigger mode is also implemented for LED calibration through the WR timing network.

For the TVT, a hardware trigger requires time-coincident SiPM hits at both ends of a single plastic-scintillator strip, and a higher-level inter-layer coincidence is formed by a software trigger in the DAQ~\cite{Zhang:2024znq}. For the WT, a global trigger is implemented in the DAQ system. Each PMT is first self-triggered with a configurable threshold, and when the number of fired PMT channels within a predefined time window of 900~ns exceeds the global trigger threshold of 10, all PMT hits are read out to form a triggered event.

The post-trigger data bandwidths are 328~Mb/s, 284~Mb/s, and 4.8~Mb/s for the CD, WT, and TVT, respectively. To keep the data-storage bandwidth below the design target of 150~Mb/s, the DAQ applies both lossless and lossy online compression~\cite{Zhang:2024znq}. Lossless compression is applied to the charge/time hit streams of the CD, WT, and TVT. Lossy compression is applied only to CD muons and their secondaries within a configurable time window of 10~\(\mu\mathrm{s}\). In this mode, only event-level quantities are stored, including the total PE count and the event vertex reconstructed using the charge-center method. The resulting data-storage rate is about 120~Mb/s.

\subsection{Offline System}

The TAO offline software system (TAOSW) is used for processing both experimental and simulated data. It is built on the SNiPER framework~\cite{Zou:2015ioy} and reuses common software packages from the JUNO offline system. TAOSW includes modules for simulation, calibration, and reconstruction, together with common services for the event data model (EDM)~\cite{Li:2017zku}, database access, detector geometry, and visualization~\cite{Liao:2024bso}. A Geant4-based detector simulation package has also been developed within TAOSW. It shares common components with the JUNO detector simulation~\cite{Lin:2022htc}, including the event generator, physics list, and external libraries. The TAO detector geometry has been implemented in detail in the simulation. In addition, a dedicated electronics simulation package has been developed to account for the effects of the SiPMs and the electronics system.

The raw data are transferred from the DAQ servers to the IHEP data center, converted into ROOT~\cite{Brun:1997pa} format, and stored in the EDM as RTRaw files. These files are then replicated to other Tier-1 data centers through the distributed computing infrastructure. A Keep-Up Production (KUP) pipeline performs channel-level calibration and reconstructs the basic event quantities, including the number of PEs and hit time for each fired channel, as well as the event-level energy and vertex. The reconstructed information is stored in event summary data (ESD) files. A dedicated physics production is carried out separately using updated calibration constants and optimized reconstruction algorithms, generating the final ESD files that serve as the primary input to subsequent physics analyses.

\subsection{Data Quality Monitoring}
\label{sec:dataquality}

A dedicated data quality monitoring (DQM) system has been established for TAO on the basis of the JUNO DQM framework. The system monitors reconstructed data from the keep-up production (KUP) pipeline and provides feedback on run conditions and detector performance with a latency of 1--2 hours. The KUP pipeline runs on the computing cluster of the IHEP Computing Center in China. Dedicated web interfaces provide run information, reconstruction results, and long-term monitoring plots for each subdetector system.

\begin{figure}[!htbp]
    \centering
    \includegraphics[width=0.9\linewidth]{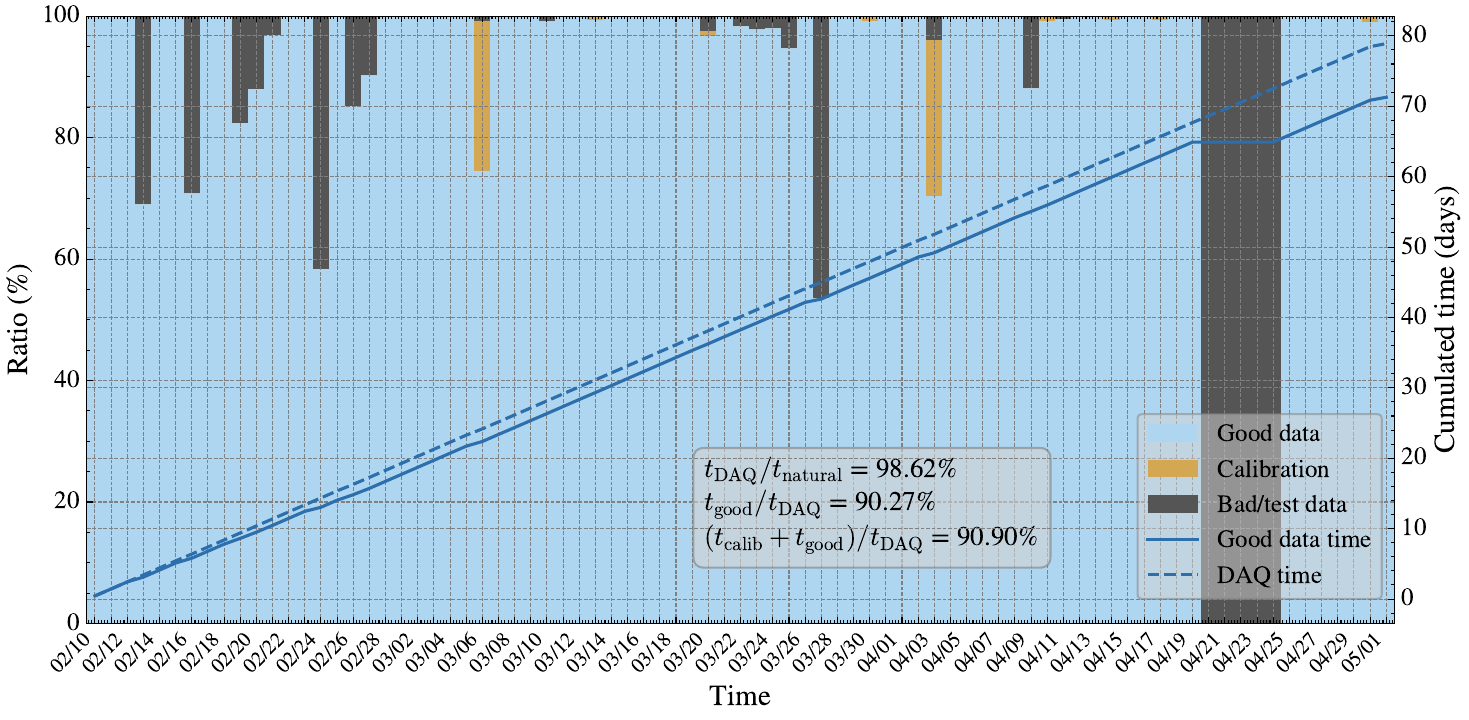}
    \caption{DAQ livetime fraction of TAO data taking as a function of time.}
    \label{fig:tao_dataquality}
\end{figure}

The DQM system performs automated monitoring of detector stability at both the channel and event levels. The monitored quantities include occupancy, time offset, gain, dark-count rate (DCR), energy distribution, vertex distribution, total-PE distribution, and muon/singles rates. For each SiPM, the occupancy in a given data file is defined as the number of signals recorded by that SiPM divided by the total number of events in the same file. Based on these monitoring quantities, lists of good files and good channels are produced for subsequent physics data production and analysis. Fig.~\ref{fig:tao_dataquality} shows the DAQ livetime fraction for the data period used in the analysis of the first TAO neutrino spectrum. The DAQ livetime reaches up to 98.62\%, and 90.27\% of the recorded data are classified as good physics data and used in the analysis. Periods of bad data are mainly associated with detector maintenance.

%% file: sec-performance-v2.tex
\subsection{Waveform Reconstruction in FPGA}

The digitized waveforms from 128 readout channels are processed by a single FPGA on each FEC. For a typical single-photon waveform, the rise and fall times are approximately 100~ns and 300~ns, respectively. The waveform amplitude is sampled continuously with a 125~MHz frequency and stored in on-board registers accessible to the DAQ.

\begin{figure}[!htbp]
    \centering
    \includegraphics[width=0.9\linewidth]{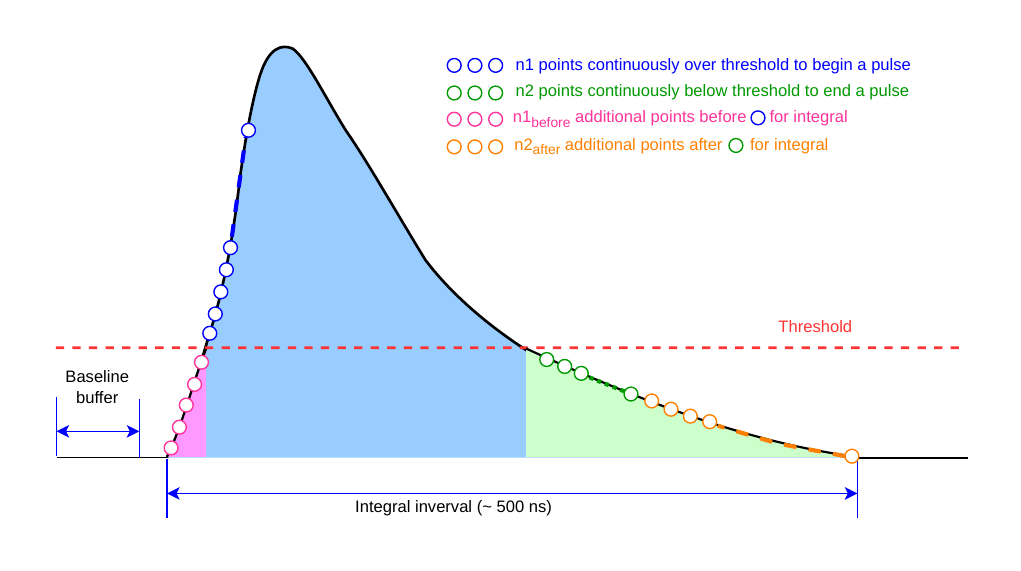}
    \caption{Schematic illustration of the COTI algorithm, showing the main configurable parameters used for pulse identification and charge integration.}
    \label{fig:tao_coti}
\end{figure}

Real-time waveform processing is performed using the COTI algorithm, as illustrated in Fig.~\ref{fig:tao_coti}. The main steps are summarized as follows:
\begin{itemize}
    \item The baseline for each channel is determined in two ways. A fixed baseline is obtained by averaging waveform samples read from the FEC registers when the SiPM bias voltage is turned off. During physics data taking, a dynamic baseline is also evaluated for each pulse using a buffer of 64 samples immediately preceding the first over-threshold point, with the average of the first 32 samples in the buffer taken as the dynamic baseline. This dynamic baseline improves charge reconstruction after high-energy events, such as muons, by accounting for waveform overshoot. In practice, the fixed baseline is used by default in the physics analysis, since it provides better charge resolution than the dynamic baseline. A configurable threshold, defined relative to the baseline, is used for pulse identification.
    \item A signal pulse is identified when \(n1\) consecutive sampling points, spaced by 8~ns, exceed the configurable threshold. The timestamp of the first point above threshold in this sequence is recorded as the pulse time. The end of the pulse is defined when \(n2\) consecutive sampling points fall below the threshold. The FEC counts such pulses for each channel and stores the resulting pulse rate in registers for DAQ readout and detector monitoring.
    \item For charge reconstruction, the integration window starts at the \(n1_{\mathrm{before}}\)-th sample before the pulse start point and ends at the \(n2_{\mathrm{after}}\)-th sample after the pulse end point. These pre-pulse and post-pulse extensions are introduced to recover sub-threshold charge components and to reduce possible channel-level non-linearity. If another above-threshold sample is encountered, the extension is truncated, and the actual integration window may therefore be shorter than the nominal \(n1_{\mathrm{before}}\) or \(n2_{\mathrm{after}}\) samples. The baseline is subtracted from each sample within the integration window to obtain the pulse charge.
    \item For each pulse, the reconstructed charge and time, together with the pulse width (defined as the number of sampling points in the integration window), the dynamic baseline, and two overflow flags for amplitude and charge, are packaged into the output raw-data stream.
\end{itemize}

The COTI parameters were optimized through offline analysis of waveform data with the goals of improving charge resolution of signal pulse, reducing charge non-linearity, and suppressing the fake hit generated by electronic-noise fluctuation. The optimized parameter set consists of a threshold of 70 ADC counts (corresponding to about 0.37 PEs in amplitude), \(n1 = 15\), \(n1_{\mathrm{before}} = 5\), \(n2 = 8\), and \(n2_{\mathrm{after}} = 11\). This configuration has been used since the start of physics data taking.

\subsection{Electronics Noise Level}

\begin{figure}[!htbp]
  \centering
  \begin{subfigure}[b]{0.9\linewidth}
    \centering
    \includegraphics[width=\linewidth]{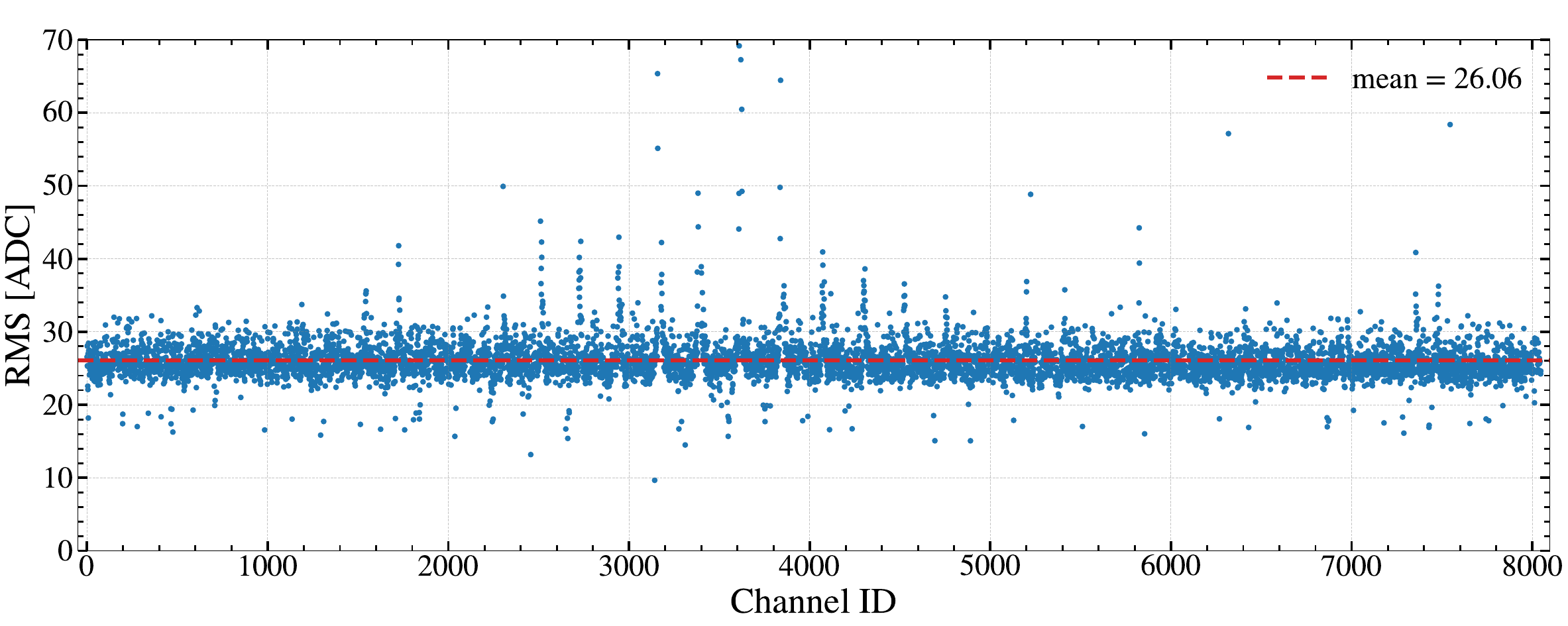}
    \caption{}
    \label{fig:pedestal_rms}
  \end{subfigure}\hfill
  \begin{subfigure}[b]{0.9\linewidth}
    \centering
    \includegraphics[width=\linewidth]{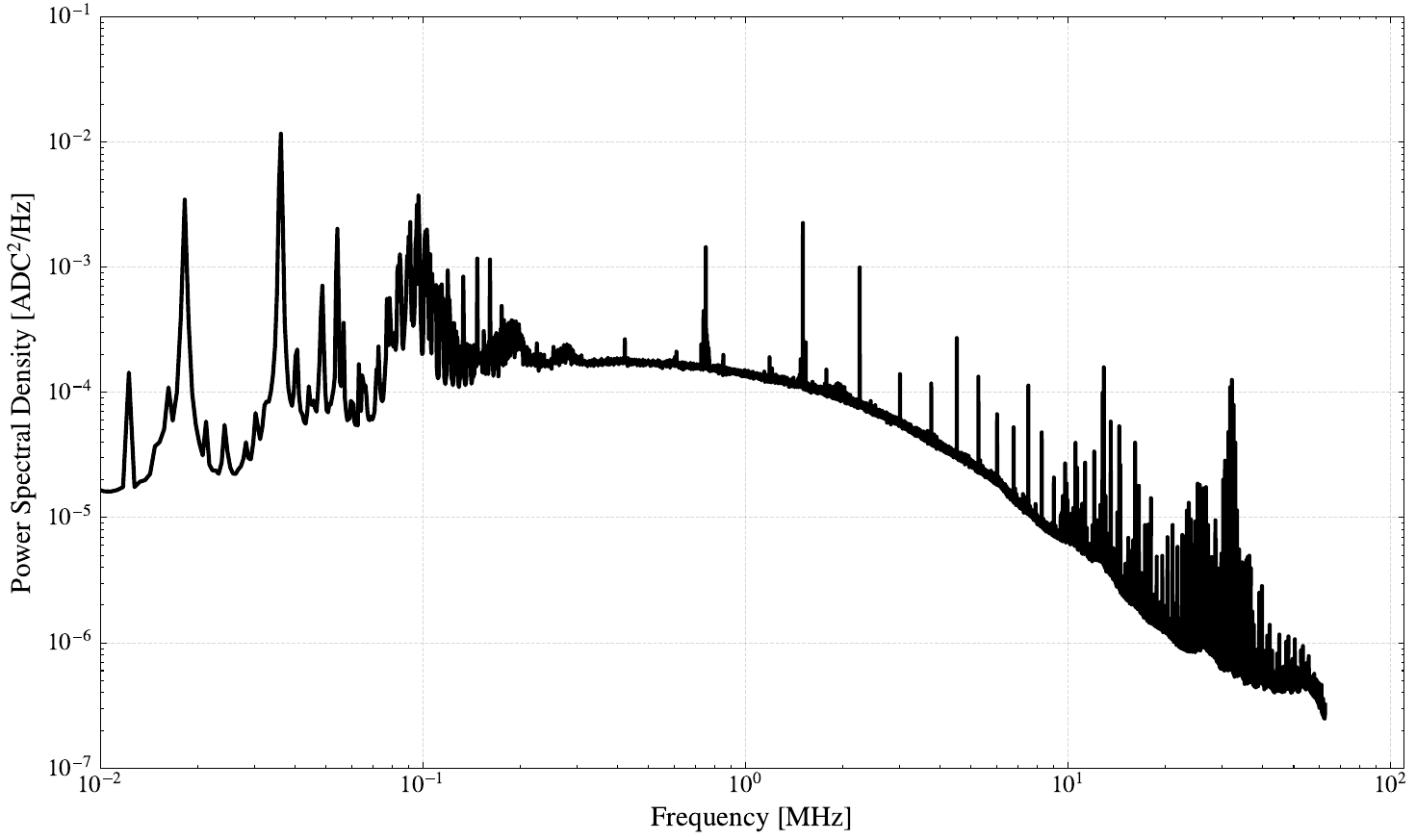}
    \caption{}
    \label{fig:baseline_noise_fft}
  \end{subfigure}
  \caption{(a) Pedestal RMS as a function of channel ID. The red line indicates the average pedestal RMS of 26 ADC counts over all channels. (b) Noise power spectral density obtained from a fast Fourier transform of recorded baseline waveforms.}
  \label{fig:electronics_noise}
\end{figure}

The electronic noise of each channel is characterized by the root-mean-square (RMS) of the pedestal. The pedestal is defined as the signal level in the absence of any pulse and is obtained from waveform samples stored in the FEC registers when the SiPM bias voltage is turned off. As shown in Fig.~\ref{fig:pedestal_rms}, the pedestal RMS as a function of channel ID has an average value of about 26 ADC counts. For SiPMs operated at the nominal overvoltage of 3.5~V, this corresponds to an equivalent noise amplitude of about 0.13~PEs. The observed noise level is slightly above the design requirement of 0.1~PEs specified in the TAO CDR~\cite{JUNO:2020ijm}. Channels with large RMS values (\(>35\) ADC counts) are found to cluster around a small number of HV cables.

To further investigate the noise characteristics, the noise power spectral density (PSD) is obtained by applying a fast Fourier transform (FFT) to recorded baseline waveforms for a small fraction of channels with the SiPM bias voltage turned off. The representative spectrum shown in Fig.~\ref{fig:baseline_noise_fft} indicates that both low- and high-frequency components contribute to the overall noise. For comparison, the signal frequency content is mainly concentrated in the range of 1--10~MHz.

\begin{figure}[!htbp]
  \centering
  \begin{subfigure}[b]{0.5\linewidth}
    \centering
    \includegraphics[width=\linewidth, height=7cm, keepaspectratio]{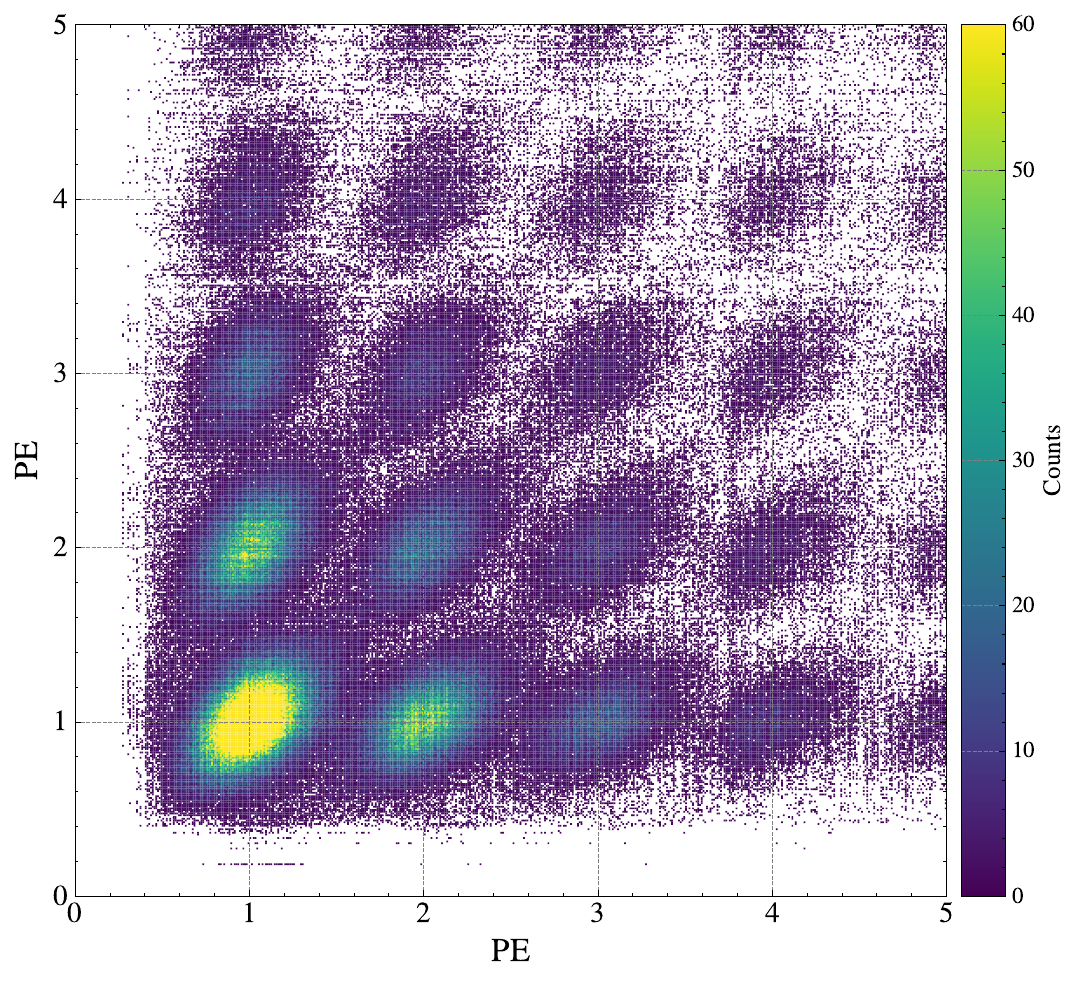}
    \caption{}
    \label{fig:corr_noise_2D}
  \end{subfigure}\hfill
  \begin{subfigure}[b]{0.5\linewidth}
    \centering
    \includegraphics[width=\linewidth, height=7cm, keepaspectratio]{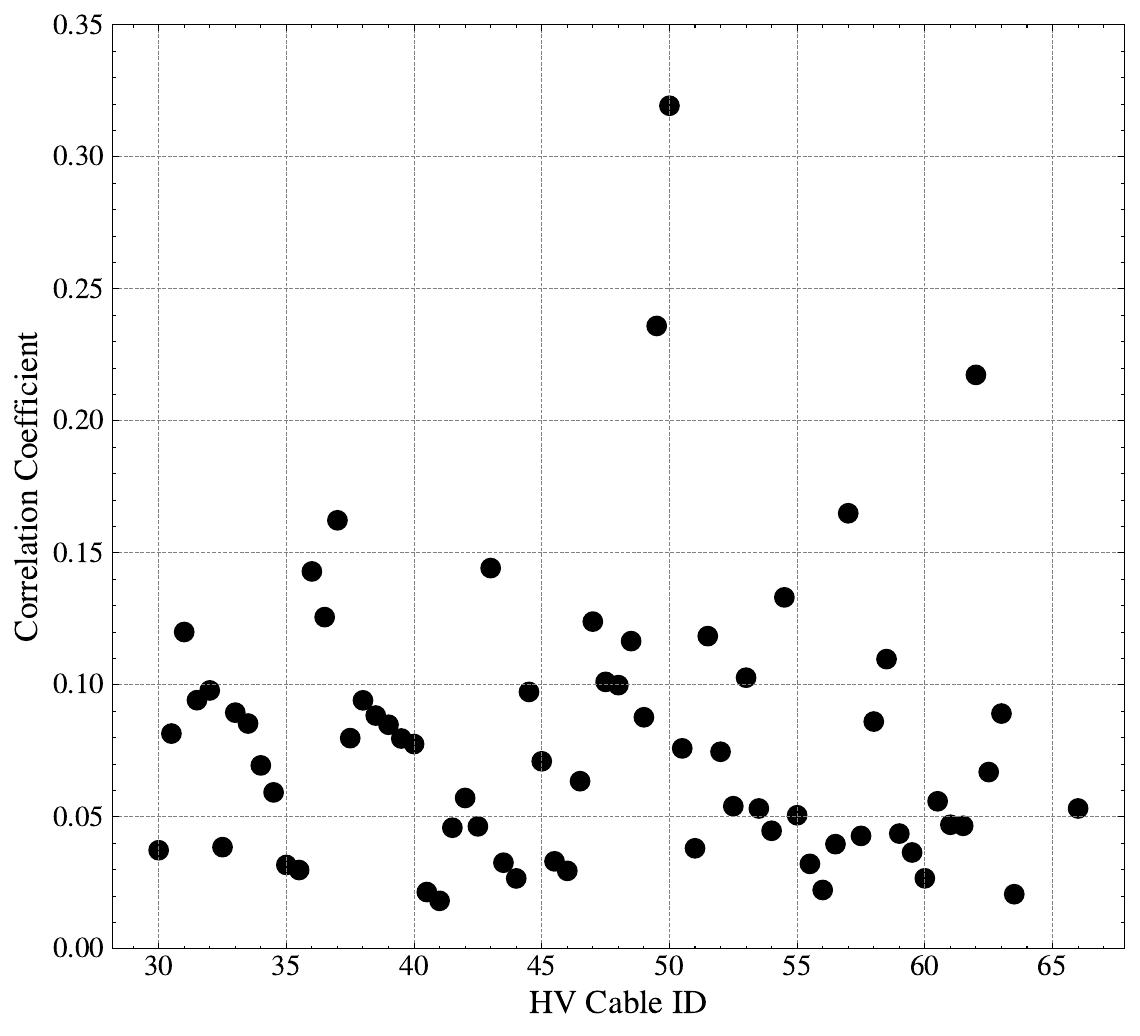}
    \caption{}
    \label{fig:corr_noise_vs_hvcable}
  \end{subfigure}
  \caption{(a) Charge correlation between channels connected to external HV cable ID 50, with a correlation coefficient of 0.32. (b) Pearson correlation coefficient among channels associated with the same HV cable as a function of HV cable ID.}
  \label{fig:electronics_corr_noise}
\end{figure}

Correlated noise is also observed in the electronics system. As shown in Fig.~\ref{fig:corr_noise_2D}, a clear correlation is present in the charge measurements between the two groups of channels sharing the same external HV cable. Detailed studies indicate that the correlated noise originates mainly from a subset of HV cables rather than from the low-voltage or signal cables. The Pearson correlation coefficients among channels associated with the same HV cable are shown as a function of HV cable ID in Fig.~\ref{fig:corr_noise_vs_hvcable}. These coefficients are significantly larger than those observed for the low-voltage and signal cables, for which the correlation coefficients are below 0.05. The charge-correlation example shown in Fig.~\ref{fig:corr_noise_2D} corresponds to HV cable ID 50, which exhibits the largest correlation coefficient among all HV cables.

Two major sources have been identified for the correlated noise: the cryogenic system used for detector cooling and the air-conditioning system used to stabilize the temperature of the electronics cabinets. This correlated noise can degrade the detector energy resolution when the energy estimator is based on PE counts with decimal precision. However, owing to the excellent single-PE charge resolution of the SiPMs, this impact is largely suppressed when integer PE counts are used in the energy reconstruction.

To mitigate the noise level, in particular the correlated noise, an upgrade is planned during the next detector maintenance period this year. The upgrade includes improvements to the grounding and shielding of the electronics system to suppress low-frequency noise induced by the cryogenic and air-conditioning systems. In addition, a filter system will be included as well to attenuate high-frequency noise components.

\subsection{CD Operation Status}

\subsubsection{Stability of the CD temperature}

\begin{figure}[!htbp]
    \centering
    \includegraphics[width=0.9\linewidth]{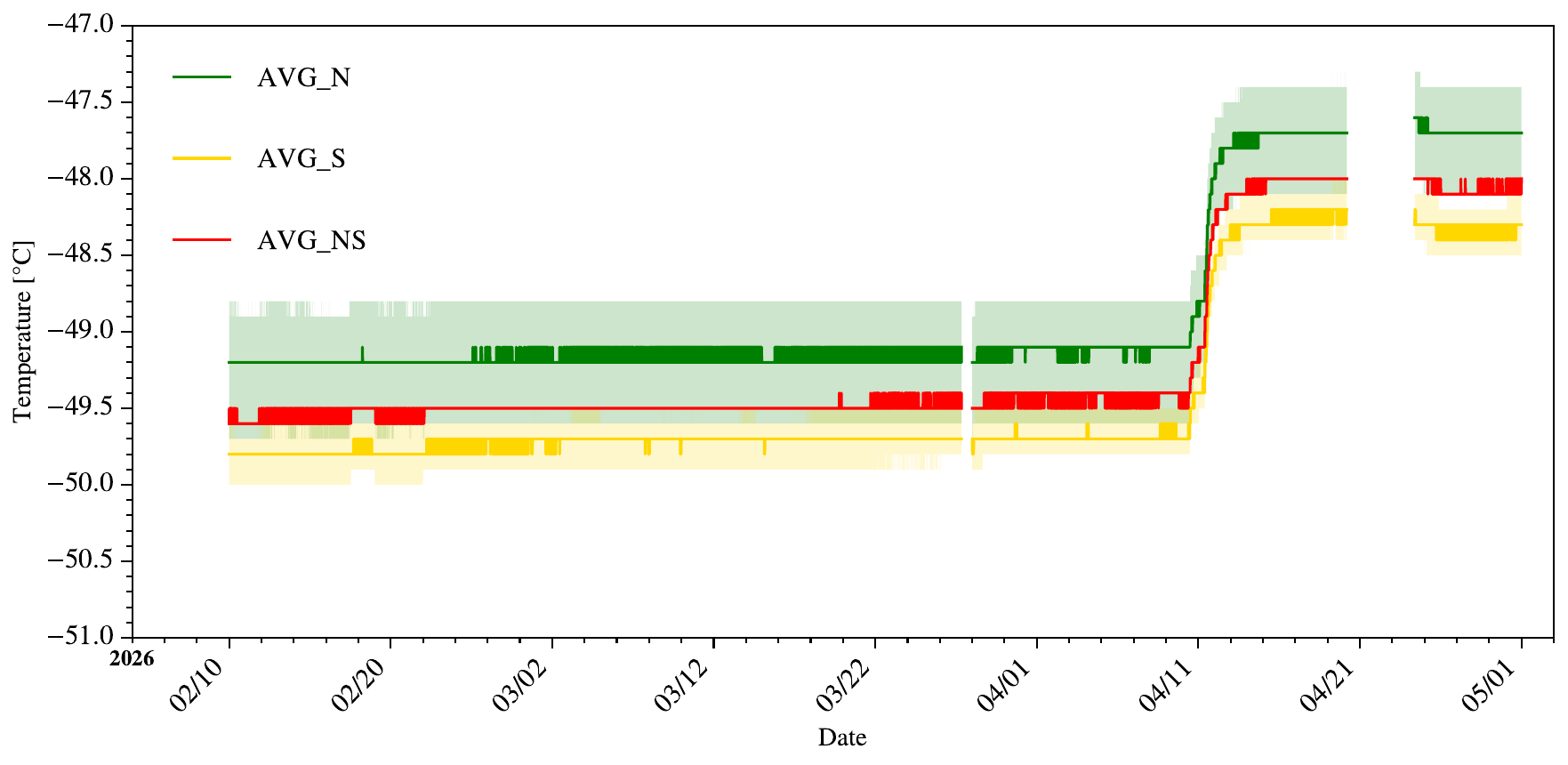}
    \caption{Average temperature of the CD in the upper hemisphere (green), lower hemisphere (orange), and full detector volume (red) since the start of physics data taking. The bands indicate the temperature spread measured by sensors located at different positions on the upper and lower copper hemispheres. The temperature step on 10 April 2026 corresponds to an intentional increase of the operating set point by \(1.5\,^{\circ}\mathrm{C}\).}
    \label{fig:cd_temperature_stability}
\end{figure}

The temperature stability of the CD is critical for stable detector operation, since the breakdown voltage of the SiPMs depends on the operating temperature. Fig.~\ref{fig:cd_temperature_stability} shows the average CD temperature in the north hemisphere (green line), south hemisphere (orange line), and full detector volume (red line) since the start of physics data taking. The bands indicate the temperature spread measured by sensors located at different positions on the north and south copper hemispheres, corresponding to a temperature uniformity within \(\pm 0.5\,^{\circ}\mathrm{C}\) inside the CD.

During the first week of physics data taking, the CD temperature increased by about \(0.1\,^{\circ}\mathrm{C}\). It then remained stable until 10 April 2026, with overall variations of less than \(0.2\,^{\circ}\mathrm{C}\) around the nominal operating value. On 10 April 2026, the operating temperature was intentionally increased by \(1.5\,^{\circ}\mathrm{C}\) with respect to the original nominal value because the redundant cooling capacity of the cryogenic system had decreased. On 21 April 2026, the cryogenic system was shut down because of an issue with the programmable logic controller and relay. After the system was recovered, the detector temperature returned to a stable state.

During physics data taking, both the temperature uniformity and stability satisfy the design requirements specified in the CDR~\cite{JUNO:2020ijm}, demonstrating the reliable performance of the cryogenic system during detector operation.

\subsubsection{Good channel list}

\begin{figure}[!htbp]
\centering
\includegraphics[width=0.9\linewidth]{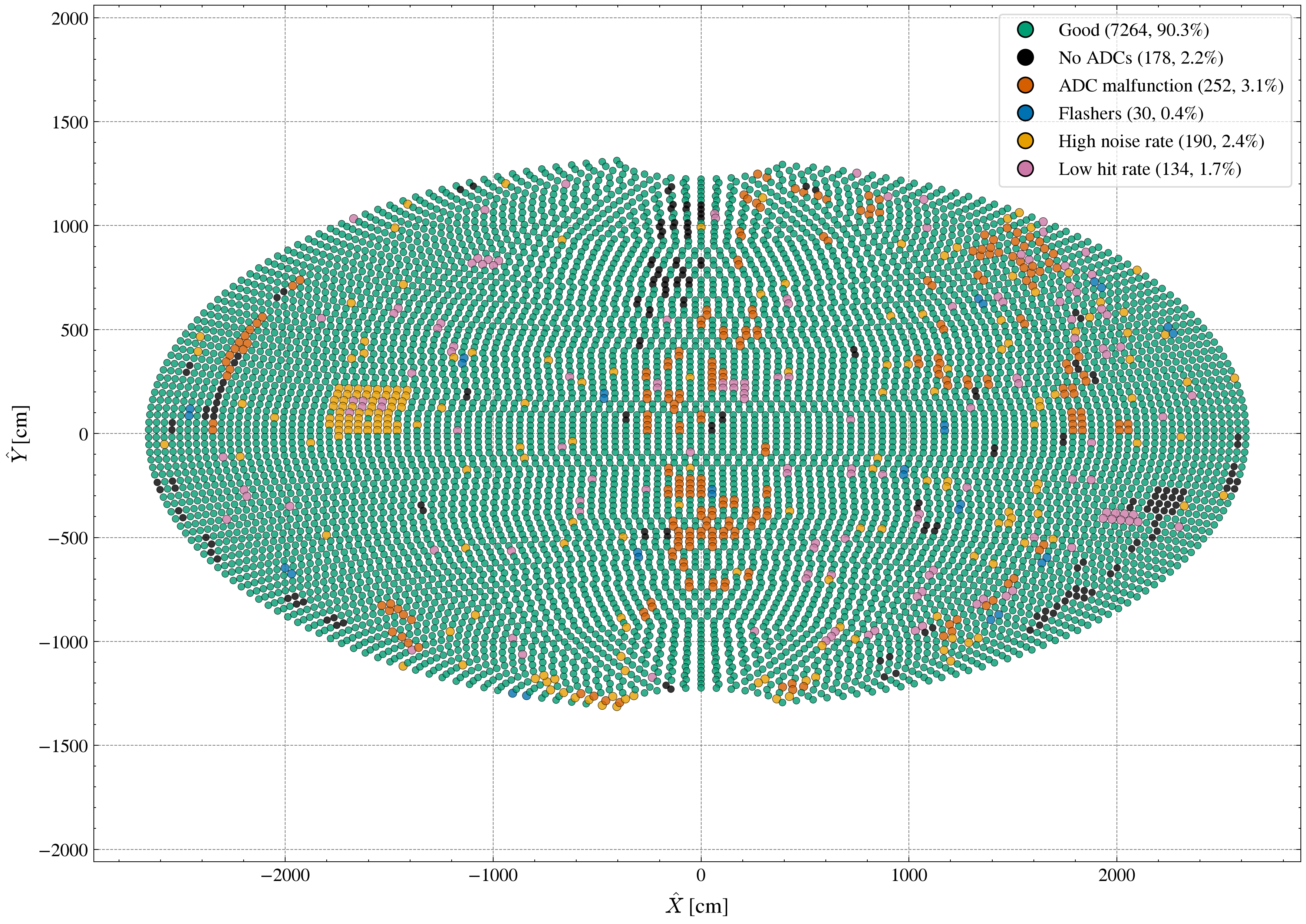}
\caption{Distribution and categorization of excluded channels in the central detector. The channel positions are shown, and the colors indicate the fault categories described in the text.}
\label{fig:bad_channel_list}
\end{figure}

The CD contains 8,048 readout channels in total. A subset of these channels is excluded from data taking or subsequent analyses because of channel-performance issues, while another subset is excluded because of missing or malfunctioning back-end electronics. These excluded channels are categorized as follows and mapped in Fig.~\ref{fig:bad_channel_list}:
\begin{itemize}
    \item \textbf{No ADC connection:} 178 channels are absent from the data stream because of unavailable ADC cards.
    \item \textbf{ADC malfunction:} 252 channels are disabled during data taking because of unstable ADC chip operation. Among them, 14 exhibit anomalous timing behavior.
    \item \textbf{Flashers:} 30 channels are identified as flashers, emitting excessive light and thereby causing a significant increase in the hit rates of other channels.
    \item \textbf{High noise rate:} 190 channels show elevated noise levels. Channels with noise-induced hit rates above 100~kHz are disabled during operation. In addition, the elevated noise prevents accurate gain calibration for some affected channels, which are therefore excluded from subsequent data analysis.
    \item \textbf{Low hit rate:} 134 channels exhibit abnormally low hit rates because of connection losses in the internal feedthrough PCBs or loss of HV/LV connections on the FEBs.
\end{itemize}

After excluding the channels listed above, 7,264 (90.3\%) of the 8,048 channels are retained in the good-channel list used for the first TAO neutrino-spectrum measurement. The number of retained channels varies by fewer than 10 from run to run, mainly because gain fitting occasionally fails for a small number of channels.

New ADC cards have been manufactured and are scheduled for installation during the next detector maintenance period. In addition, the planned electronics upgrade is expected to mitigate a significant fraction of the high-noise channel issues. The stability of the gain-fitting procedure will be further improved.

\subsubsection{Bias voltage of the SiPMs}

The SiPM bias voltages were determined in two steps to ensure optimal detector performance. First, a uniform overvoltage was established across all channels. The overvoltage is defined as the difference between the applied bias voltage and the SiPM breakdown voltage. The breakdown voltage of each SiPM tile was determined from pre-deployment mass-testing results obtained in the laboratory~\cite{Rybnikov_2024}. Based on these measurements, the initial bias voltages were set to provide a common overvoltage for all channels. The SiPM gains were then measured channel by channel to validate the settings, and the bias voltages of channels with outlying gain values were further adjusted to improve gain uniformity.

\begin{figure}[!htbp]
    \centering
    \includegraphics[width=0.8\linewidth]{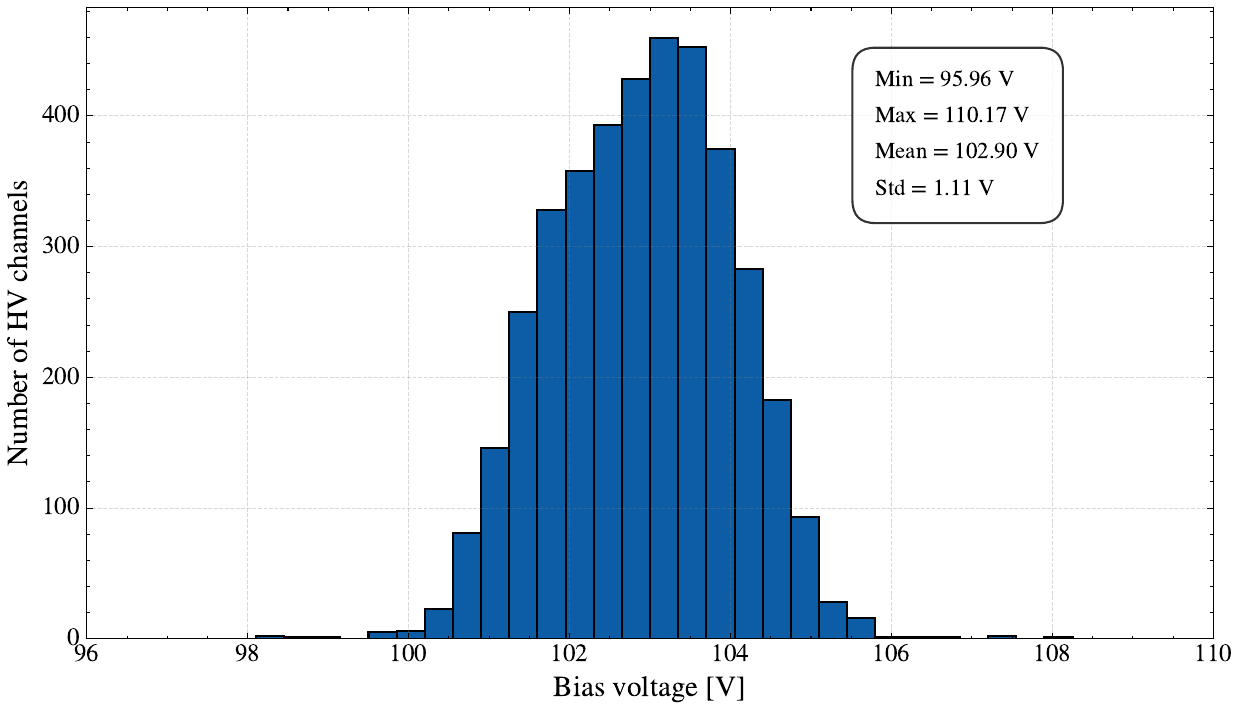}
    \caption{Distribution of the SiPM bias voltages applied during physics data taking. The bias voltage of each channel is set to achieve a uniform overvoltage of 3.5~V, which was determined to be optimal from the energy resolution of the 511-keV gamma rays in \(^{68}\)Ge calibration data.}
    \label{fig:bias_volt_sipms}
\end{figure}

Second, the optimal overvoltage was determined on the basis of the overall detector performance. Since key SiPM characteristics, including gain, dark-count rate (DCR), photon-detection efficiency, and optical crosstalk probability, depend strongly on the overvoltage and affect detector performance in different ways, this parameter was optimized using calibration data. A series of calibration runs with a \(^{68}\)Ge source deployed at the detector center was performed for overvoltages between 2.5~V and 5.0~V in steps of 0.5~V. At each overvoltage, the energy resolution for the two 511-keV gamma rays from \(^{68}\)Ge was evaluated. The best energy resolution was obtained at 3.5~V, which was therefore chosen as the operating overvoltage. The resulting distribution of SiPM bias voltages used during physics data taking is shown in Fig.~\ref{fig:bias_volt_sipms}.

\subsection{Channel-level Performance Results}

This section presents the channel-level calibration methods and performance results for the SiPMs in CD. These calibrations provide the basis for precise energy and vertex reconstruction, and their performance is critical for achieving TAO's target energy resolution.

\subsubsection{Gain of the SiPMs}

To extract the gain of each channel, the measured ADC spectrum is fitted with a multi-component model that includes the pedestal, the single PE (SPE) peak, and higher-order PE peaks with contributions from after-pulses. The complete fit function \(F(x)\) is defined as
\begin{equation}
\label{eq:fit_model}
\begin{aligned}
F(x) &= A_1 \cdot \erf(x; \mu_0, \sigma_0) \cdot \sum_{i=0}^{1} G(i; 1, \alpha) \cdot N(x; 1, i, \sigma_1) \\
     &\quad + \sum_{n=2}^{N} A_n \sum_{i=0}^{n} G(i; n, \alpha) \cdot N(x; n, i, \sigma_n) \, ,
\end{aligned}
\end{equation}
where \(A_n\) is the normalization constant. The first term describes the tail of the pedestal and SPE region, while the second term models the multi-PE peaks. The error function \(\erf(x; \mu_0, \sigma_0)\) parameterizes the transition from the pedestal to the first PE peak, centered at the pedestal mean \(\mu_0\) with width \(\sigma_0\). 
For events with \(n\) initially detected photo-electrons, after-pulses can produce \(i\) additional charge contributions, each carrying an effective charge \(Q_{\rm ap}\). The probability of observing \(i\) after-pulses is modeled by
\begin{equation}
\label{eq:after-pulse_prob}
G(i; n, \alpha) = (1 - n\alpha)(n\alpha)^i \, , \qquad n\alpha < 1 \, ,
\end{equation}
where \(\alpha\) is the after-pulse probability per SPE. The condition \(n\alpha < 1\) ensures proper normalization of the distribution. Each combination of \(n\) photo-electrons and \(i\) after-pulse contributions is represented by a Gaussian peak,
\begin{equation}
\label{eq:gaussian_peak}
N(x; n, i, \sigma_n) =
\frac{1}{\sqrt{2\pi}\,\sigma_n}
\exp\!\left[
-\frac{\Bigl( x - \bigl( \mu_1 + (n-1)\cdot {\rm Gain} + i\cdot Q_{\rm ap} \bigr) \Bigr)^2}
{2\sigma_n^2}
\right] \, ,
\end{equation}
where \(\mu_1\) is the position of the SPE peak, \({\rm Gain}\) is the SiPM gain in ADC units per photo-electron, and \(Q_{\rm ap}\) is the effective after-pulse charge. The width \(\sigma_n\) describes the resolution of the \(n\)-photo-electron peak and generally increases with \(n\). The primary parameter of interest is the SiPM gain, extracted from the spacing between successive photo-electron peaks, while the after-pulse probability \(\alpha\) and the effective after-pulse charge \(Q_{\rm ap}\) are determined simultaneously from the amplitudes and positions of the satellite peaks.

\begin{figure}[!htbp]
    \centering
    \includegraphics[width=0.9\linewidth]{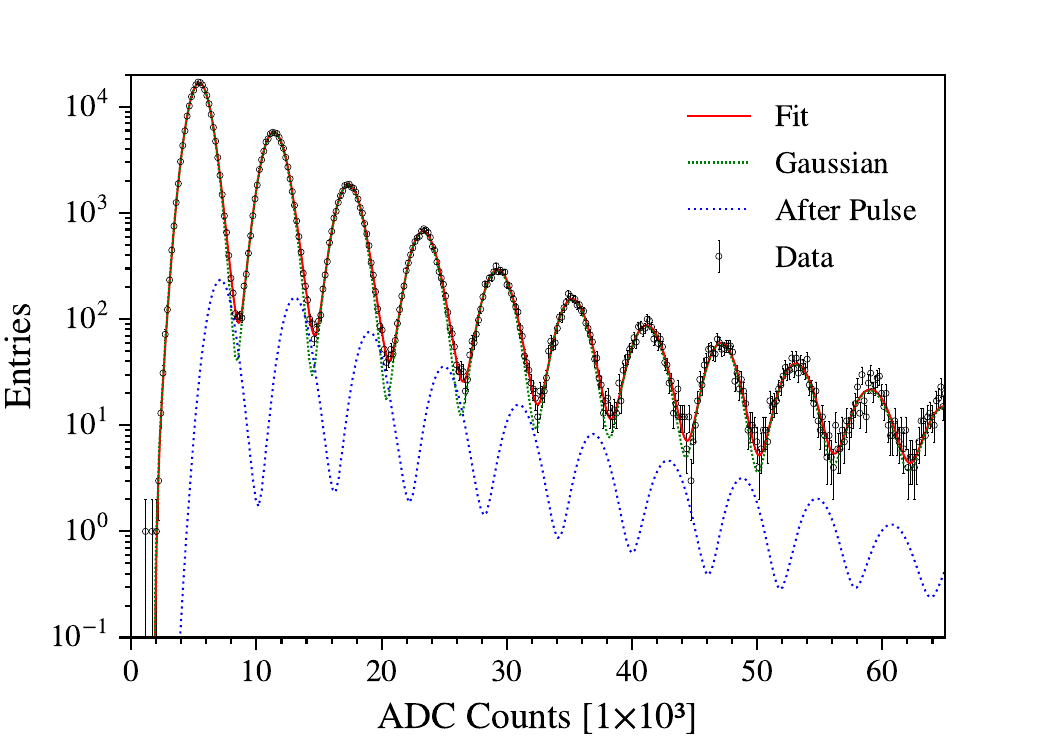}
    \caption{Example of the gain fit for a single SiPM channel. The ADC spectrum (black points) is described by a fit (red curve) with photo-electron (green dashed) and after-pulse (blue dashed) components indicated. The SiPM gain is extracted from the spacing between successive photo-electron peaks.}
    \label{fig:gain_fit}
\end{figure}

\begin{figure}[!htbp]
  \centering
  \begin{subfigure}[b]{0.49\linewidth}
    \centering
    \includegraphics[width=\linewidth]{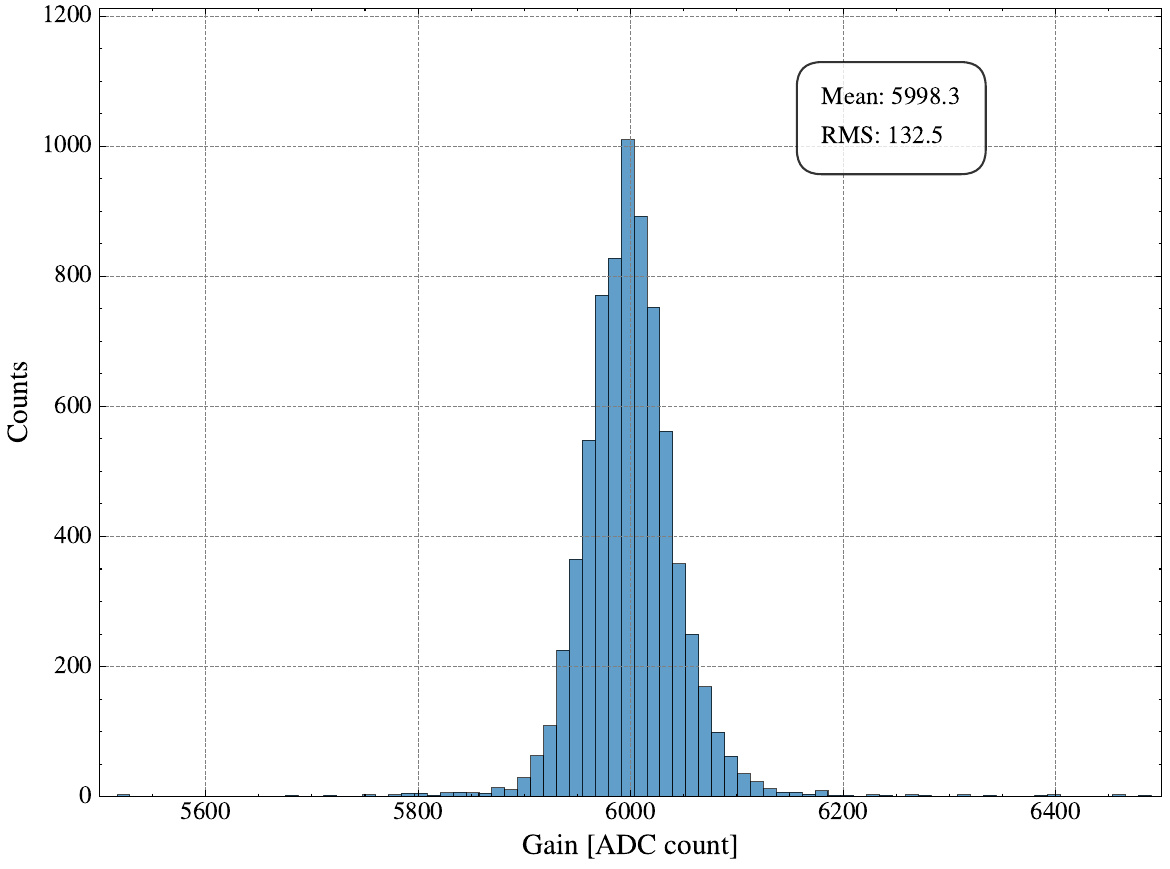}
    \caption{}
    \label{fig:gain_results}
  \end{subfigure}\hfill
  \begin{subfigure}[b]{0.49\linewidth}
    \centering
    \includegraphics[width=\linewidth]{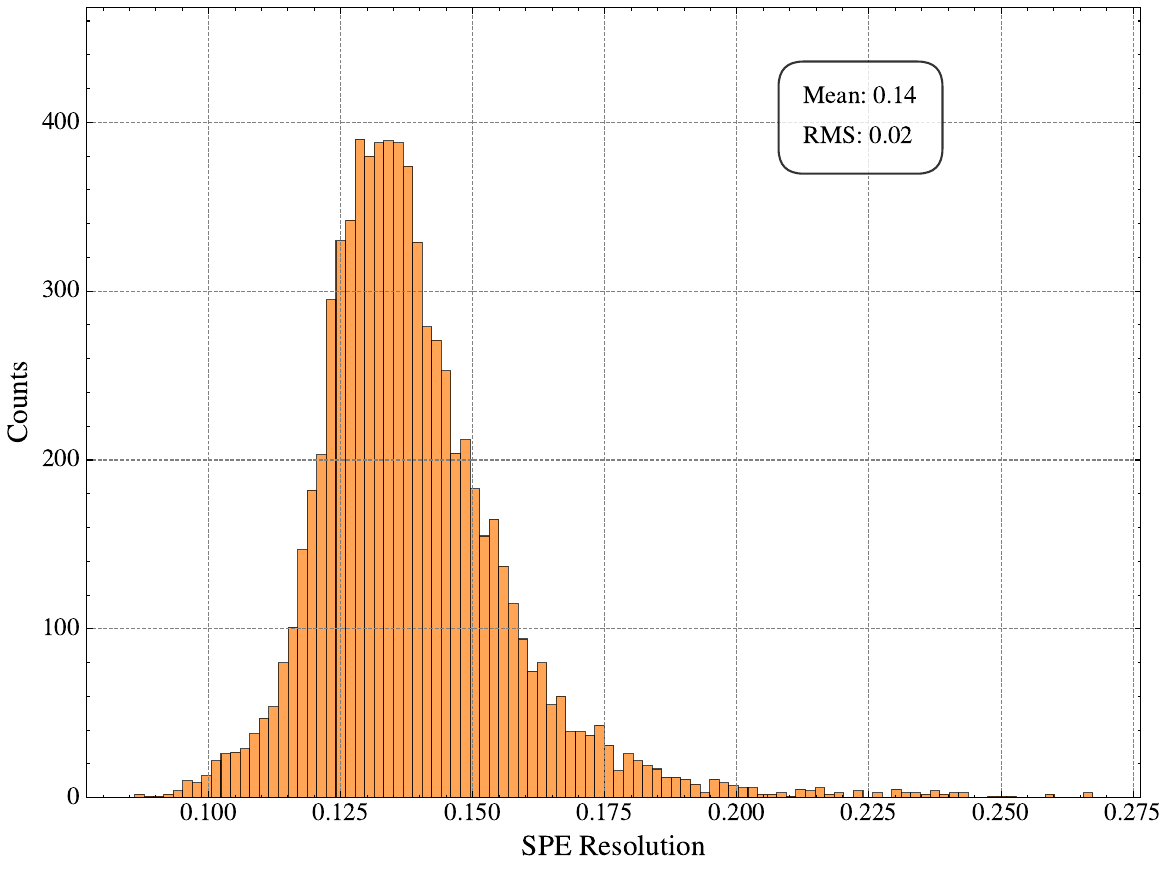}
    \caption{}
    \label{fig:spe_results}
  \end{subfigure}
  \caption{(a) Distribution of the SiPM gain for all good channels in the first physics run (RUN1382). (b) Distribution of the SPE resolution for all good channels in RUN1382, defined as the ratio \(\sigma_1/{\rm Gain}\).}
  \label{fig:cd_evt_rates}
\end{figure}

The gain calibration is performed on a run-by-run basis using the ADC-spectrum fitting procedure described above. A typical fit result is shown in Fig.~\ref{fig:gain_fit}, together with the photo-electron (green dashed line) and after-pulse (blue dashed line) components. For each detector channel and each data-taking run, the gain is extracted independently, allowing the temporal evolution of the SiPM response to be monitored throughout the full data-taking period. The gain values obtained from dedicated \(^{68}\)Ge calibration runs are consistent with those obtained from physics runs. The gain distribution for all good channels in the first physics run (RUN1382) is shown in Fig.~\ref{fig:gain_results}. Cross-checks between independent analysis groups show excellent agreement, confirming the robustness and reproducibility of the fitting procedure.

The SPE resolution is a key performance metric, defined as \(\sigma_1/{\rm Gain}\). This quantity characterizes the ability of the detector to resolve individual photo-electrons and therefore affects the precision of the energy reconstruction. The SPE resolution is extracted for each channel from the ADC-spectrum fit. The measured SPE resolution for all good channels in RUN1382 is shown in Fig.~\ref{fig:spe_results}, with an average value of 13.9\% and good consistency across different runs, satisfying the TAO requirement of less than 15\% specified in the CDR~\cite{JUNO:2020ijm}.

\begin{figure}[!htbp]
    \centering
    \includegraphics[width=0.9\linewidth]{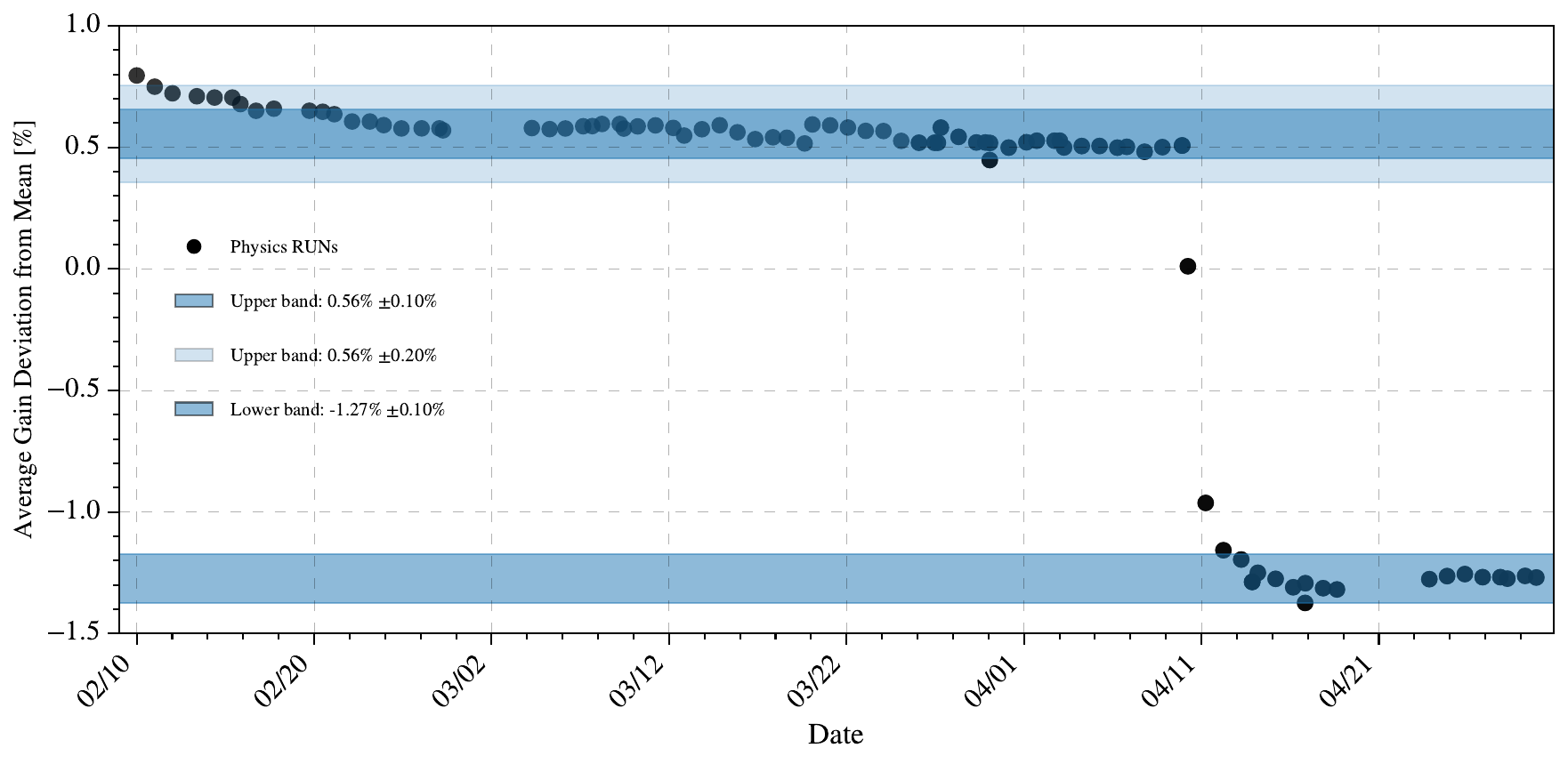}
    \caption{The relative variation of the average gain, with respect to the average value over the full period, as a function of time since the start of physics data taking. The reduction near 10 April 2026 is attributed to a deliberate \(1.5\,^{\circ}\mathrm{C}\) increase in the detector operating temperature.}
    \label{fig:gain_stability}
\end{figure}

The long-term gain stability is evaluated using the run-by-run gain calibration. The relative variation of the average gain, with respect to the average value over the full period, is shown in Fig.~\ref{fig:gain_stability}. Before 10 April 2026, the relative gain variation was better than 0.2\%. At the beginning of the physics data taking, a small gain decrease of about 0.1\% was observed, which was mainly attributed to a slight increase in detector temperature during that period. Around 10 April 2026, the gain decreased by about 1.9\%, corresponding to the intentional increase of the detector operating temperature by about \(1.5\,^{\circ}\mathrm{C}\). As indicated in Fig.~\ref{fig:cd_temperature_stability}, several days were required for the detector to reach thermal equilibrium after this temperature adjustment. Once the detector temperature stabilized, the gain variation remained within 0.1\%. After applying the run-by-run gain calibration, the residual impact of gain drifts on physics data has been further reduced.

\subsubsection{Time Offset}

The time offset between channels arises primarily from two sources: fixed differences in cable lengths, which remain constant over time, and variable latency introduced by the ADC chips. The latter changes after ADC reconfiguration or power cycling because the current operating mode does not provide deterministic latency. Within a single ADC chip, however, all 16 channels typically share the same ADC-induced latency under stable operating conditions.

Time-offset corrections are applied both online and offline. The online correction is implemented in the FEC firmware during waveform reconstruction whenever a change in the time offset is identified, while the offline correction is performed on a run-by-run basis. The corresponding calibration parameters are extracted from selected events with reconstructed energies in the range of 0.7--3~MeV and passing the veto requirements for muon events and consecutive triggers. These events are dominated by radioactive decays with an approximately spherically symmetric spatial distribution, so that their average interaction points are close to the detector center and geometric timing biases are minimized.

\begin{figure}[!htbp]
\centering
\includegraphics[width=0.47\linewidth]{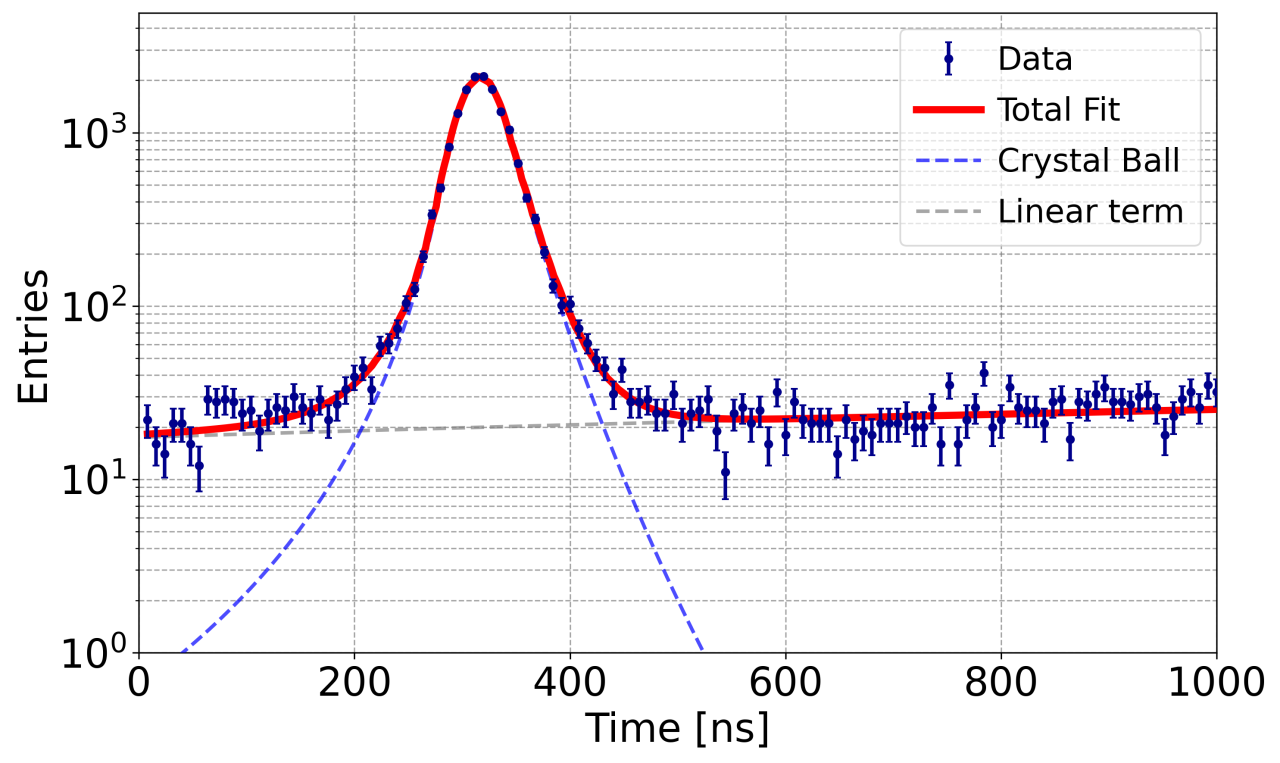}
\hfill
\includegraphics[width=0.47\linewidth]{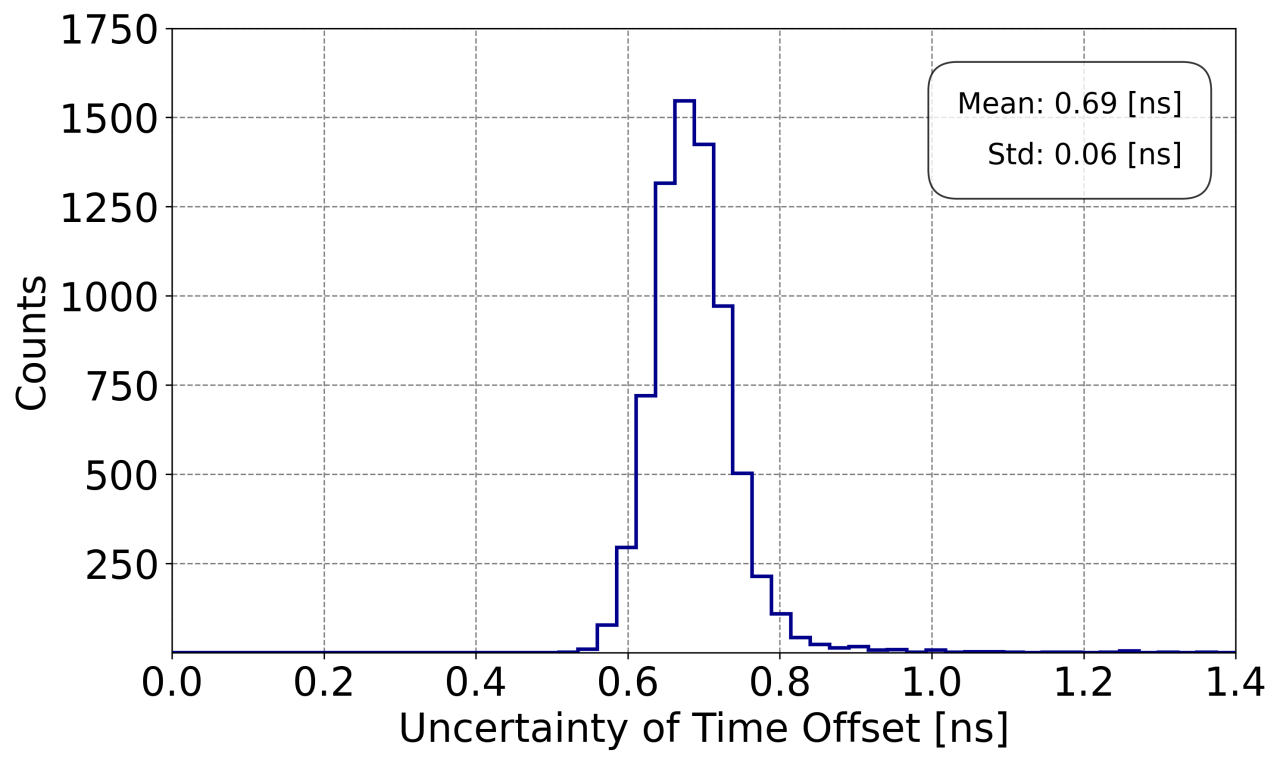}
\caption{Left: Representative hit-time distribution of photons that produce a signal in one channel. The red curve shows the fit with a double-sided Crystal Ball function plus a linear term. Right: Uncertainty of the time offsets for all good channels.}
\label{fig:time_offset}
\end{figure}

The left panel of Fig.~\ref{fig:time_offset} shows a representative hit-time distribution of photons that produce a signal in one channel after event selection. The peak position \(\mu\) is determined by fitting the distribution with a double-sided Crystal Ball function plus a linear term,

\begin{equation}
f(x)=ax+b+N\times
\begin{cases}
A_L \left( B_L + \dfrac{\mu - x}{\sigma} \right)^{-n_L},
& \text{if } \dfrac{\mu - x}{\sigma} \ge \alpha_L, \\[1.2ex]
\exp\!\left( -\dfrac{(x-\mu)^2}{2\sigma^2} \right),
& \text{if } -\alpha_L < \dfrac{x-\mu}{\sigma} < \alpha_R, \\[1.2ex]
A_R \left( B_R + \dfrac{x - \mu}{\sigma} \right)^{-n_R},
& \text{if } \dfrac{x - \mu}{\sigma} \ge \alpha_R,
\end{cases}
\label{eq:double_CB}
\end{equation}

\begin{equation}
A_L = \left( \frac{n_L}{|\alpha_L|} \right)^{n_L}
\exp\!\left( -\frac{\alpha_L^2}{2} \right), \qquad
B_L = \frac{n_L}{|\alpha_L|} - |\alpha_L|,
\label{eq:double_CB_left}
\end{equation}

\begin{equation}
A_R = \left( \frac{n_R}{|\alpha_R|} \right)^{n_R}
\exp\!\left( -\frac{\alpha_R^2}{2} \right), \qquad
B_R = \frac{n_R}{|\alpha_R|} - |\alpha_R|.
\label{eq:double_CB_right}
\end{equation}
where the left and right tail coefficients \(A_L\), \(A_R\), \(B_L\), and \(B_R\) are defined in Eqs.~\eqref{eq:double_CB_left} and \eqref{eq:double_CB_right}, respectively. Here, \(\mu\) represents the peak time, \(\sigma\) the width of the Gaussian core, and \(\alpha_{L,R}\) and \(n_{L,R}\) the tail parameters. The linear term is introduced to describe the background shape.

The uncertainty of the time offset for each channel is determined from repeated measurements of the peak time difference relative to a reference channel (channel 0). As shown in the right plot of Fig.~\ref{fig:time_offset}, the median timing uncertainty over all good channels is 0.69~ns. This precision is mainly limited by the timestamp assignment in the waveform reconstruction and by electronic noise. The extracted time offsets are validated using weekly \(^{68}\)Ge calibration runs with the source deployed at the detector center and are found to be consistent. After applying the time-offset correction, the relative timing of all channels remains stable throughout physics data taking.

%

\subsubsection{SiPM Dark-count Rate}

The DCR of each channel is evaluated using events in the energy range of 1--3~MeV. Events below 1~MeV are excluded to suppress trigger-time fluctuations and sharpen the channel-level timing peak around 310~ns. A pre-signal time window of width \(\Delta T = 104\)~ns is used to count the number of pulses, \(n\). The DCR is then calculated as \( {\rm DCR} = n/(N \times \Delta T)\), where \(N\) is the total number of selected events. This calibration is performed on a run-by-run basis during offline data processing.

\begin{figure}[!htbp]
\centering
\includegraphics[width=0.95\linewidth]{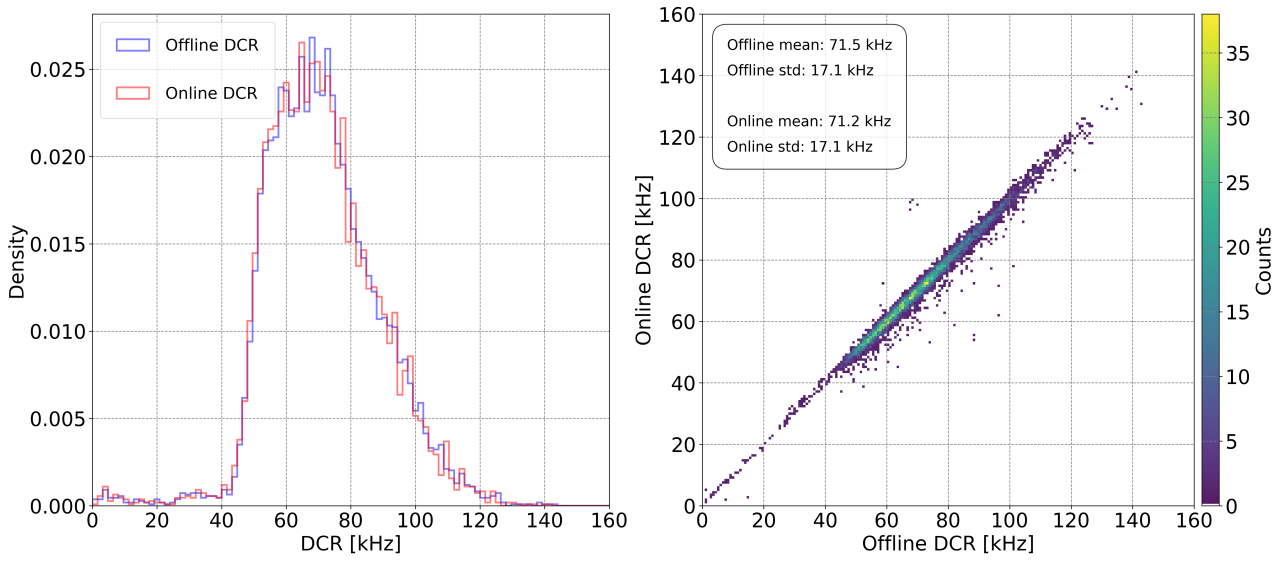}
\caption{Dark-count rate before the optical crosstalk correction for RUN1382, the first physics run. Left: distribution of the offline-measured DCR, with the red line indicating the DCR measured online by the DAQ monitor through the FEC readout. Right: comparison between the DCR measured offline and that measured online for individual channels.}
\label{fig:dcr}
\end{figure}

Fig.~\ref{fig:dcr} shows the DCR results for the first physics run. In the left panel, the blue curve represents the hit rate measured online by the DAQ monitor through the FEC readout, which is in good agreement with the offline-derived result. A similar agreement is also observed for individual channels, as indicated in the right panel. With this method, however, the contribution from external optical crosstalk cannot be separated. As discussed in the following subsection, approximately 25\% of this rate is attributed to that effect. After correcting for optical crosstalk, the average DCR over all channels is 58~kHz per channel, corresponding to 50~Hz/mm\(^2\) when normalized to the active area of the SiPMs.

\begin{figure}[!htbp]
\centering
\includegraphics[width=0.9\linewidth]{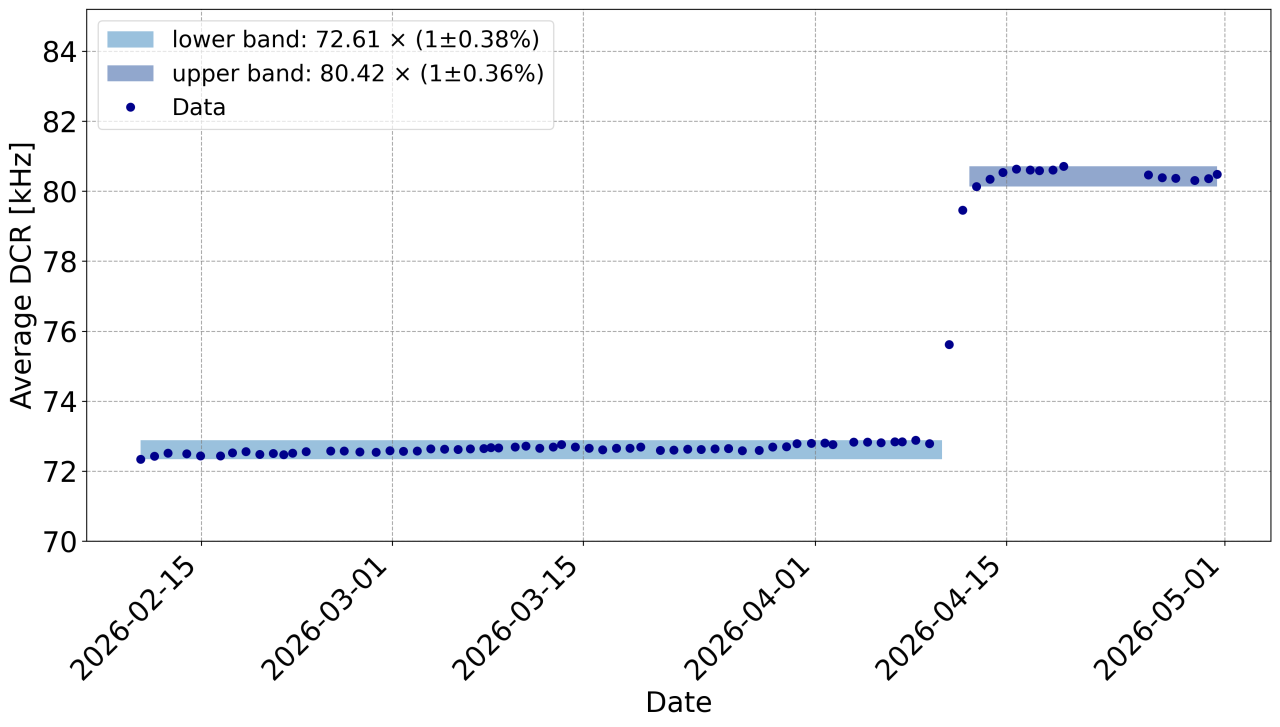}
\caption{Average dark-count rate over all channels as a function of time. The increase observed around 10 April coincides with a rise of about \(1.5\,^{\circ}\mathrm{C}\) in the CD temperature. When the detector temperature is stable, the variation of the average DCR remains within 0.4\%.}
\label{fig:dcr_stability}
\end{figure}

Under stable thermal conditions, the DCR exhibits excellent long-term stability, as shown in Fig.~\ref{fig:dcr_stability}. The increase observed around 10 April coincides with a rise of about \(1.5\,^{\circ}\mathrm{C}\) in the CD temperature.

\subsubsection{Optical crosstalk}
\label{sec:oct}

Infrared photons can be emitted during the Geiger-mode avalanche process in SiPMs~\cite{PhysRev.100.700}. These photons may either induce secondary avalanches within the same SiPM (internal optical crosstalk) or escape from the emitting SiPM and trigger avalanches in neighboring devices (external optical crosstalk). In TAO, the SiPMs are arranged in a dense spherical geometry with high optical coverage, which enhances the propagation of infrared photons between channels. As a result, external optical crosstalk dominates the total crosstalk probability, while the internal component remains below 1\% when the SiPMs are operated in LAB~\cite{Guan:2023uaa}.

\begin{figure}[!htbp]
  \centering
  \begin{subfigure}[b]{0.5\linewidth}
    \centering
    \includegraphics[width=\linewidth]{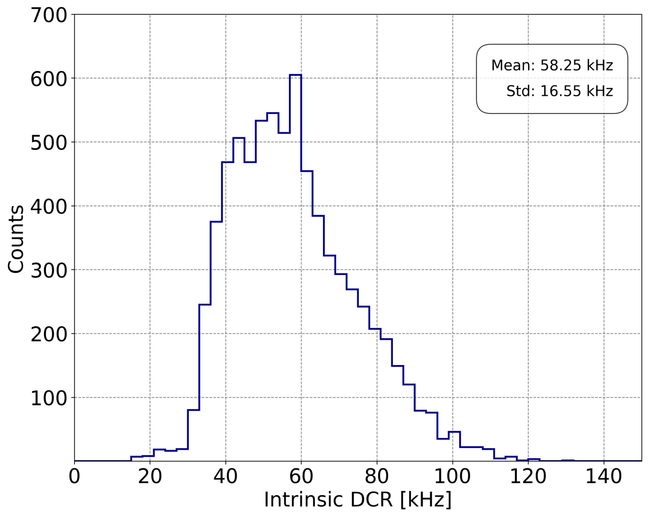}
    \caption{}
    \label{fig:hit_rate_off}
  \end{subfigure}\hfill
  \begin{subfigure}[b]{0.5\linewidth}
    \centering
    \includegraphics[width=\linewidth]{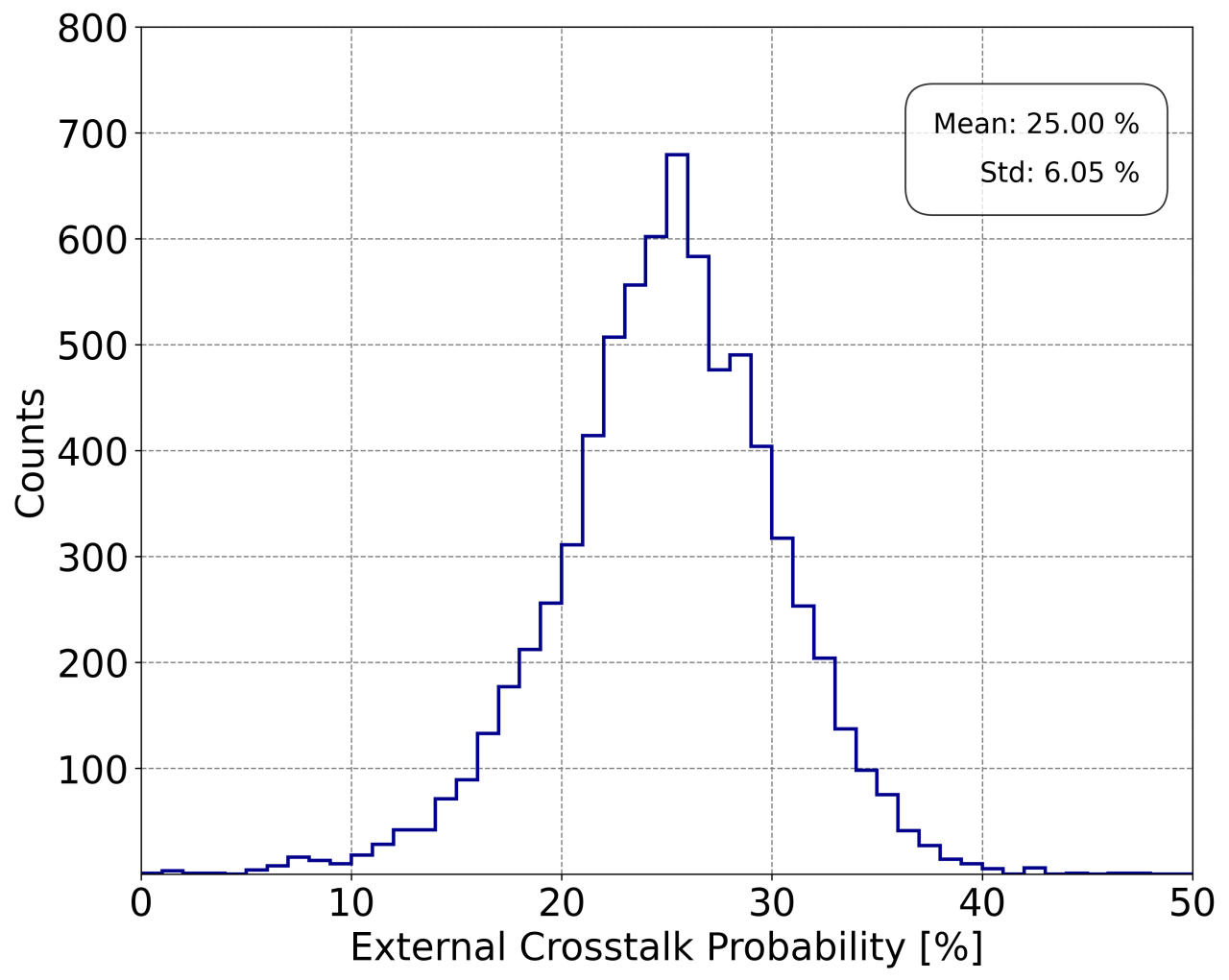}
    \caption{}
    \label{fig:prob_eCT}
  \end{subfigure}
  \caption{(a) Distribution of the hit rates measured for each channel with all other channels turned off. (b) Distribution of the external optical cross-talk probabilities measured for all channels.}
  \label{fig:oCT}
\end{figure}

To quantify the external optical crosstalk probability, dedicated hit-rate measurements are performed. For each channel \(i\), the hit rate \(R_{i,{\rm off}}\) is measured with channel \(i\) biased at its nominal operating voltage while all other channels are turned off. Under this condition, the observed rate is dominated by intrinsic dark noise. The distribution of \(R_{i,{\rm off}}\) over all good channels is shown in Fig.~\ref{fig:hit_rate_off}. This baseline rate is then compared with the rate \(R_{i,{\rm on}}\) measured with all channels active. The external optical crosstalk probability for channel \(i\), contributed by all other channels, is defined as
\begin{equation}
    P_{i,{\rm oct}} = \frac{(R_{i,{\rm on}} - R_{i,{\rm off}})\times (N-1)}{\sum_{j\neq i}^{N} R_{j,{\rm off}}} \, ,
    \label{eq:prob_oct}
\end{equation}
where \(N\) is the total number of good channels. The resulting external optical crosstalk probabilities are shown in Fig.~\ref{fig:prob_eCT}, with an average value of 25\%. However, the contribution of external optical crosstalk depends on the event hit pattern and is therefore expected to be position-dependent. A more detailed study of this effect and its impact on detector performance is ongoing.

\subsubsection{Relative Photon-detection Efficiency}

The \(^{68}\)Ge calibration source deployed at the detector center is used to determine the relative photon-detection efficiency (rPDE) of each channel. A clean sample of \(^{68}\)Ge events is selected by requiring a reconstructed vertex radius smaller than 150~mm and a total reconstructed energy between 0.75 and 1.0~MeV, corresponding to the full-energy peak from the absorption of the two 511-keV gamma rays. Owing to the spherical symmetry of the detector geometry and the central source position, all channels are expected to receive the same average incident light intensity.

The number of photons impinging on each channel follows Poisson statistics, and the mean number of detected PEs per event, denoted by \(\mu_i\), is proportional to the photon-detection efficiency of channel \(i\). Since the absolute mean number of incident photons is not known precisely, only the relative photon-detection efficiency can be determined. After correcting for the DCR contribution, the rPDE of channel \(i\) is defined as
\begin{equation}
    \mathrm{rPDE}_i = \frac{\mu_i}{\langle \mu \rangle} = \frac{\mu_i \cdot N}{\sum_{j=1}^{N} \mu_j}, \quad {\rm with} \quad \mu_i = \mu^0_i - \mu_{\mathrm{DCR}} \, ,
    \label{eq:rPDE}
\end{equation}
where \(\mu^0_i\) is the observed mean PE yield including both signal and DCR contributions, and \(\mu_{\mathrm{DCR}}\) is the DCR-induced offset. The quantity \(\mu^0_i\) is obtained from the fraction of events with zero detected pulses,
\[
\mu^0_i = -\ln\left( \frac{N^0_{{\rm evt},i}}{N_{{\rm evt},i}} \right),
\]
where \(N^0_{{\rm evt},i}\) is the number of zero-pulse events and \(N_{{\rm evt},i}\) is the total number of selected \(^{68}\)Ge events for channel \(i\).

\begin{figure}[!htbp]
    \centering
    \includegraphics[width=0.95\linewidth]{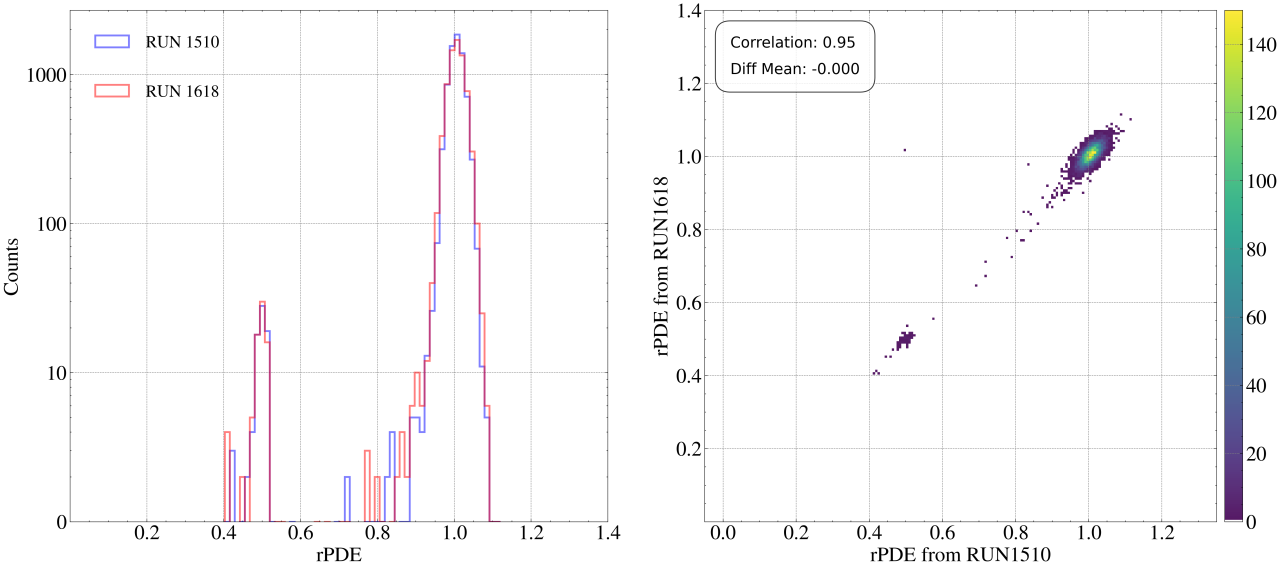}
    \caption{Distribution of the relative photon-detection efficiency for all channels, obtained from RUN1510 (blue) and RUN1618 (red). The rPDE remains stable throughout physics data taking.}
    \label{fig:rPDE}
\end{figure}

The resulting rPDE distribution over all channels is shown in Fig.~\ref{fig:rPDE} for RUN1510 and RUN1618, the last run in the dataset. Most channels exhibit a consistent response, while a small subset ($\sim$81 channels) shows rPDE values about 50\% below the average. This deficit is most likely caused by degraded electrical connections between specific SiPM tiles and their FEBs, leading to partial signal loss. The rPDE is observed to remain stable throughout physics data taking. 

\subsection{Light Yield}

\begin{figure}[!htbp]
    \centering
    \begin{subfigure}[b]{0.45\linewidth}
        \centering
        \includegraphics[width=\linewidth]{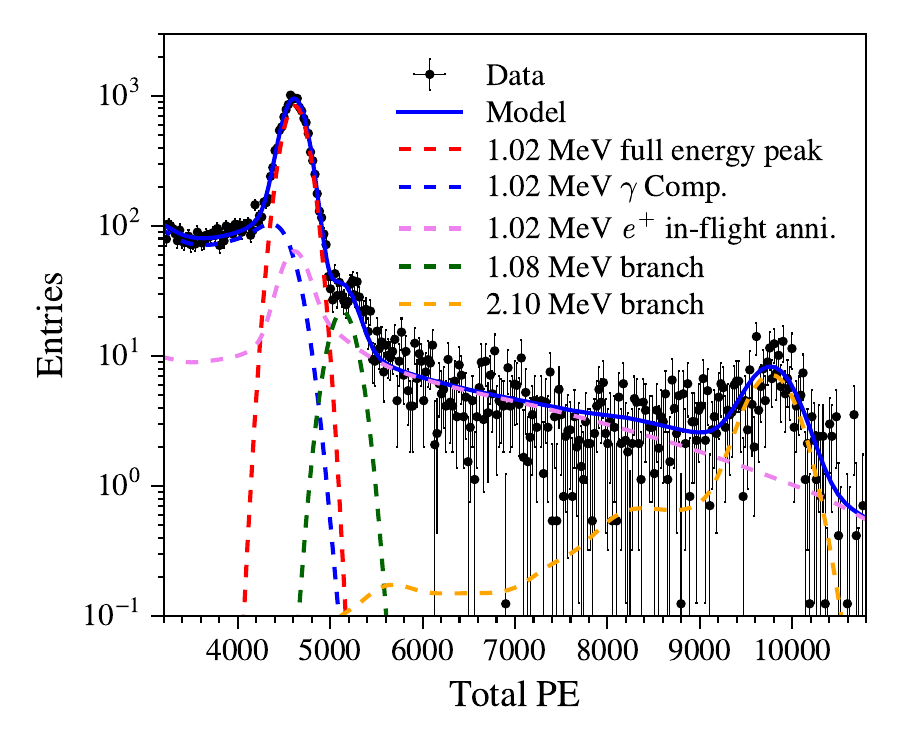}
        \caption{}
        \label{fig:ly_ge68_center}
    \end{subfigure}\hfill
    \begin{subfigure}[b]{0.49\linewidth}
        \centering
        \includegraphics[width=\linewidth]{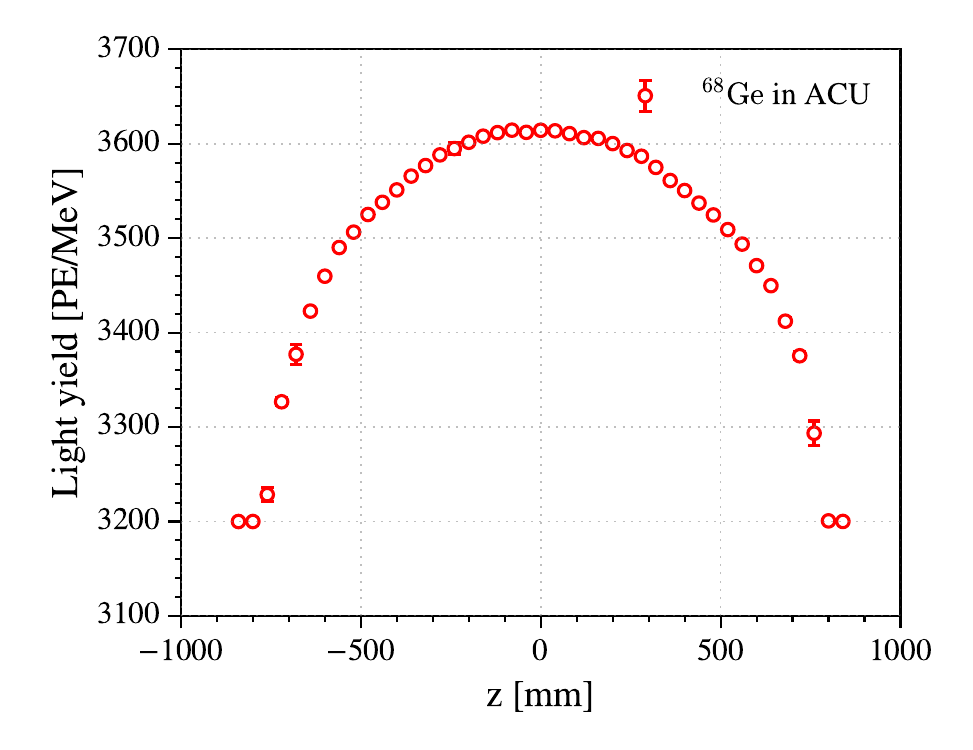}
        \caption{}
        \label{fig:ly_ge68_scan}
    \end{subfigure}
    \caption{(a) PE spectrum of \(^{68}\)Ge calibration events with the source deployed at the detector center. The solid blue line shows the total fit, and the dashed red component denotes the Gaussian contribution from the full-absorption peak of the two 511-keV gamma rays. The remaining components, indicated in the legend, are taken from the detector simulation. (b) Light yield of \(^{68}\)Ge as a function of position along the ACU axis.}
    \label{fig:ge68_light_yield}
\end{figure}

Fig.~\ref{fig:ly_ge68_center} shows the PE spectrum of \(^{68}\)Ge events with the source deployed at the detector center after subtraction of the SiPM dark-noise contribution. For each event, the total PE count is obtained by summing over all fired channels with hit times in the range of 200--600~ns, and the run-dependent DCR values are used to subtract the noise contribution channel by channel. The solid blue line shows the fit to the spectrum. The dashed red component represents the full-absorption peak of the two 511-keV gamma rays, modeled by a Gaussian function, while the other contributing processes, indicated in the legend, are taken from the detector simulation with free normalization factors.

The fit yields a full-absorption peak corresponding to 4518~PEs/MeV at the detector center. As discussed in Section~\ref{sec:oct}, external optical crosstalk contributes 25\% of the physical PEs on average. After correcting for this contribution, the light yield for \(^{68}\)Ge events at the detector center is 3614~PEs/MeV. Fig.~\ref{fig:ly_ge68_scan} shows the variation of the \(^{68}\)Ge light yield along the ACU axis. The light yield decreases near the top and bottom chimneys because of photon losses.

In JUNO, the reference light yield is defined using the 2.2-MeV gamma ray from neutron capture on hydrogen, produced by an AmC source deployed at the detector center. TAO does not include a neutron source in its calibration system; instead, neutrons from IBD events distributed within a radius of less than 400~mm are used to estimate the reference light yield. Since the detector energy response is non-linear, as discussed in Section~\ref{sec:nonlinearity}, the light yield is energy dependent. The value derived from the \(^{68}\)Ge full-absorption peak at the detector center is therefore not expected to be identical to the reference light yield obtained from neutron-capture gamma rays in IBD events. After correcting for non-linearity, the reference light yield in TAO is determined to be $\sim$4000~PEs/MeV. Recovery of the currently bad channels is expected to increase the corresponding light yield by about 10\%. 

The light yield observed in TAO is significantly higher than that reported for previous liquid-scintillator detectors, including the value of 1785~PEs/MeV measured by JUNO~\cite{JUNO:2025fpc}. This improvement is primarily attributed to the high photon-detection efficiency of the SiPMs, their compact arrangement within the TAO detector, and the reduced photon absorption due to the small detector size. In addition, operating the liquid scintillator at low temperature increases the light yield by about 9\% relative to room temperature~\cite{Xie2022TAO}.

\subsection{Singles Rate in CD}

The singles rate in the CD is evaluated using the first physics data, and its stability is monitored. Singles are selected by applying the following criteria: reconstructed energies in the range of 0.7--12~MeV, a multiplicity cut of no other CD events within a \(\pm 100~\mu\mathrm{s}\) time window, a veto window of 50~\(\mu\mathrm{s}\) for muons tagged by the veto detectors, and a veto window of 200~\(\mu\mathrm{s}\) for muons identified in the CD. 

After correcting for the livetime loss caused by the muon veto and multiplicity selection, the singles rates in different detector volumes are obtained for each run, as shown in Figure~\ref{fig:cd_evt_rates}. In the full detector volume, the singles rate is about 210~Hz and shows a slight decrease over time. In the fiducial volumes \(R < 800~\mathrm{mm}\) and \(R < 650~\mathrm{mm}\), the corresponding rates are approximately 125~Hz and 47~Hz, respectively, and exhibit similar decreasing trends. The observed singles rate is lower than the expectation of less than 100~Hz for \(R < 650~\mathrm{mm}\) reported in Ref.~\cite{JUNO:2020ijm}, indicating good control of material screening and detector cleanliness during installation.

\begin{figure}[!htbp]
  \centering
  \includegraphics[width=0.8\linewidth]{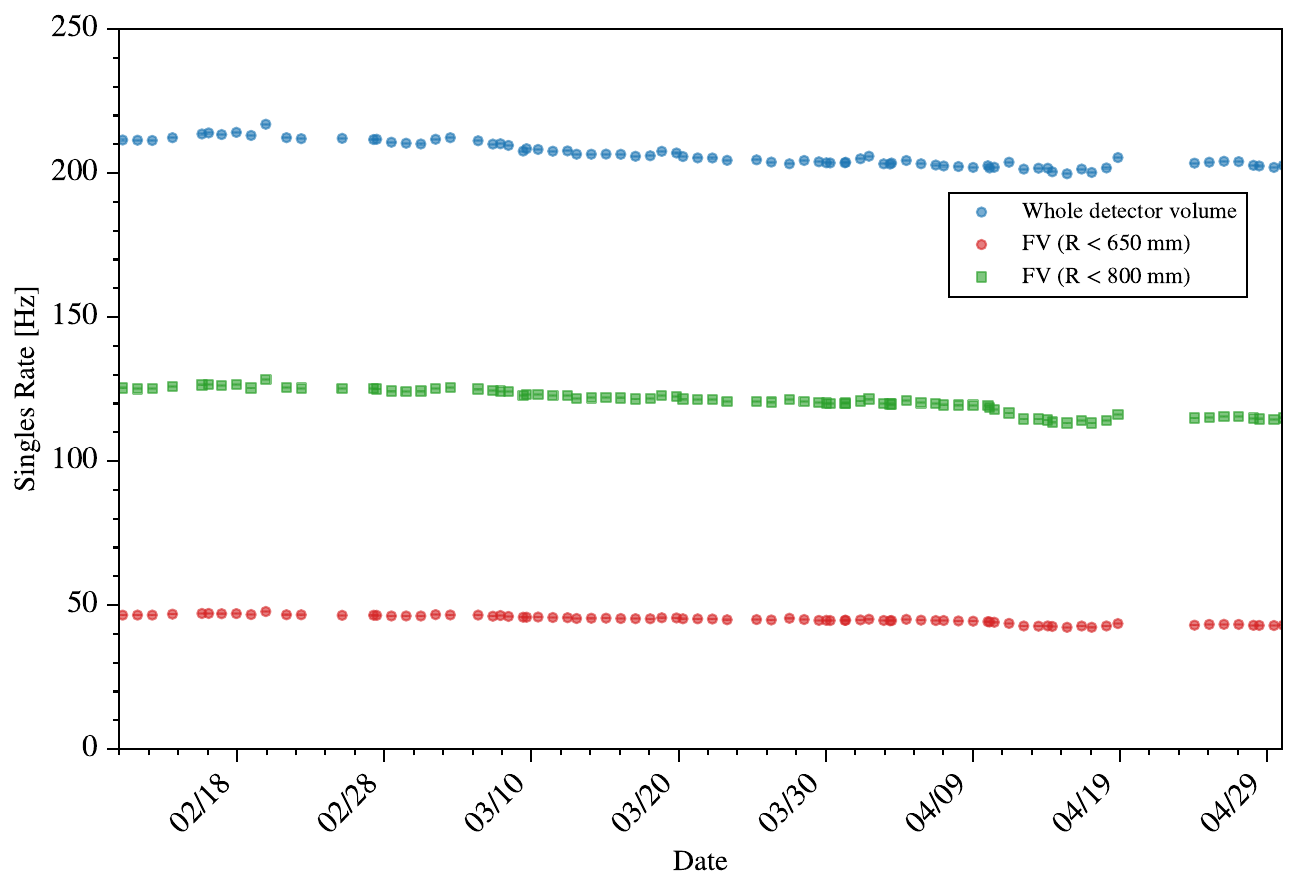}
  \caption{Run-by-run singles rates in the full detector volume (blue), in the detector volume \(R < 800~\mathrm{mm}\) (green), and in the detector volume \(R < 650~\mathrm{mm}\) (red).}
  \label{fig:cd_evt_rates}
\end{figure}

%% file: sec-energy_cal_rec-v3.tex
The reconstruction of event vertices and energies in TAO is based on the Charge Center (CC) method, which provides a fast and robust algorithm well suited to the compact detector geometry and high photon-detection coverage. A comprehensive simulation study of this method was presented in Ref.~\cite{Shi:2025zut}. In the present work, the CC reconstruction chain is implemented, optimized, and validated using calibration data from the ACU and CLS systems, together with natural radioactive peaks such as \(^{208}\mathrm{Tl}\).

\subsection{Calibration Data}

\begin{figure}[!htbp]
    \centering
    \includegraphics[width=0.7\linewidth]{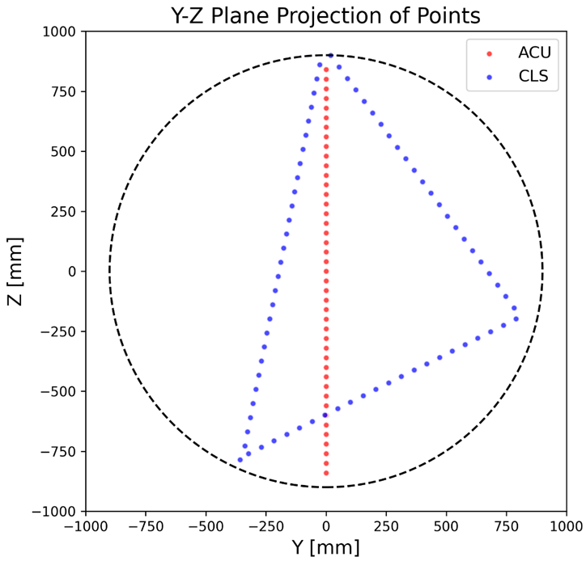}
    \caption{Projection onto the \(y\)-\(z\) plane of the calibration points used in the ACU (red) and CLS (blue) scans. The ACU points are spaced in 40-mm intervals along the vertical axis, and the CLS points in 60-mm intervals along the cable trajectory.}
    \label{fig:CalibPos}
\end{figure}

Weekly and monthly calibration campaigns are carried out during physics data taking. A weekly 10-minute calibration run is performed with the \textsuperscript{68}Ge source deployed at the detector center. In addition, detailed monthly calibration scans are performed with both the \textsuperscript{68}Ge and \textsuperscript{137}Cs sources. For the \textsuperscript{68}Ge scan, the source is deployed at 45 positions along the vertical axis, from +880~mm to -880~mm in steps of 40~mm. A 3-minute run is taken at each position, except at the two endpoints (\(\pm 880\)~mm), where a 5-minute run is used. For the \textsuperscript{137}Cs scan, 77 calibration points are defined along the CLS wire trajectory, starting from the bottom of the top acrylic-vessel chimney. The source is moved at a speed of 6~mm/s with a step length of 60~mm. At each position, the source remains stationary for a 2-minute data-taking period, with a 10-second interval allocated for movement between consecutive points. The Y-Z plane projection of the calibration points are shown in Fig.~\ref{fig:CalibPos}.

\subsection{Vertex Reconstruction}

For each event, the CC algorithm determines the vertex by calculating the PE-weighted centroid of the SiPM channels. Before the centroid calculation, the PE in each channel is corrected for the expected dark-noise contribution:
\begin{equation}
  q_i = Q_i^{\rm int} - \mathrm{DCR}_i \cdot \Delta t,
\end{equation}
where \(Q_i^{\rm int}\) is the integer-rounded charge sum in the \(i\)-th channel within the selected signal time window, with the rounding applied to suppress correlated electronic noise and improve the charge resolution. \(\mathrm{DCR}_i\) is the dark-count rate of that channel, including the contribution from external optical crosstalk, and \(\Delta t = 400~\mathrm{ns}\) is the width of the selected signal time window. To mitigate the impact of non-functional or poorly calibrated channels, the detector surface is divided into 192 equal-area pixels using a HEALPix-inspired scheme. The local response of each pixel is represented by the average PE of the good channels within that pixel. The vertex is then calculated as the PE-weighted centroid of the pixels. A geometric correction factor of 1.5 is applied to the reconstructed radius to compensate for the intrinsic inward bias of the charge centroid in a spherical detector. The radius is further refined using an empirical correction derived from ACU calibration data (\(^{68}\)Ge deployed along the \(z\)-axis) and CLS calibration data (\(^{137}\)Cs scanned along off-axis chords), as follows:
\begin{equation}
  R_{\rm corr} = a R_{\rm raw}^2 + b R_{\rm raw},
\end{equation}
where the raw radius \(R_{\rm raw}\) is mapped to the corrected radius \(R_{\rm corr}\) using a quadratic function, with the coefficients \(a\) and \(b\) determined from a combined fit to the calibration data. This correction accounts for the radial non-uniformity caused by the non-uniform light collection. To further reduce residual biases, particularly in the outer detector region and near the north and south polar openings, a second-stage radius-only correction is applied. This second-stage correction is derived from the residual vertex bias observed in the calibration sources, and is implemented as a two-dimensional function of the reconstructed radius and polar angle, \(\Delta R(R,\theta)\), with a piecewise parameterization for different detector regions.

\begin{figure}[!htbp]
    \centering
    \begin{minipage}[b]{0.48\linewidth}
        \centering
        \includegraphics[width=\linewidth]{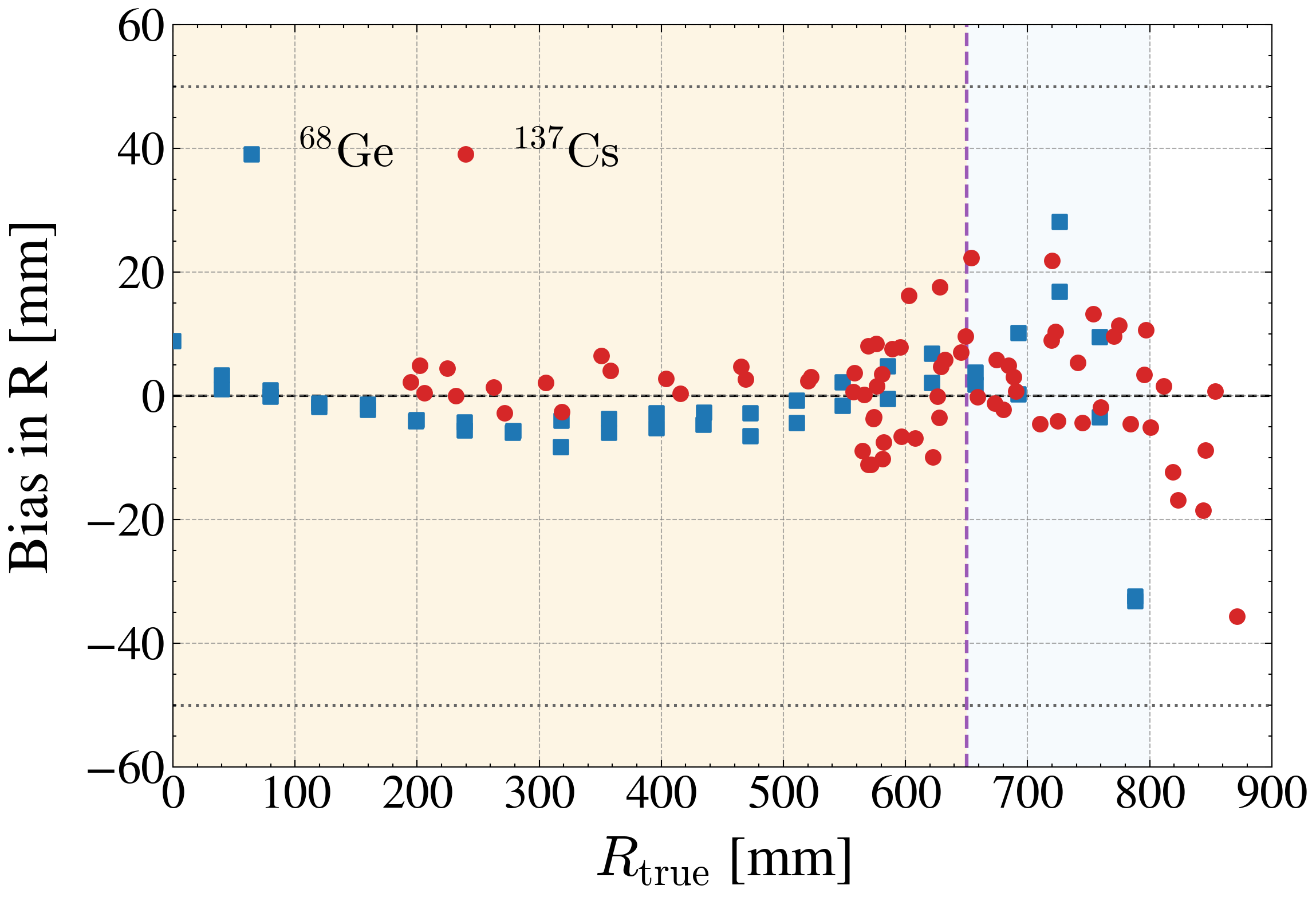}
    \end{minipage}
    \hfill
    \begin{minipage}[b]{0.48\linewidth}
        \centering
        \includegraphics[width=\linewidth]{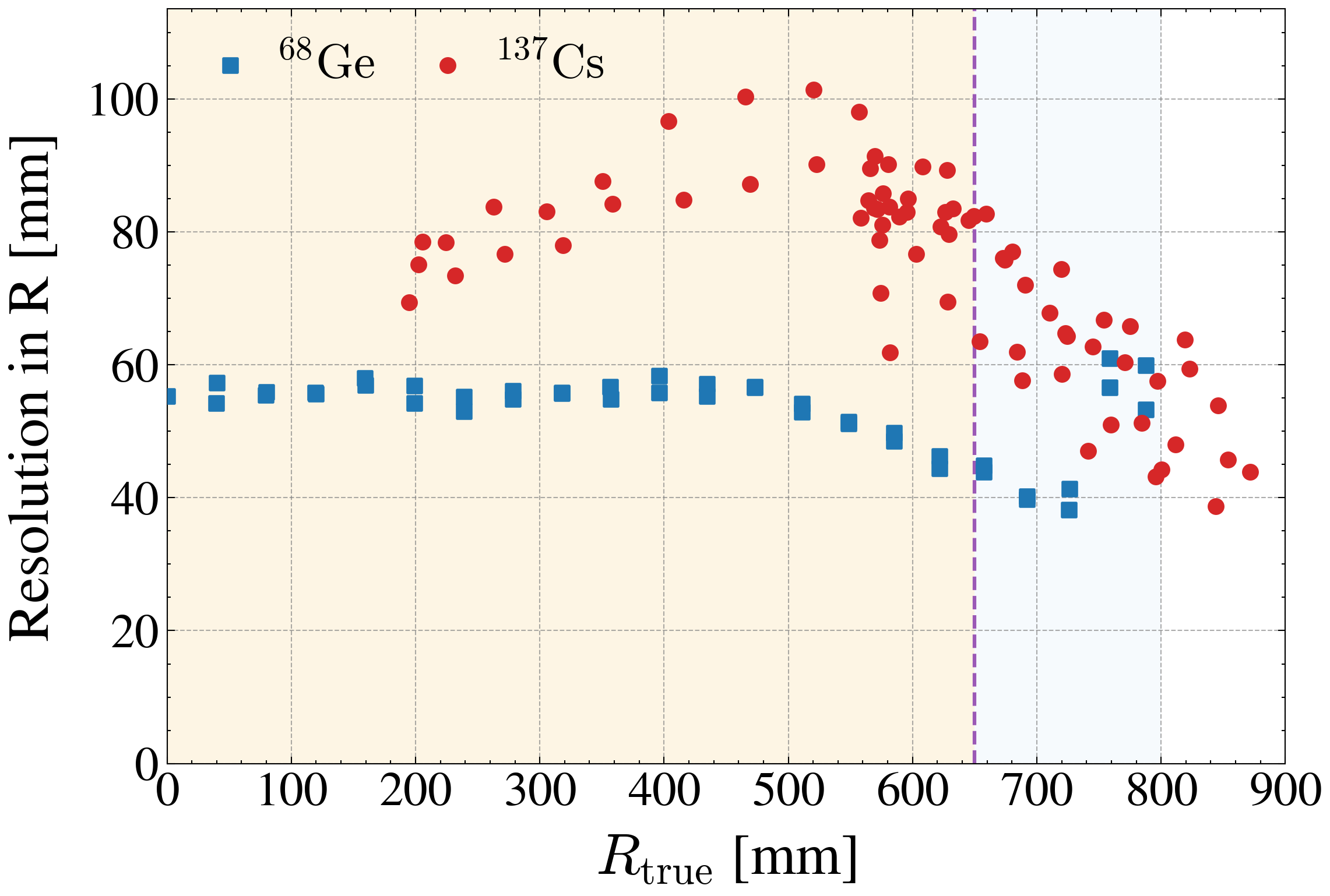}
    \end{minipage}
    \caption{Left: residual radial bias as a function of the true radius, evaluated using ACU \(^{68}\)Ge and CLS \(^{137}\)Cs calibration data. The vertical dashed lines indicate the fiducial-volume boundaries at \(R=650~\mathrm{mm}\) and \(R=800~\mathrm{mm}\). Right: radial vertex resolution as a function of the true radius. Here, \(R_{\rm true}\) denotes the reference radius of the energy-deposition position obtained from simulation.}
    \label{fig:cc_vertex_performance}
\end{figure}

Fig.~\ref{fig:cc_vertex_performance} shows the vertex reconstruction performance of the CC algorithm after the full correction chain. Within the volume \(R < 650~\mathrm{mm}\), the radial bias is controlled within \(\pm 20~\mathrm{mm}\), and the radial resolution is approximately \(80~\mathrm{mm}\) around the detector center. In the spherical shell \(650~\mathrm{mm} < R < 800~\mathrm{mm}\), the radial bias remains within \(\pm 30~\mathrm{mm}\) over most of the volume, while the radial resolution is about \(70~\mathrm{mm}\) near the boundary. For \(R < 800~\mathrm{mm}\), the overall vertex resolution is dominated by events in the outer region, where energy leakage and boundary effects become more significant. The poorer vertex resolution for \(^{137}\)Cs is attributed to the larger fluctuation in the deposited-energy position of its single \(\gamma\) ray, compared with \(^{68}\)Ge, for which the two \(\gamma\) rays are emitted back to back.

\subsection{Energy Reconstruction}

The energy estimator is derived from the total PE of all the channels, corrected for the missing charge of the bad channels using the same pixel-level averaging procedure (\(\mathrm{PE}_{\rm corr}\)) as that used in vertex reconstruction. To correct for the position-dependent detector response, a two-dimensional non-uniformity correction map, \(g(R,\theta)\), is constructed from the ACU and CLS calibration data and interpolated using the Clough--Tocher method. This map is evaluated at the reconstructed vertex \((R_{\rm corr},\theta_{\rm corr})\) to obtain the position-corrected visible energy,
\begin{equation}
  E_{\rm vis} = \frac{\mathrm{PE}_{\rm corr}}{g(R_{\rm corr}, \theta_{\rm corr}) \cdot Y_0},
\end{equation}
where \(Y_0\) is the light yield at the detector center after bad-channel correction, calibrated using the \(^{68}\mathrm{Ge}\) full-absorption peak.

\begin{figure}[!htbp]
    \centering
    \includegraphics[width=0.48\linewidth]{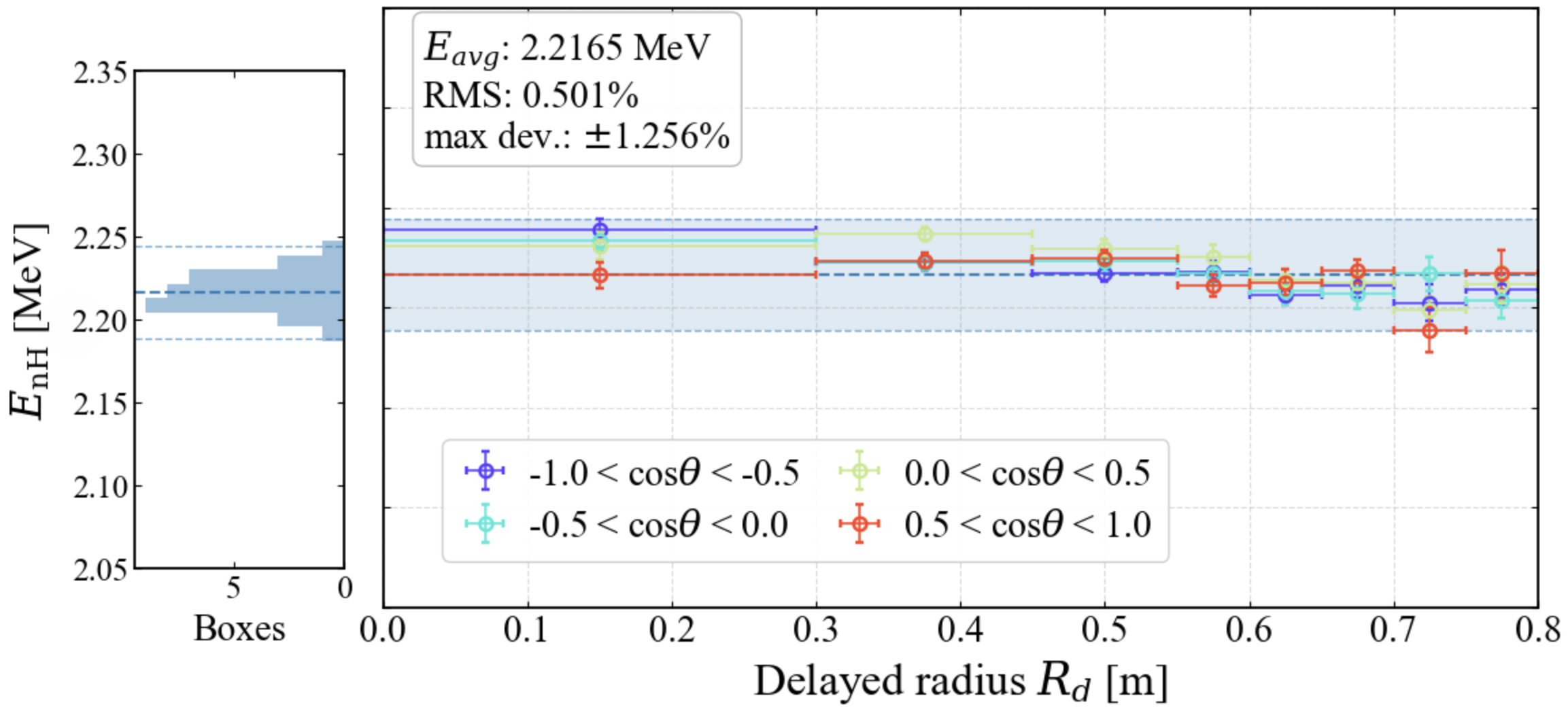}
    \hfill
    \includegraphics[width=0.48\linewidth]{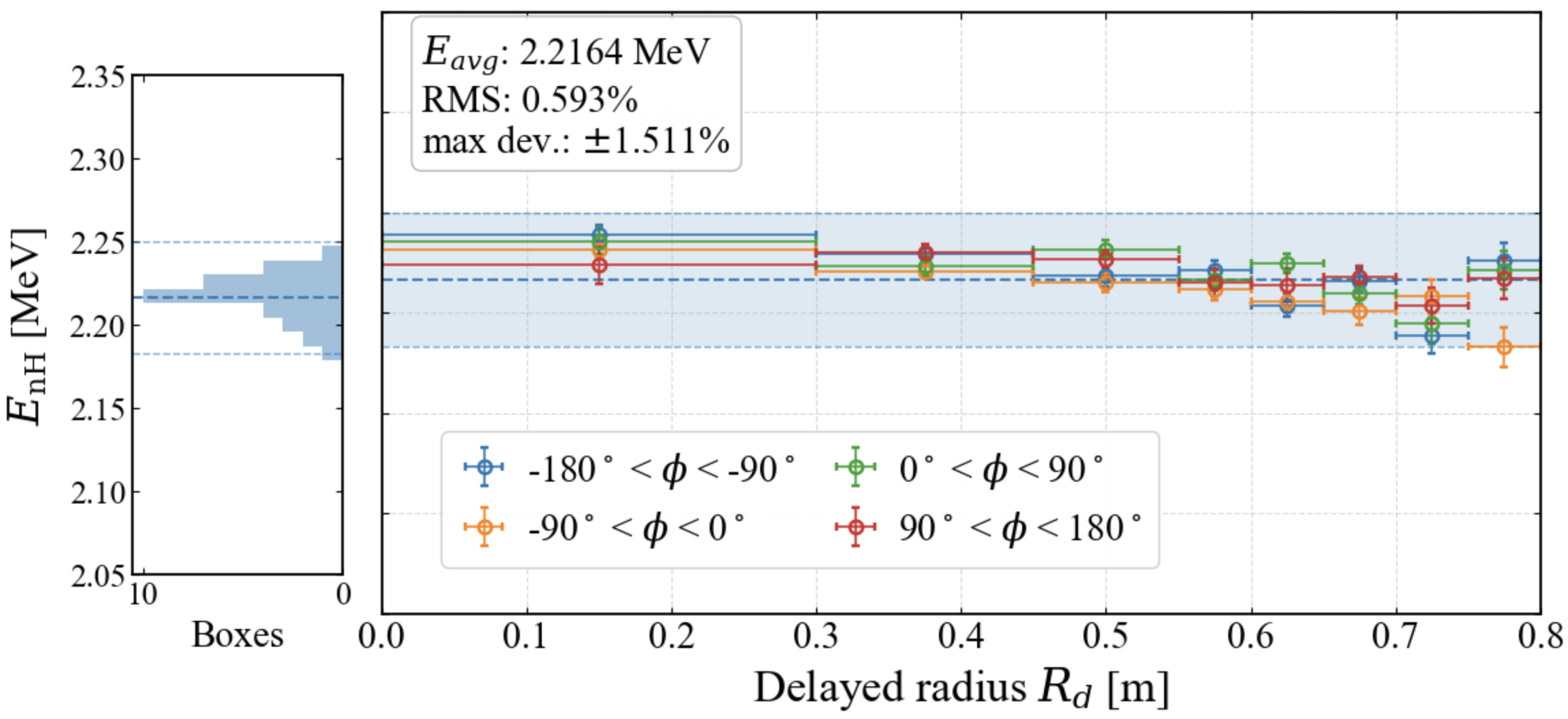}
    \caption{Residual energy non-uniformity evaluated using the \(\gamma\) ray from neutron capture on hydrogen in IBD events after the non-uniformity correction. Left: detector volume divided into voxels in the \(R\)-\(\cos\theta\) plane. Right: detector volume divided into voxels in the \(R\)-\(\phi\) plane.}
    \label{fig:nonuniformity-ibd-both}
\end{figure}

Although this correction removes the dominant large-scale non-uniformity, residual non-uniformities at the level of a few percent remain and show a clear dependence on both \(R\) and \(\theta\). To suppress these effects, a second-stage correction is applied. The detector volume is divided into bins in \(R^3\) and \(\cos\theta\), and in each sub-volume the 2.615-MeV \(\gamma\)-ray peak from \(^{208}\mathrm{Tl}\) is used to derive a local energy correction factor. This procedure reduces the RMS of residual non-uniformity to below 1\%, with a maximum deviation of 1.5\% over the detector volume of \(R < 800~mm\). The performance is evaluated using the \(\gamma\) ray from neutron capture on hydrogen in IBD events, as shown in Fig.~\ref{fig:nonuniformity-ibd-both}.

\begin{figure}[!htbp]
    \centering
    \includegraphics[width=0.8\linewidth]{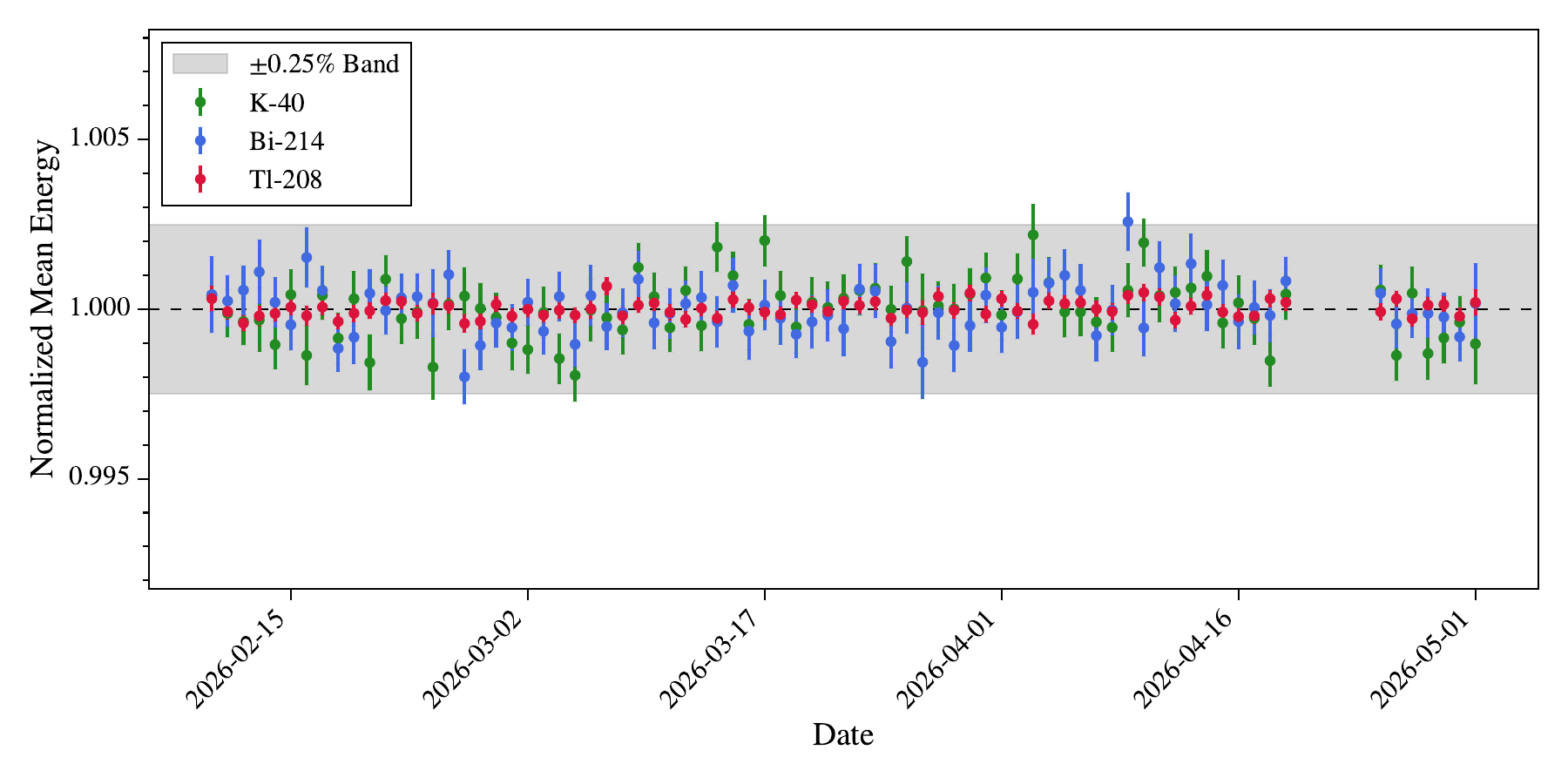}
    \caption{Energy-scale stability after the temporal correction, evaluated using the \(\gamma\)-ray peaks from \(^{208}\mathrm{Tl}\), \(^{40}\mathrm{K}\), and \(^{214}\mathrm{Bi}\).}
    \label{fig:escale_stability}
\end{figure}

In addition to the spatial non-uniformity, the detector energy scale exhibits slow temporal variations. Part of this variation is associated with changes in the detector operating temperature, including the intentional temperature adjustment during data taking, while the remaining contributions are still under investigation. To correct for these time-dependent effects, the 2.615-MeV \(\gamma\)-ray peak from \(^{208}\mathrm{Tl}\) is continuously monitored as a function of time. The peak is fitted in 0.5-hour time bins, and the resulting peak positions are smoothed with a Gaussian kernel weighted by the fit uncertainties. During periods of stable detector operation, a smoothing width of 24 hours is used, while a shorter width of 3 hours is applied during periods with rapid energy-scale variations. For each event, the reconstructed energy is corrected according to the smoothed \(^{208}\mathrm{Tl}\) peak position at the event time. This correction reduces the time-dependent variation of the energy scale to within 0.25\%, as evaluated using the \(\gamma\)-ray peaks from \(^{208}\mathrm{Tl}\), \(^{40}\mathrm{K}\), and \(^{214}\mathrm{Bi}\) shown in Fig.~\ref{fig:escale_stability}.

Finally, the absolute energy scale is anchored by fitting the neutron-capture-on-hydrogen peak in IBD delayed events within the fiducial volume of \(R < 650~\mathrm{mm}\) and scaling its reconstructed energy to 2.223~MeV. Together with the vertex-reconstruction performance discussed above, these results show that the CC reconstruction chain provides a robust basis for TAO physics analysis.

%% file: sec-energy_response-v2.tex
\subsection{Energy Non-linearity}
\label{sec:nonlinearity}
The detector energy non-linearity arises from ionization quenching and Cherenkov radiation in the liquid scintillator~\cite{DayaBay:2019fje,JUNO:2025gmd}. For electrons, the scintillation light yield is modeled using Birks' law,
\begin{equation}
    \frac{dN_S}{dx}=\frac{S\, dE/dx}{1+kB\, dE/dx},
\end{equation}
where \(S\) is the scintillation yield and \(kB\) is Birks' constant. The Cherenkov contribution is described by the Frank--Tamm formula and is parameterized in this work as
\begin{equation}
    N_C = a\int_{b<\beta^2}^{E}\frac{1-b/\beta^2}{dE/dx}\,dE,
\end{equation}
where \(a\) and \(b\) are model parameters. The generated scintillation and Cherenkov photons may undergo various optical processes inside the detector, such as Rayleigh scattering, absorption and potential re-emission, reflection and refraction at the boundary between different detector components. Eventually, a fraction of them will be detected by SiPMs, contributing to the total number of PEs. Here, we use $Y_{S}$ and $Y_{C}$ to denote the scintillation- and Cherenkov-induced PE yields, respectively. They are governed by the same expressions as $N_{S}$ and $N_{C}$, with $S$ and $a$ scaled in terms of detected PE. The non-linearity in the total number of PEs propagates directly into the reconstructed energy. The resulting electron energy non-linearity is written as
\begin{equation}
    f_{\rm NL}^{e^-}(E_{\rm true}) = \frac{E_{\rm rec}}{E_{\rm true}} = \frac{Y_S + Y_C}{Y^{\rm nH}}\frac{E_{\rm true}^{\rm nH}}{E_{\rm true}},
\end{equation}
where \(Y^{\rm nH}\) is the PE yield calibrated with \(\gamma\)-ray from neutron capture on hydrogen in IBD events with FV of 400~mm and \(E_{\rm true}^{\rm nH}=2.223~\mathrm{MeV}\).

\subsubsection{\(\gamma\) and \(e^+\) energy non-linearity}
\label{sec:gamma_nl}

Gamma rays do not deposit energy directly but transfer their energy to secondary electrons and positrons through Compton scattering, the photoelectric effect, and pair production. The \(\gamma\)-ray energy non-linearity is therefore obtained by convolving the electron non-linearity with the \(e^\pm\) energy distributions from Geant4 simulation:
\begin{equation}
f_{\rm NL}^{\gamma}(E_{\rm true}) =
\frac{\int_{0}^{E^{e^\pm}_{\rm max}} P(E^{e^\pm}) \times E^{e^\pm} \times f_{\rm NL}^{e^-}(E^{e^\pm})\, dE^{e^\pm}}
{\int_{0}^{E^{e^\pm}_{\rm max}} P(E^{e^\pm}) \times E^{e^\pm}\, dE^{e^\pm}},
\label{eq:gammaScintNL}
\end{equation}
where \(P(E^{e^\pm})\) is the number of secondary \(e^\pm\) with kinetic energy of $E^{e\pm}$ for the corresponding \(\gamma\)-ray energy, $E_\mathrm{max}^{e^\pm}$ is the maximum kinetic energy of the induced $e^\pm$. The denominator accounts for the missing energy in the simulation due to production thresholds~\cite{DayaBay:2019fje}.

For positrons, the difference in stopping power relative to electrons is neglected. After losing their kinetic energy, positrons annihilate and produce two 0.511-MeV \(\gamma\) rays. The positron energy non-linearity is therefore given by
\begin{equation}
    f_{\rm NL}^{e^{+}}(E_{\rm true})=\frac{T^{e^+}f_{\rm NL}^{e^-}(T^{e^+})+2\times 0.511 \times f_{\rm NL}^{\gamma}(0.511~\mathrm{MeV})}{E_{\rm true}},
\end{equation}
where \(T^{e^+}\) is the positron kinetic energy.

\subsubsection{Model fit and non-linearity prediction}
\label{sec:nl_positron}

\begin{figure}
    \centering
    \begin{subfigure}[b]{0.45\linewidth}
        \centering
        \includegraphics[width=\linewidth]{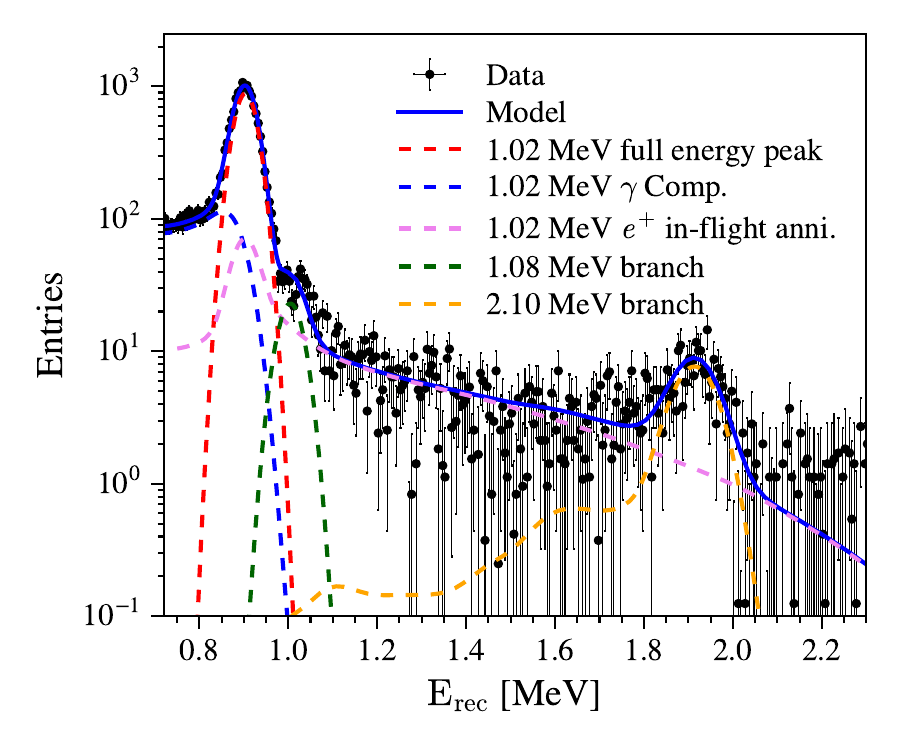}
        \caption{\(^{68}\mathrm{Ge}\) at the detector center}
        \label{fig:Ge68_fit}
    \end{subfigure}
    \begin{subfigure}[b]{0.45\linewidth}
        \centering
        \includegraphics[width=\linewidth]{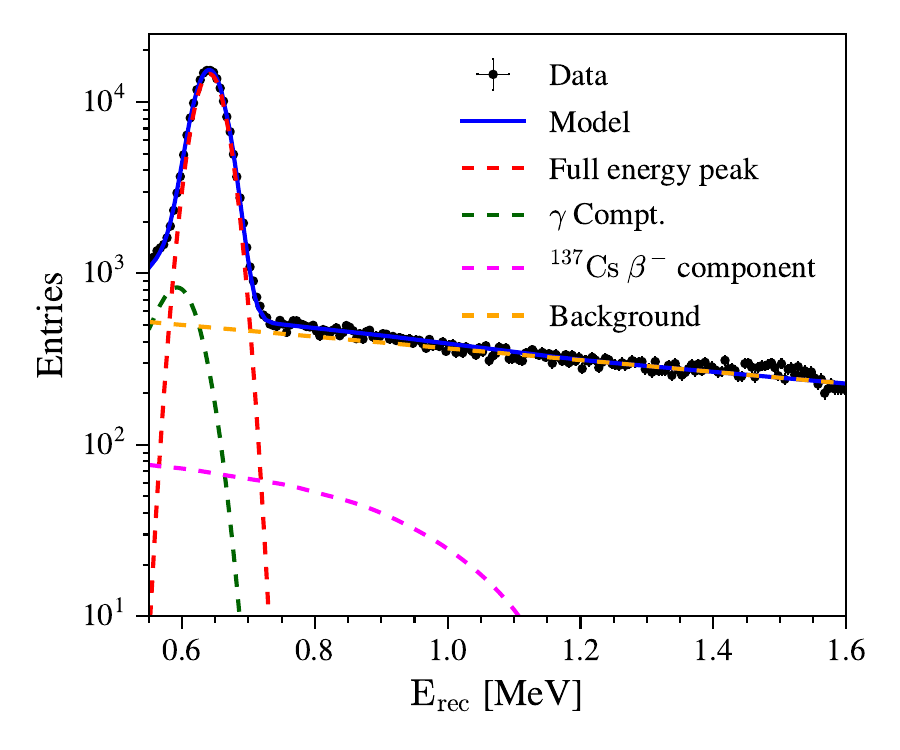}
        \caption{\(^{137}\mathrm{Cs}\) at \(R=193~\mathrm{mm}\)}
        \label{fig:Cs137_fit}
    \end{subfigure}
    \hfill
    \begin{subfigure}[b]{0.45 \linewidth}
        \centering
        \includegraphics[width=\linewidth]{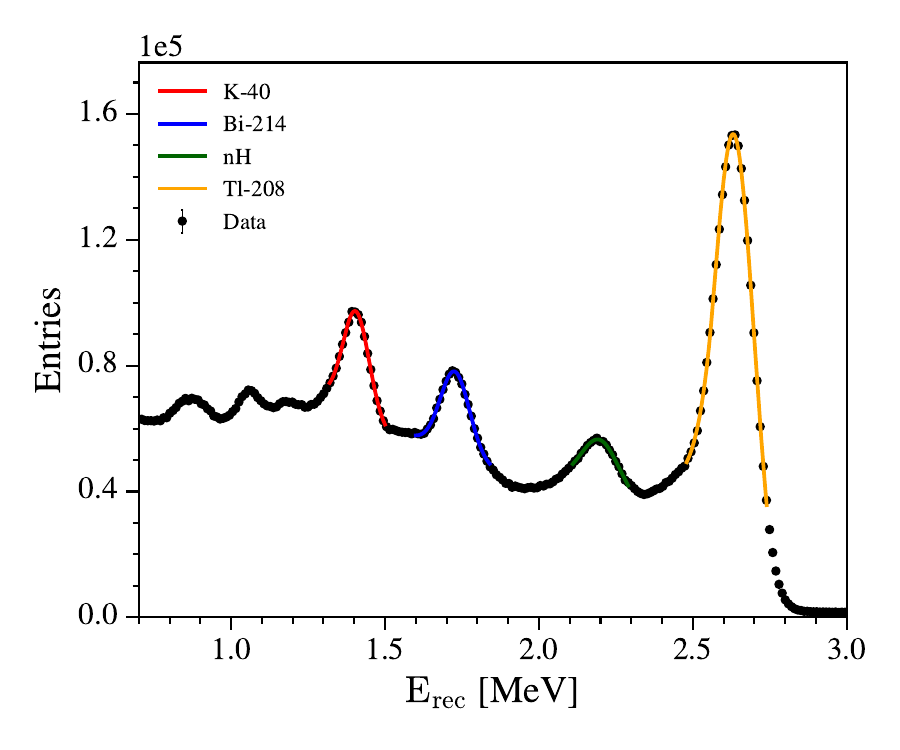}
        \caption{Singles energy spectrum}
        \label{fig:fit_singles}
    \end{subfigure}
    \begin{subfigure}[b]{0.45 \linewidth}
        \centering
        \includegraphics[width=\linewidth]{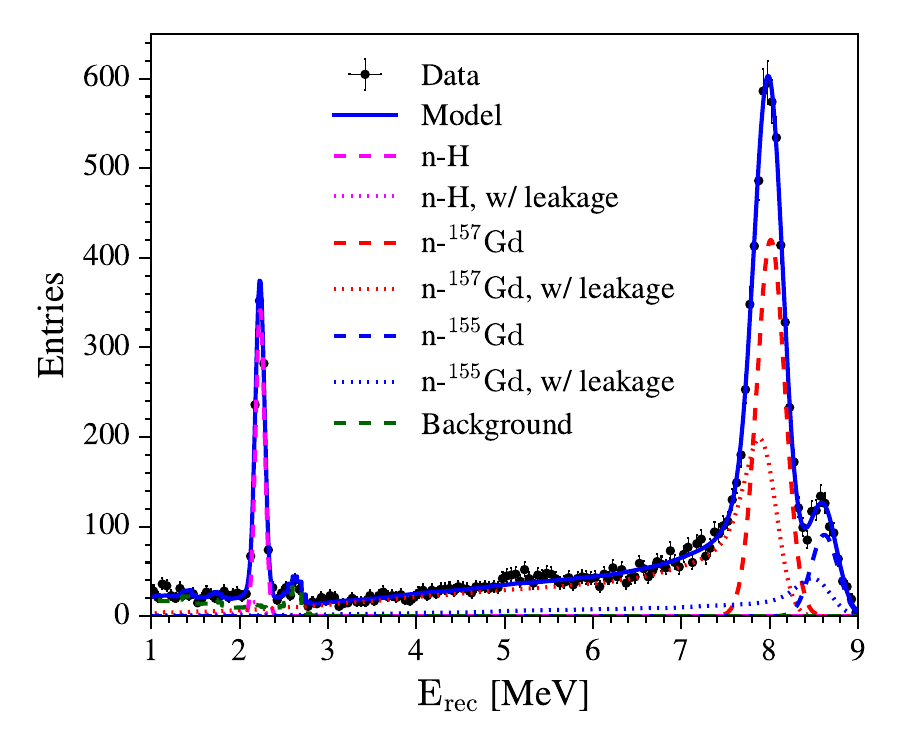}
        \caption{IBD delayed signals}
        \label{fig:ibd_delay_fit_400}
    \end{subfigure}
    \caption{Fits to energy spectra of different \(\gamma\)-ray sources: (a) \(^{68}\mathrm{Ge}\) at the detector center; (b) \(^{137}\mathrm{Cs}\) at \(R=193~\mathrm{mm}\); (c) single events within the central 20\% region; and (d) the IBD delayed spectrum within \(R<400~\mathrm{mm}\).}
    \label{fig:fit_gamma_spec}
\end{figure}

The non-linearity models for \(e^-\), \(e^+\), and \(\gamma\) share the common parameters \(S\), \(kB\), \(a\), and \(b\). In the present analysis, these parameters are constrained using \(\gamma\)-ray data only, as the current statistics of cosmogenic isotopes such as \(^{12}\mathrm{B}\) are insufficient. The reference \(\gamma\)-ray points are obtained from calibration sources (\(^{137}\mathrm{Cs}\) and \(^{68}\mathrm{Ge}\)), mono-energetic ambient backgrounds (\(^{40}\mathrm{K}\), \(^{214}\mathrm{Bi}\), and \(^{208}\mathrm{Tl}\)), and neutron-capture peaks in IBD delayed events. Representative fits to the energy spectra of these calibration and background \(\gamma\)-ray samples are shown in Fig.~\ref{fig:fit_gamma_spec}. An uncorrelated uncertainty of 0.5\% is assigned to each \(\gamma\)-ray point to account for residual spatial variations in the energy non-linearity, which is characterized using the $\gamma$ peaks from singles in different detector volumes. Note that the neutron-capture peak in Fig.~\ref{fig:fit_singles} is excluded in the energy non-linearity fit due to its overlap with the 2.20~MeV peak from $^{214}\mathrm{Bi}$. 

\begin{figure}[!htbp]
    \centering
    \begin{subfigure}[b]{0.48\linewidth}
        \centering
        \includegraphics[width=\linewidth]{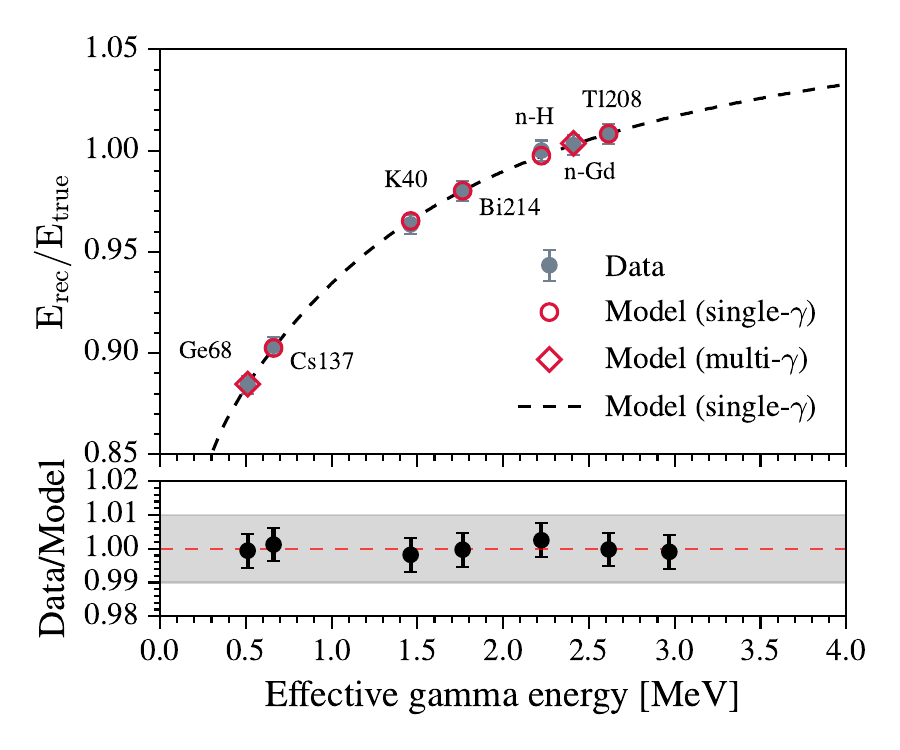}
        \caption{}
        \label{fig:gamma_fit}
    \end{subfigure}\hfill
    \begin{subfigure}[b]{0.5\linewidth}
        \centering
        \includegraphics[width=\linewidth]{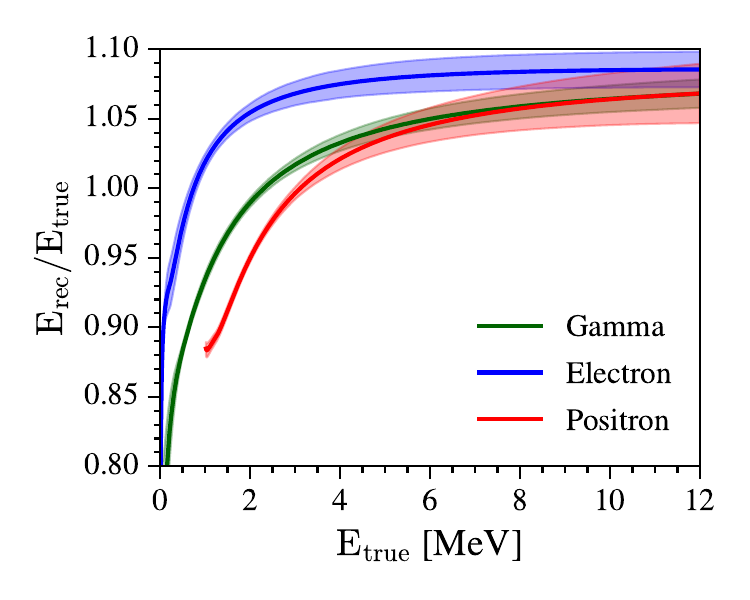}
        \caption{}
        \label{fig:nl_curves}
    \end{subfigure}
    \caption{(a) Fit to the \(\gamma\)-ray peaks using the non-linearity model. The lower panel shows the ratio of the model to the data. (b) Predicted non-linearity curves for \(e^-\), \(e^+\), and \(\gamma\), together with their uncertainties.}
    \label{fig:energy_nl_results}
\end{figure}

The reconstructed mean energies of the characteristic \(\gamma\)-ray peaks are fitted using the $\gamma$ energy non-linearity model in Section~\ref{sec:gamma_nl}. For a single-\(\gamma\) source, the effective gamma energy is defined as the deposited gamma energy. For multi-\(\gamma\) sources, such as \(^{68}\mathrm{Ge}\) and neutron-capture events on Gd, it is defined as the energy of an equivalent single-\(\gamma\) source that produces the same modeled nonlinear response. The resulting fit is shown in Fig.~\ref{fig:gamma_fit}, where the gray markers denote the input \(\gamma\)-ray points and the red markers the best-fit values. The dashed line is the best-fit \(\gamma\)-ray non-linearity curve. The corresponding predictions for \(e^-\), \(e^+\), and \(\gamma\) are shown in Fig.~\ref{fig:nl_curves}, together with the 68\% confidence-level uncertainty bands. The maximum uncertainty in the positron non-linearity is about 1.0\% in the high-energy region and is expected to be reduced in the future by incorporating cosmogenic isotope constraints. To be conservative, the positron energy non-linearity uncertainty is enlarged to 2\% in the high-energy region in the reactor neutrino analysis in Sec.~\ref{sec:ibd}, and the positron energy non-linearity curve from JUNO~\cite{JUNO:2025gmd} is introduced as an additional pull term.

\subsection{Energy Leakage}
\label{sec:energyleakage}

Because of the small dimensions of the TAO detector, a non-negligible fraction of the energy deposited by positron from IBD can escape from the GdLS target volume. This effect must be modeled accurately to construct the detector response matrix.

\subsubsection{Energy Leakage Matrix with Non-linearity Convolution}
\label{sec:leakage_convolution}

The energy leakage matrix \(R(E_{\mathrm{true}}, E_{\mathrm{vis}})\) describes the probability that a positron with true energy \(E_{\mathrm{true}}\) yields a visible energy \(E_{\mathrm{vis}}\) inside the GdLS target. Here, \(E_{\mathrm{true}} = T_{e^+} + 1.022~\mathrm{MeV}\), where \(T_{e^+}\) is the kinetic energy of the primary positron and 1.022~MeV accounts for positron annihilation. The matrix is constructed from Monte Carlo IBD simulation data using 5-keV binning over the range 0--12~MeV on both axes and with the same delayed-energy selection as in the standard IBD analysis. The fiducial-volume definition used in the analysis, \(R \leq 800~\mathrm{mm}\), is applied.

The energy non-linearity models presented in Secs.~\ref{sec:nl_positron} are derived under the assumption that the full particle energy is deposited in the GdLS. In TAO, however, the non-negligible energy leakage renders this assumption inadequate for an accurate prediction of the detector response. Dedicated Monte Carlo studies show that the leakage of the positron kinetic energy is almost negligible within the fiducial volume \(R \leq 800~\mathrm{mm}\), whereas the annihilation gamma rays exhibit a substantially larger escape probability, especially near the detector boundary. Consequently, applying a single positron non-linearity correction to the total deposited energy, without distinguishing between the positron kinetic and gamma contributions, can bias the predicted prompt-energy spectrum.

To account for this effect, the energy leakage and the energy non-linearity are convolved at the level of individual Monte Carlo events. For each IBD event, the truth information provides the initial positron kinetic energy \(T_{e^+}\), the initial energies of the \(\gamma\) rays produced along the positron chain, \(E^{\gamma}_{\rm ini}\), and the corresponding deposited energies in the GdLS, \(E^{e^+}_{\rm dep}\) and \(E^{\gamma}_{\rm dep}\). Since the non-linearity correction depends on the initial particle energy, the visible energy is constructed by applying the non-linearity correction to the initial energies and subtracting the contribution from the escaped energy:
\begin{equation}
\begin{split}
E_{\rm vis} =\;&
\left[
T_{e^+} f_{\rm NL}^{e^-}(T_{e^+})
+ \sum_{\gamma} E^{\gamma}_{\rm ini} f_{\rm NL}^{\gamma}(E^{\gamma}_{\rm ini})
\right] \\
&-
\left[
E^{e^+}_{\rm leak} f_{\rm NL}^{e^-}(E^{e^+}_{\rm leak})
+ \sum_{\gamma} E^{\gamma}_{\rm leak} f_{\rm NL}^{\gamma}(E^{\gamma}_{\rm leak})
\right]
\end{split}
\label{eq:Evis_mixedNL}
\end{equation}
where \(E^{e^+}_{\rm leak} = T_{e^+} - E^{e^+}_{\rm dep}\) and \(E^{\gamma}_{\rm leak} = E^{\gamma}_{\rm ini} - E^{\gamma}_{\rm dep}\). The positron kinetic-energy component is described using the electron non-linearity curve \(f_{\rm NL}^{e^-}\), since the difference in stopping power between \(e^+\) and \(e^-\) is neglected, as discussed in Sec.~\ref{sec:nl_positron}.

\begin{figure}[!htbp]
    \centering
    \includegraphics[width=0.6\linewidth]{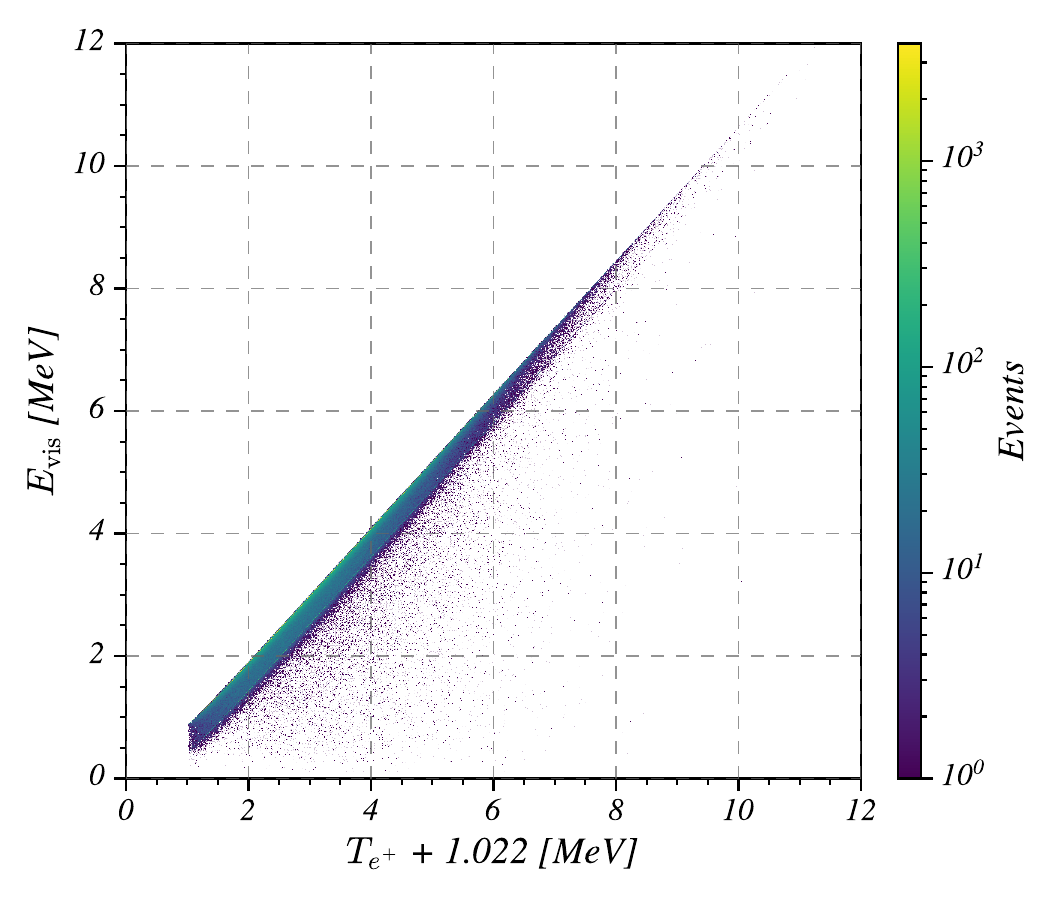}
    \caption{Energy leakage matrix for the fiducial volume \(R \leq 800~\mathrm{mm}\). The horizontal axis represents the true positron energy \(T_{e^+} + 1.022~\mathrm{MeV}\), and the vertical axis represents the visible energy with non-linearity correction \(E_{\mathrm{vis}}\) in the GdLS.}
    \label{fig:energy_leakage_matrix}
\end{figure}

The detector response matrix from the true positron energy \(E_{\rm true}=T_{e^+}+1.022~\mathrm{MeV}\) to the visible energy \(E_{\rm vis}\) is then constructed by filling
\begin{equation}
M_{\rm leak+NL}(E_{\rm true}, E_{\rm vis}) =
P(E_{\rm vis} \mid E_{\rm true}),
\label{eq:MleakNL}
\end{equation}
using the same Monte Carlo IBD sample, delayed-energy selection, and fiducial-volume cut \(R \leq 800~\mathrm{mm}\) as in the standard analysis. This matrix replaces the separate treatment of energy leakage and total-energy non-linearity in the response chain. The resulting matrix is shown in Fig.~\ref{fig:energy_leakage_matrix}. A prominent diagonal band is observed for fully contained events, together with a low-energy tail associated with partial energy leakage.

This matrix is an essential ingredient for the forward folding of the predicted positron spectrum in the model prediction described in Sec.~\ref{subsec:ibd_prompt_spectrum}.

\subsubsection{Energy leakage validation using \(^{137}\)Cs calibration data}

The energy leakage effect is measured using \(^{137}\)Cs calibration data obtained at 74 off-center positions defined by the CLS calibration system. At each position, the measured 0.662-MeV \(\gamma\)-ray spectrum is fitted with a model including a full-energy peak, a leakage component taken from Monte Carlo simulation, and a \(\beta^-\) component from the \(^{137}\)Cs decay. The fitted leakage fraction is then compared with the Monte Carlo results.

\begin{figure}[!htbp]
    \centering
    \includegraphics[width=0.6\linewidth]{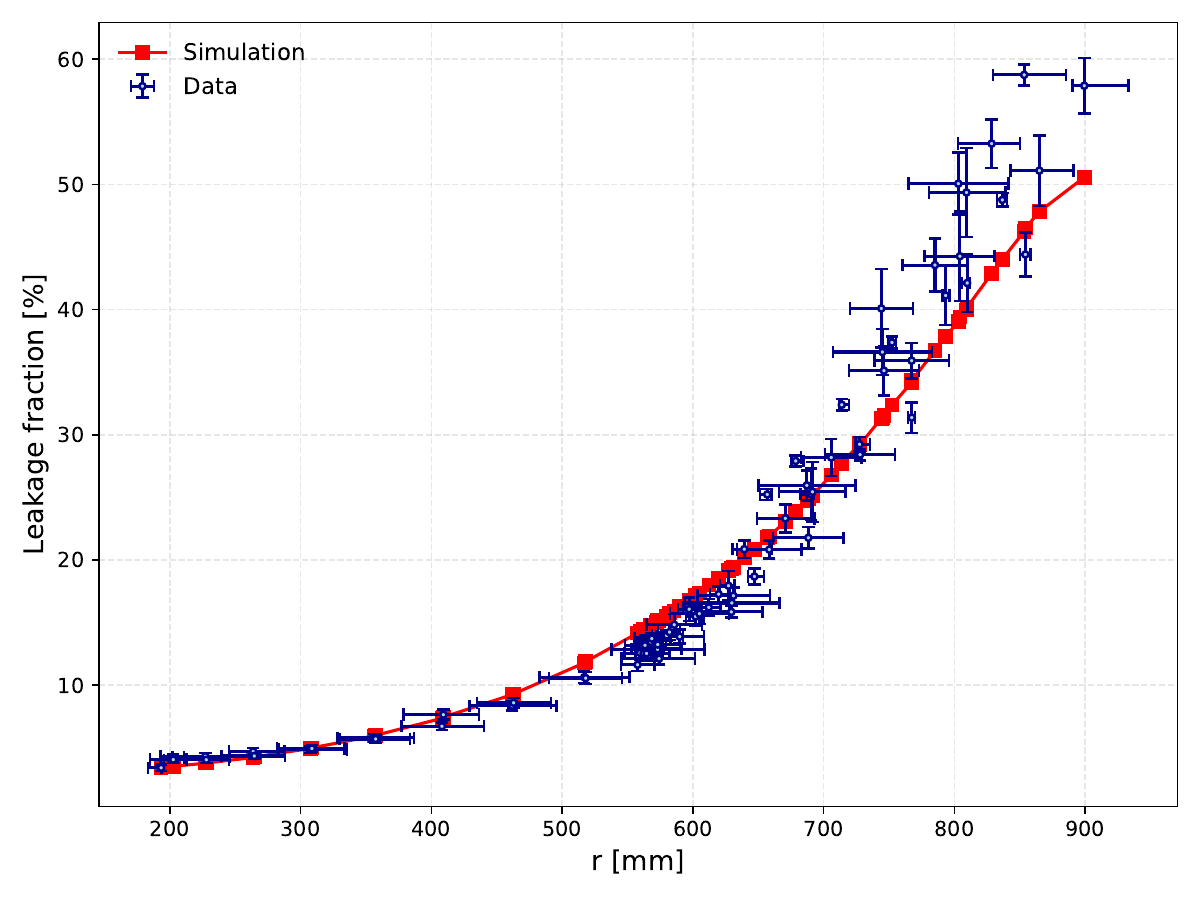}
    \caption{Leakage fraction as a function of radial distance \(r\) from the detector center, obtained from \(^{137}\)Cs calibration data (blue markers) and Monte Carlo simulation (red markers).}
    \label{fig:leakage_fraction_vs_r}
\end{figure}

Fig.~\ref{fig:leakage_fraction_vs_r} shows the leakage fraction as a function of radial position. The leakage fraction increases from about 4\% near the detector center to about 63\% near the detector boundary. 
The comparison of the fit results between calibration data and the Monte Carlo prediction near the 800~mm fiducial cut validates the modeling of energy leakage effects.

\subsubsection{Uncertainty of the energy leakage matrix}
\label{sec:leakage_uncertainty}

The energy leakage matrix and the corresponding deposited energy spectrum depend on the fiducial-volume cut applied to the reconstructed vertex radius. Uncertainties in the vertex bias and resolution therefore propagate into the leakage matrix through event migration across the fiducial-volume boundary. This effect is evaluated using a toy Monte Carlo method based on the measured charge center vertex performance shown in Fig.~\ref{fig:cc_vertex_performance}. 
For each toy experiment, the fiducial-volume selection is reapplied after smearing the event vertices according to the measured bias and resolution, and the corresponding deposited energy spectrum is obtained.

The resulting spectral systematic covariance matrix is obtained with those toy experiments after taking into account the full detector response model in the IBD analysis. For the fiducial-volume cut \(R \leq 800~\mathrm{mm}\), the relative uncertainty (diagonal terms of the covariance matrix) on the prompt-energy spectrum remains below 1\% in the energy region of interest (2--6~MeV) and increases toward lower energies, where the leakage tail becomes more significant. This uncertainty is included in the total systematic covariance matrix together with the energy-scale, energy non-linearity, and background uncertainties in Session~\ref{subsec:ibd_uncertainty}.

\subsection{Energy Resolution}
\label{sec:resolution}

The energy resolution is defined as \(\sigma_{E_{\rm rec}}/\bar{E}_{\rm rec}\), where \(E_{\rm rec}\) is the reconstructed energy and \(\sigma_{E_{\rm rec}}\) is the standard deviation of the \(E_{\rm rec}\) distribution. It depends on both particle type and energy. In liquid-scintillator detectors, the energy resolution receives contributions from many sources, which have been comprehensively summarized in Ref.~\cite{JUNO:2024fdc}. In TAO, in addition to the contributions associated with the LS itself, SiPM- and electronics-related effects, such as dark noise, optical crosstalk, after-pulsing, and charge smearing, also make important contributions.

Because the energy resolution is position dependent, a volume-averaged resolution is characterized using \(\gamma\)-ray sources at different locations. For the central region, the resolutions of the \(\gamma\)-ray peaks are extracted directly from the fits in Fig.~\ref{fig:Ge68_fit}, \ref{fig:Cs137_fit}, and \ref{fig:ibd_delay_fit_400}. In these fits, the energy-leakage shapes are constrained using detector simulation for different \(\gamma\)-ray sources at different positions. For larger fiducial volume (\(R<800~\mathrm{mm}\)), different procedures are used for different \(\gamma\)-ray sources. For the \(^{68}\)Ge calibration source, only a finite number of calibration points are available along the \(z\)-axis. The measured energy resolutions at these positions are interpolated to obtain the radial dependence of the resolution, and the corresponding volume-averaged resolutions are then derived by applying a volume-weighted average over the relevant fiducial volume. For the n-H and n-Gd capture peaks, the resolutions at larger radii are obtained by enlarging the radius cut in the event selection. 

\begin{table}[htbp]
\centering
\caption{Reconstructed energies and energy resolutions for representative \(\gamma\)-ray sources in different detector volumes.}
\label{tab:energy_res}
\begin{tabular}{ccc}
\toprule
Source & E$_{\rm rec}$ [MeV] & Resolution [\%] \\
\midrule
\multirow{3}{*}{$^{68}$Ge} & \multirow{3}{*}{0.904} & 2.82$\pm$0.01 (center) \\
                           &                           & 3.12\% (\(R<800~mm\)) \\
\hline
$^{137}$Cs & 0.598 (\(R=193~mm\)) & 3.65$\pm$0.01 \\
\hline
\multirow{3}{*}{nH} & 2.234 (\(R<400~mm\)) & 2.24$\pm$0.11 \\
                    & 2.229 (\(R<800~mm\)) & 2.493$\pm$0.006 \\
\hline
\multirow{3}{*}{nGd} & 8.003 (\(R<400~mm\)) & 1.91$\pm$0.06 \\
                     & 8.011 (\(R<800~mm\)) & 2.019$\pm$0.005 \\
\bottomrule
\end{tabular}
\end{table}

\begin{figure}[!htbp]
    \centering
    \begin{subfigure}[b]{0.48\linewidth}
        \centering
        \includegraphics[width=\linewidth]{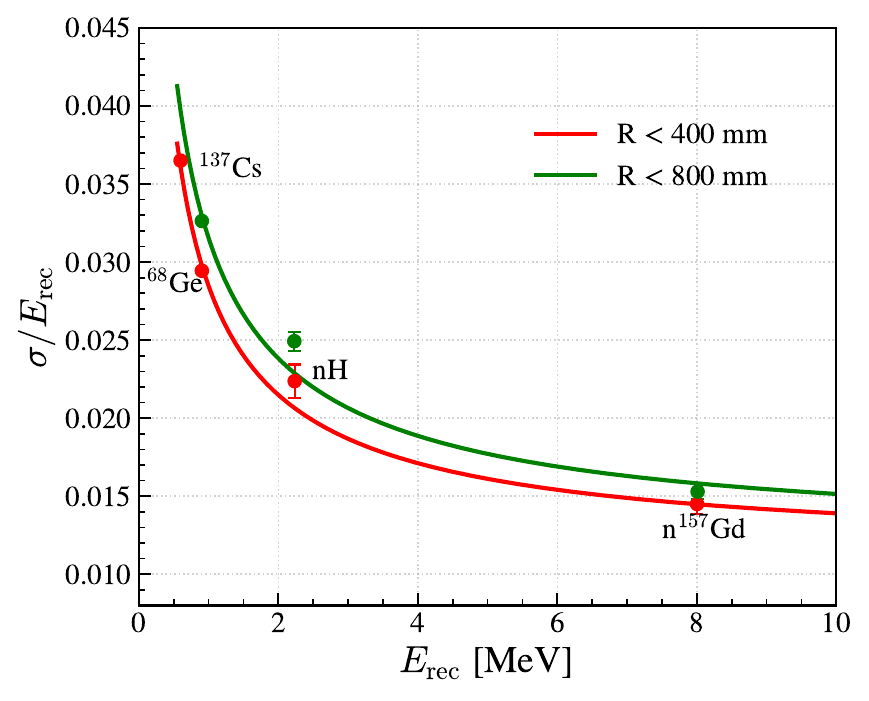}
        \caption{}
        \label{fig:gamma_res_fit}
    \end{subfigure}\hfill
    \begin{subfigure}[b]{0.48\linewidth}
        \centering
        \includegraphics[width=\linewidth]{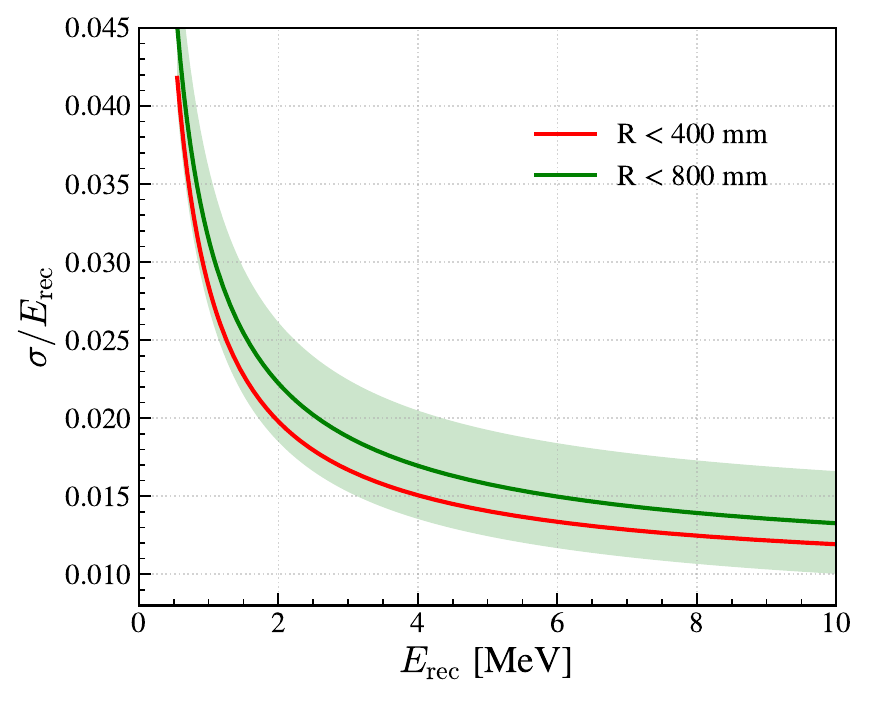}
        \caption{}
        \label{fig:positron_res_curves}
    \end{subfigure}
    \caption{Energy-resolution results. (a) Fit to the measured \(\gamma\)-ray energy resolutions. (b) Predicted positron energy resolution after applying the data-driven correction, with the shaded band indicating the systematic uncertainty.}
    \label{fig:energy_res_results}
\end{figure}

The measured \(\gamma\)-ray energy resolutions in different fiducial volumes are summarized in Table~\ref{tab:energy_res}. The energy-dependent resolution is fitted with
\begin{equation}
    \frac{\sigma_{E_\mathrm{rec}}}{E_{\rm rec}} = \sqrt{\left(\frac{a}{\sqrt{E_{\rm rec}}}\right)^2+b^2+\left(\frac{c}{E_{\rm rec}}\right)^2},
\end{equation}
where \(a\), \(b\), and \(c\) are the stochastic, constant, and noise terms, respectively. The fit results are shown in Fig.~\ref{fig:gamma_res_fit}, in which the red and blue lines represent the fitted energy-resolution curves for \(\gamma\)-ray events within \(R<400~\mathrm{mm}\) and \(R<800~\mathrm{mm}\), respectively. It should be noted that \(^{68}\)Ge and n-\(^{157}\)Gd are multi-\(\gamma\)-ray sources, and their measured resolutions are therefore not directly equivalent to those of single-\(\gamma\)-ray sources because of the different contributions from quenching and Cherenkov processes. The corresponding corrections are evaluated using detector simulation, and the corrected values are shown in the figure.

To obtain the positron energy-resolution curve, the variance difference \(\sigma_{\rm corr}^{2}(E_{\rm rec},\gamma)\) between data and Monte Carlo simulation is first extracted from the \(\gamma\)-ray samples. Under the assumption that this energy-dependent discrepancy is independent of particle type, the positron energy resolution is obtained by combining the Monte Carlo prediction with the extracted discrepancy:
\begin{equation}
\begin{split}
\sigma_{\rm corr}^{2}(E_{\rm rec},\gamma) &= \sigma_{\rm data}^{2}(E_{\rm rec},\gamma) - \sigma_{\rm MC}^{2}(E_{\rm rec},\gamma), \\
\mathrm{Res}(E_{\rm rec},e^{+}) &= \frac{\sqrt{\sigma_{\rm corr}^{2}(E_{\rm rec},\gamma)+\sigma_{\rm MC}^{2}(E_{\rm rec},e^{+})}}{E_{\rm rec,e^{+}}}.
\end{split}
\label{eq:MC/data diff}
\end{equation}
The resulting positron energy-resolution curves are shown in Fig.~\ref{fig:positron_res_curves} for \(R<400~\mathrm{mm}\) (red) and \(R<800~\mathrm{mm}\) (blue). A conservative uncertainty is assigned for the energy resolution curve, which is estimated from the difference of the curves obtained for different fiducial volumes.

%% file: sec-muon_detection-v2.tex
\begin{figure}[!htbp]
    \centering
    \includegraphics[width=0.75\linewidth]{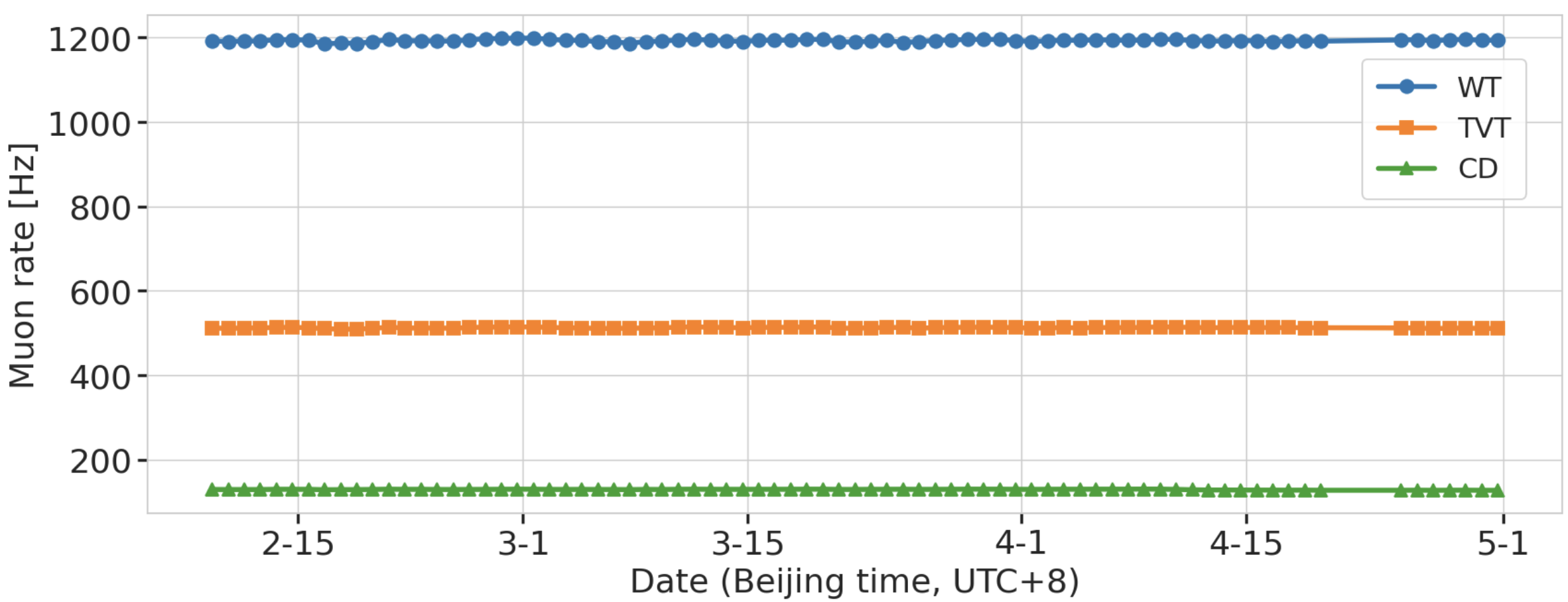}
    \caption{Cosmic-muon rates measured by the CD, WT, and TVT systems.}
    \label{fig:muon-rate}
\end{figure}

Figure~\ref{fig:muon-rate} shows the stability of the muon rates measured by the CD, WT, and TVT systems. CD muons are defined as events with reconstructed energies greater than 20~MeV, and the corresponding muon rate is measured to be 121~Hz. Muon candidates in the WT are selected by requiring at least 18 fired PMTs in one GCU. The total muon rate measured by the WT system is about 1.7~kHz. The rates measured in WT\#1, WT\#2, and WT\#3 are about 700~Hz, 510~Hz, and 685~Hz, respectively. The differences are primarily caused by the different volumes of the water tanks. Muon candidates in the TVT are selected by requiring hits in at least three of the four PS layers. The muon rate measured by the TVT is 512~Hz and remains stable over time. The corresponding muon flux, estimated from the TVT, is 35.3~Hz/m\(^2\), which is about a factor of two lower than the predicted value. This discrepancy is mainly attributed to the imperfect modeling of the laboratory building structure and the absence of nearby mountains in the detector simulation used to estimate the muon rate at the TAO site.

\begin{figure}[!htbp]
    \centering
    \includegraphics[width=0.75\linewidth]{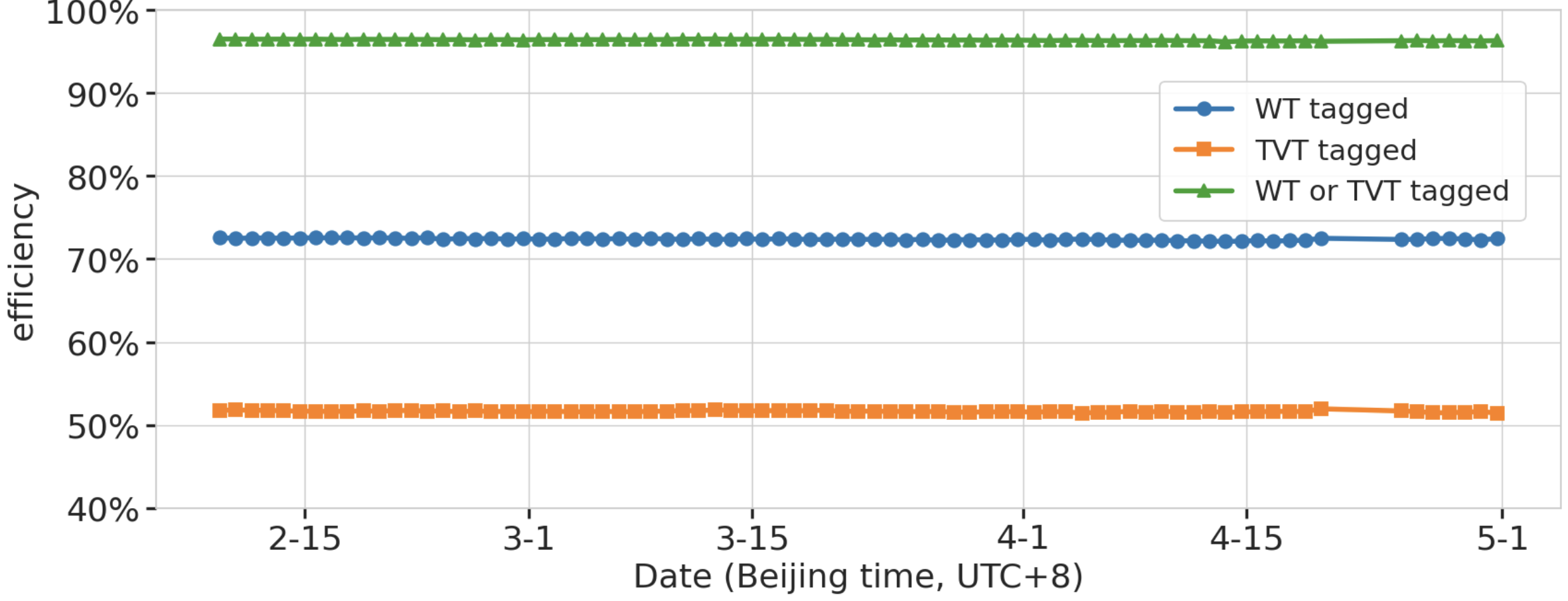}
    \caption{Muon detection efficiency of the WT and TVT systems for CD-identified muons.}
    \label{fig:cd-muon-eff}
\end{figure}

The combined muon detection efficiency of the full veto system, including the WT and TVT, is calibrated using CD-identified muons, defined as events with reconstructed energies greater than 20~MeV in the CD. The resulting combined efficiency is about 97\%, as shown in Fig.~\ref{fig:cd-muon-eff}. The residual inefficiency is mainly attributed to: 1) offset of the centroid of the TVT with respect to that of the CD, caused by an interference of the cable tray of the power plant above WT\#2; 2) the reduced height of WT\#2 with respect to the other WTs, also due to the interference of the cable tray. The blue and orange curves denote the muon detection efficiencies of the WT and TVT, respectively, both of which are strongly affected by the limited geometric acceptance of the two detectors.

%% file: sec-IBD-analysis-v2.tex
The first TAO reactor antineutrino sample is selected by exploiting the characteristic prompt--delayed topology of the IBD interaction in the central detector. The prompt signal is produced by the positron, while the delayed signal arises from neutron capture on Gd after thermalization. The correlations of the prompt--delayed pair in time, energy, and reconstructed position provide strong rejection of single-event radioactive backgrounds and allow a clean extraction of the prompt energy spectrum.

The analysis presented below uses 31.3 days of reactor-on data, corresponding to a period of full-power operation of a single Taishan reactor core, and 29.4 days of reactor-off livetime recorded during a shutdown period of both reactor cores. The IBD selection was conducted independently by three analysis groups starting from a common dataset calibrated and reconstructed with the methods described above. The groups employed separate selection criteria and estimations of their efficiencies and associated uncertainties, with the final results being consistent between the three groups. One set of results is shown in the present section. The background rejection and IBD pairing procedures follow those of previous JUNO and TAO analyses~\cite{JUNO:2015zny,JUNO:2021vlw,JUNO:2025fpc,JUNO:2025gmd,JUNO:2020ijm}, and are summarized below. 

\begin{table}[t]
    \centering
    \caption{Summary of background rejection and IBD selection criteria.}
    \begin{tabularx}{\textwidth}{llX}
        \hline
        Category & Selection & Requirement \\
        \hline
        Muon veto
        & WT muon
        & $\geq 18$ fired PMTs/GCU, veto $[-2,30]~\mu\mathrm{s}$ \\
        & TVT muon
        & $\geq 3$ PS layers, veto $[-2,30]~\mu\mathrm{s}$ \\
        & CD muon
        & $E_{\rm rec}>20~\mathrm{MeV}$, veto $[0,100]~\mu\mathrm{s}$ \\
        \hline
        IBD selection
        & Pre-selection
        & $0.7<E_{\rm rec}<12~\mathrm{MeV}$, $R<875~\mathrm{mm}$ \\
        & Prompt--delayed time
        & $2<\Delta t_{pd}<100~\mu\mathrm{s}$ \\
        & Multiplicity
        & No pre-selected CD events 100~$\mu$s before prompt, 100~$\mu$s after delayed, or between them \\
        & Delayed energy
        & $4<E_{d}<10~\mathrm{MeV}$ \\
        & Prompt--delayed distance
        & $\Delta R_{pd}<500~\mathrm{mm}$ \\
        & Fiducial volume
        & $R_{p}<800~\mathrm{mm}$ \\
        \hline
    \end{tabularx}
    \label{tab:ibd_selection_ihep}
\end{table}

\subsection{IBD Candidate Selection}
\label{subsec:ibd_candidate_selection}

The IBD selection is summarized in Table~\ref{tab:ibd_selection_ihep}. The selection starts from CD events reconstructed in runs satisfying the data-quality and detector-stability requirements discussed in Section~\ref{sec:dataquality}. The first step is designed to reject muon-induced backgrounds, such as spurious coincidences from cosmogenic radioactive isotopes and neutrons produced by muon interactions in the CD or its surroundings. These backgrounds are suppressed by full-detector time vetoes following muons tagged in the WT, TVT, or CD. A WT muon candidate is identified when at least one GCU records 18 or more fired PMTs. A TVT muon candidate is defined by hits in at least three plastic-scintillator layers. Events within $[-2,30]~\mu\mathrm{s}$ of a WT or TVT muon candidate are vetoed. For CD muons, defined as CD events with reconstructed energy above 20~MeV, events in the following $[0,100]~\mu\mathrm{s}$ interval are vetoed.

After the muon veto, a loose CD pre-selection is applied to reduce the accidental background rate before prompt--delayed pairing. This pre-selection requires $0.7 < E_{\rm rec} < 12~\mathrm{MeV}$ and $R < 875~\mathrm{mm}$, where \(E_{\rm rec}\) and \(R\) are the reconstructed energy and distance from the detector center. Only events passing this pre-selection are used in the subsequent selection stages. The lower energy bound is close to the physics-trigger threshold and retains the prompt-energy region relevant for this first positron-spectrum measurement. The radial requirement is loose with respect to the final fiducial-volume cut; it mainly rejects events reconstructed near or outside the CD boundary, reducing the accidental-pair rate while having negligible impact on the final IBD sample.

\begin{figure}[t]
    \centering

    \begin{subfigure}[t]{0.48\linewidth}
        \centering
        \includegraphics[width=\linewidth]{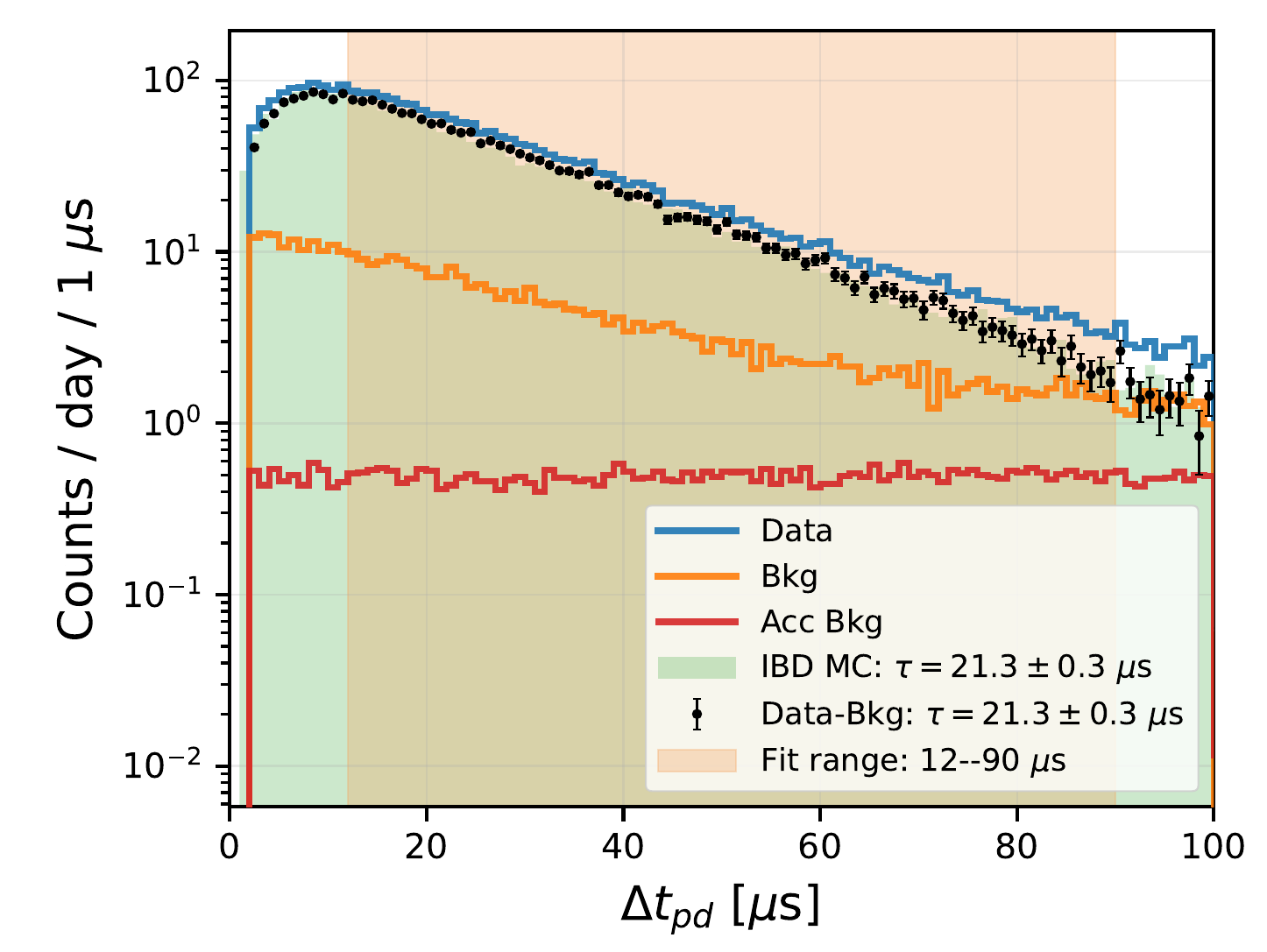}
        \subcaption{\small Prompt--delayed time difference}
        \label{fig:dt_ibd}
    \end{subfigure}
    \hfill
    \begin{subfigure}[t]{0.48\linewidth}
        \centering
        \includegraphics[width=\linewidth]{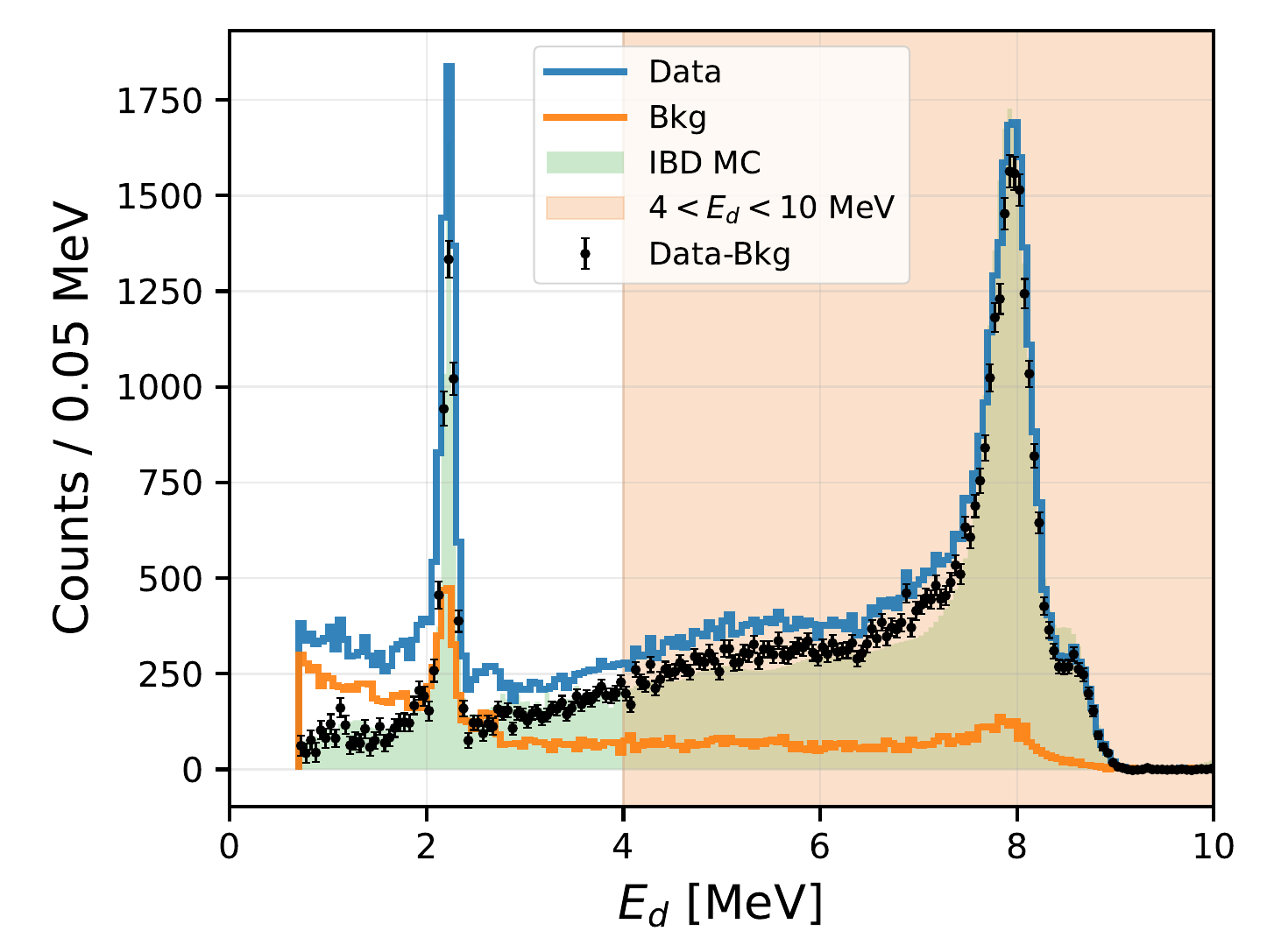}
        \subcaption{\small Delayed energy}
        \label{fig:Edel_ibd}
    \end{subfigure}

    \par\vspace{7mm} 

    \begin{subfigure}[t]{0.48\linewidth}
        \centering
        \includegraphics[width=\linewidth]{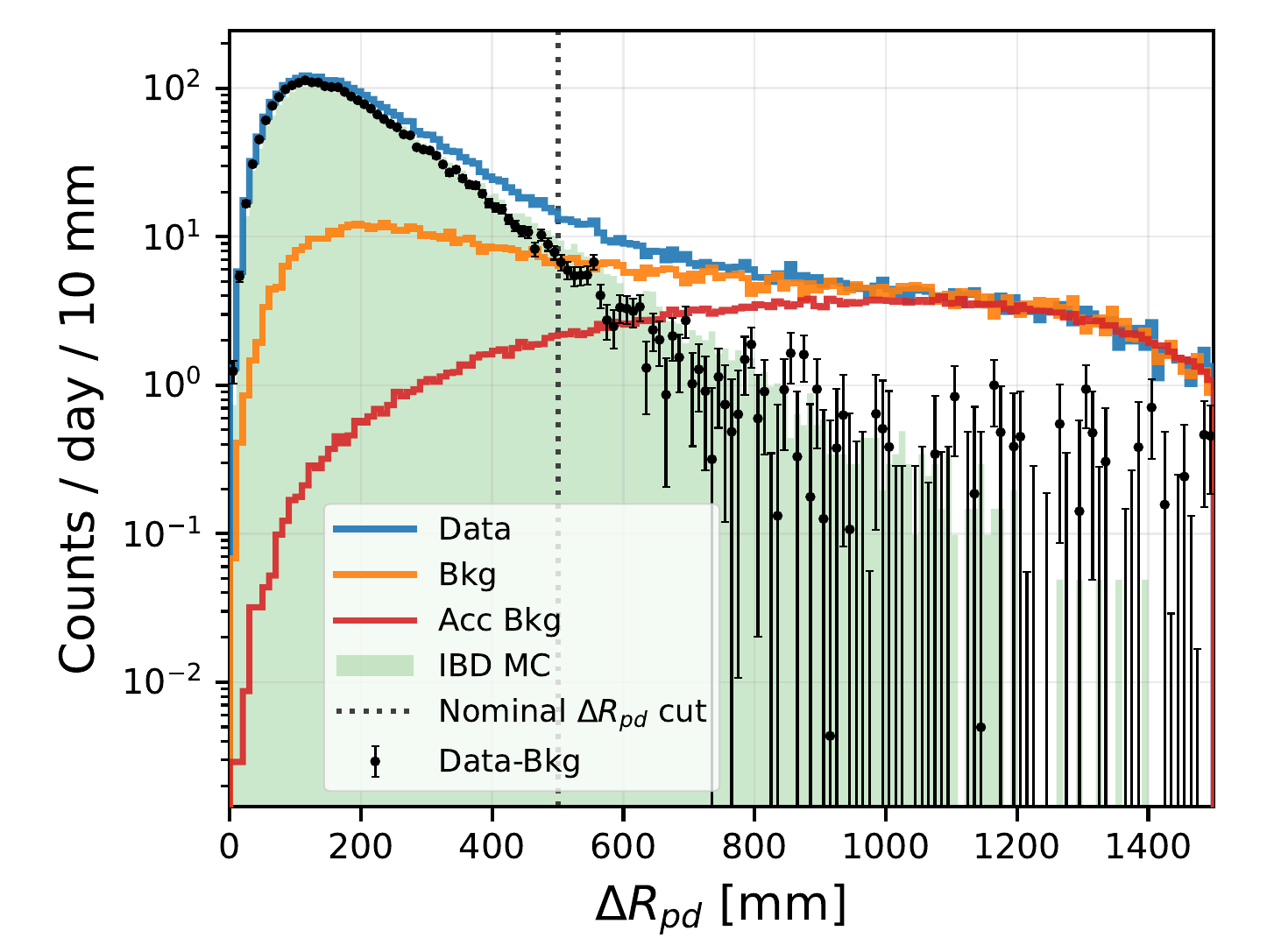}
        \subcaption{\small Prompt--delayed distance}
        \label{fig:deltaR_ibd}
    \end{subfigure}
    \hfill
    \begin{subfigure}[t]{0.48\linewidth}
        \centering
        \includegraphics[width=\linewidth]{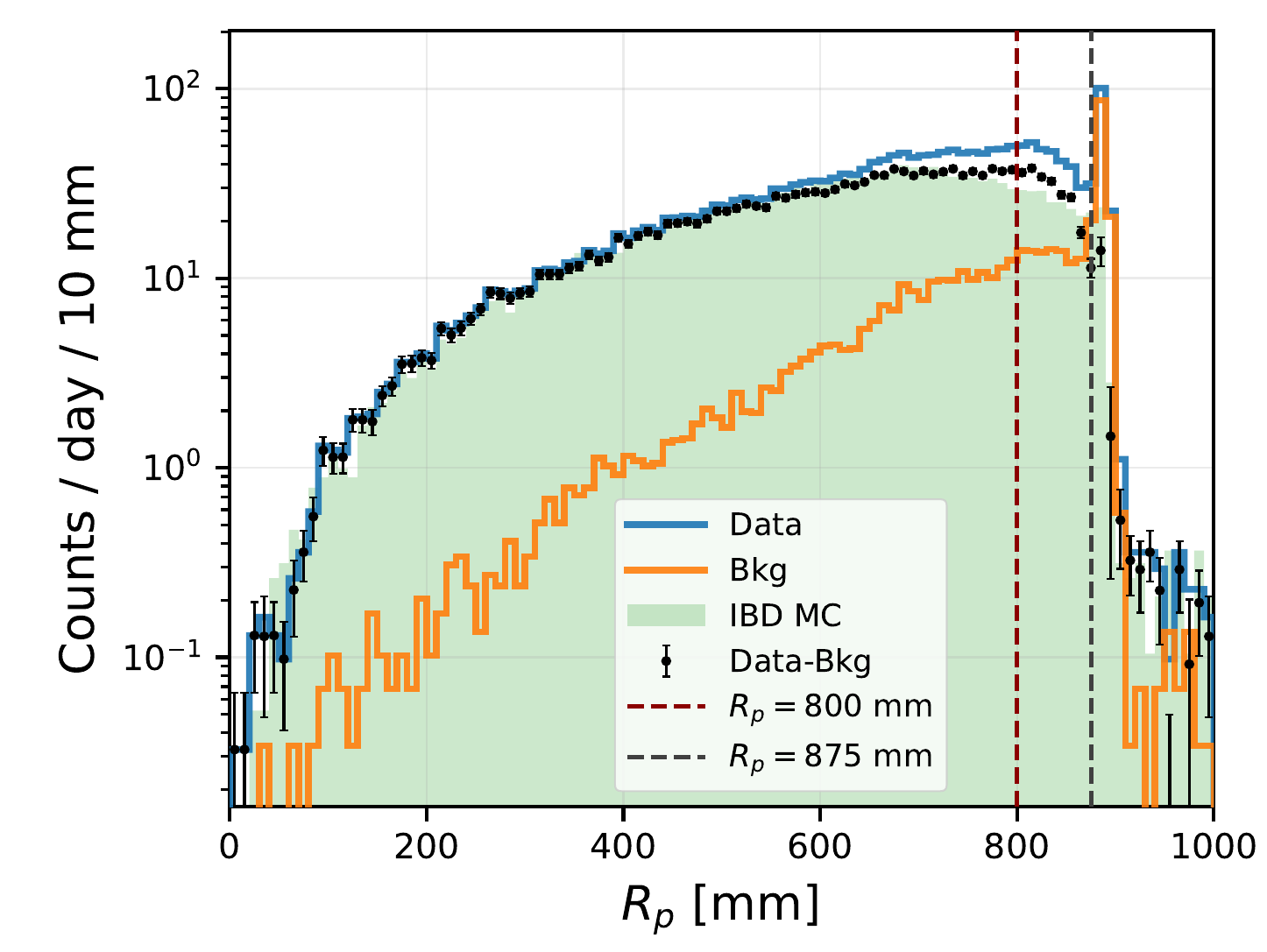}
        \subcaption{\small Prompt radius}
        \label{fig:Rp_ibd}
    \end{subfigure}

    \caption{Representative distributions of the main observables used in the IBD
selection: (a) prompt--delayed time difference, (b) delayed energy
($E_p > 3~\mathrm{MeV}$), (c) prompt--delayed distance, and
(d) reconstructed prompt distance from the detector center.
Data, background templates, and Monte Carlo simulation are shown where
appropriate. The data-background distribution is obtained by
subtracting the reactor-off data from the reactor-on data after scaling
the reactor-off sample to the reactor-on live time. The small population of events reconstructed at
$R>900~\mathrm{mm}$ originates from the current reconstruction
algorithm, which occasionally reconstructs events near the detector
boundary at radii beyond the physical detector radius.}
    \label{fig:ibd_selection_distributions}
\end{figure}

Each pre-selected event is first treated as a delayed-like candidate. A prompt candidate is then searched for in the preceding time window defined by \(2 < \Delta t_{pd} < 100~\mu\mathrm{s}\), where \(\Delta t_{pd}\) is the time interval between the prompt and delayed signals. The time distribution for selected prompt-delayed pairs is shown in Fig.~\ref{fig:dt_ibd}. After the neutron thermalization region, the data are well described by an exponential distribution. A fit in the range \(12<\Delta t_{pd}<90~\mu\mathrm{s}\) yields an effective neutron-capture time constant of \(21.3\pm0.3~\mu\mathrm{s}\), consistent with the Monte Carlo result of \(21.3\pm0.3~\mu\mathrm{s}\).


A multiplicity requirement is then applied to reject pairs for which more than one pre-selected event is present in the relevant time interval. In particular, no additional pre-selected CD event is allowed within 100~\(\mu\mathrm{s}\) before the prompt candidate, between the prompt and delayed candidates, or within 100~\(\mu\mathrm{s}\) after the delayed candidate. This selection suppresses accidental coincidences and ambiguous multi-event configurations while retaining the characteristic IBD time correlation.

Neutron capture on \(^{157}\)Gd and \(^{155}\)Gd in the GdLS is followed by the emission of cascades of \(\gamma\) rays, with full energy depositions around 7.9~MeV and 8.5~MeV, respectively~\cite{JUNO:2020ijm}. 
Because part of the cascade energy may escape the active volume or be lost in non-scintillating materials, the delayed-like signal is selected over a broad energy range, \(4 < E_d < 10~\mathrm{MeV}\), primarily to efficiently retain neutron captures on Gd. This delayed-energy window can also include a small contribution from neutron captures on carbon and other nuclei whose reconstructed energies fall in the same range. Neutron capture on hydrogen, which produces a \(2.223\)-MeV \(\gamma\) ray, is not used for IBD selection because this energy region overlaps with prominent radioactive backgrounds from \(^{214}\)Bi and \(^{208}\)Tl present in the detector.
The delayed-energy distribution for the selected sample is shown in Fig.~\ref{fig:Edel_ibd}. No additional prompt-energy selection is applied beyond the pre-selection range.

The reconstructed prompt--delayed distance is then required to satisfy
\(\Delta R_{pd}<500~\mathrm{mm}\). This requirement provides additional
rejection of accidental pairs, whose prompt and delayed vertices are largely
uncorrelated, while retaining most of the genuine IBD signal. The corresponding
distribution is shown in Fig.~\ref{fig:deltaR_ibd}. The accidental-coincidence
component, which dominates at large \(\Delta R_{pd}\), is modeled using
prompt--delayed pairs formed from 11 off-time windows corresponding to time
separations of 500--1500 ms.

Finally, a fiducial-volume requirement of \(R_p<800~\mathrm{mm}\) is applied to the prompt candidate. This cut increases the statistics of the first TAO prompt-spectrum measurement while retaining the detector region where reconstruction performance and detector-response corrections are sufficiently controlled. At the same time, it suppresses accidental coincidences from external radioactive backgrounds near the detector boundary, as illustrated by the prompt-radius distribution in Fig.~\ref{fig:Rp_ibd}. After applying all the selections described above, the total selection efficiency is about 46.6\% (Table~\ref{tab:ibd_efficiency_ihep}).


\subsection{Background Estimation}
\label{subsec:background_estimation}
The residual background after the IBD selection is estimated using the reactor-off data sample. During the reactor-off period, the same detector conditions, reconstruction procedure, and event-selection criteria are applied as for the reactor-on data, including the muon veto, pre-selection, prompt--delayed pairing, delayed-energy, prompt--delayed distance, and fiducial-volume requirements. Since the reactor was off during this period, the selected events provide a direct measurement of the non-reactor background under the IBD selection, dominated by cosmogenic backgrounds induced by cosmic-ray muons.

After the full selection, the reactor-off sample yields a background rate of \(419 \pm 4\) events per day. The corresponding prompt-energy spectrum is used as a data-driven background template for the reactor-on sample after normalization to the reactor-on livetime. This method provides an estimate of both the rate and the spectral shape of the residual background while minimizing the reliance on simulation.




\subsection{Measurement of the Positron Spectrum}
\label{subsec:ibd_prompt_spectrum}

The measured IBD prompt spectrum is obtained by subtracting the reactor-off prompt spectrum after scaling by the accumulated DAQ livetime ratio:
\begin{equation}
    N_{\rm IBD}(E_{p})
    =
    N_{\rm on}(E_{p})
    -
    \alpha N_{\rm off}(E_{p}),
    \qquad
    \alpha = \frac{T_{\rm on}}{T_{\rm off}},
    \label{eq:ibd_on_off_subtraction}
\end{equation}
where \(N_{\rm on}\) and \(N_{\rm off}\) are the reactor-on and reactor-off selected candidate spectra, and \(T_{\rm on}\) and \(T_{\rm off}\) are the corresponding DAQ livetimes. This subtraction removes backgrounds that are independent of the reactor status, including accidental and cosmogenic components. An estimate indicates that the residual SNF and long-lived non-equilibrium contributions after the reactor-on/off subtraction are negligible. The accidental background is also checked independently with a sliding time-window method, which provides a validation of its spectral shape and rate.

\begin{figure}[t]
    \centering
    \includegraphics[width=0.9\linewidth]{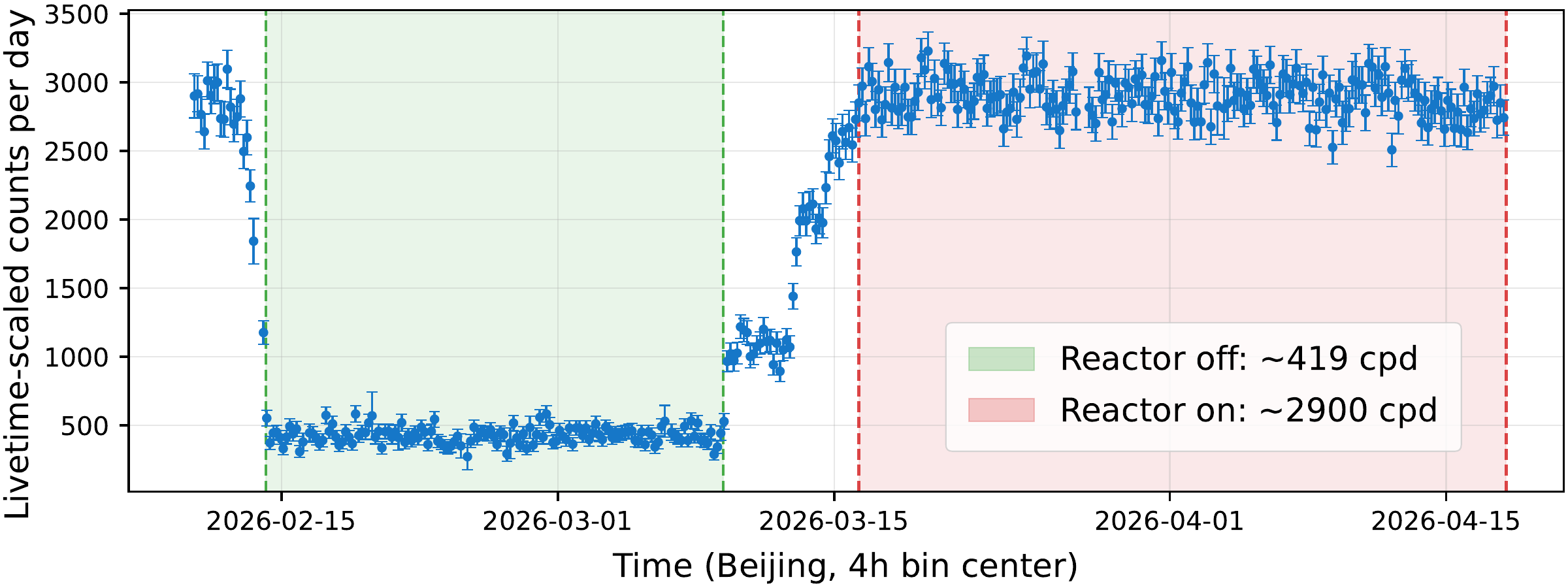}
    \caption{Time dependence of the selected IBD candidate rate. Each point corresponds to a 4-hour interval, with the rate scaled to one day. The reduction during the reactor-off period confirms the reactor-correlated origin of the selected sample and supports the use of reactor-off data for background estimation.}
    \label{fig:ibd_rate_time_ihep}
\end{figure}

The time dependence of the selected IBD candidate rate in 4-hour time bins is shown in Fig.~\ref{fig:ibd_rate_time_ihep}. The rate follows the reactor operation status closely, with a clear reduction during the reactor-off period. In this analysis, spectra from two independent periods of reactor shutdown are combined to form the background template. Their shapes and rates are found to be consistent within statistical uncertainties. Time-modulation effects on cosmogenic backgrounds are found to be negligible. Data collected during periods of transient reactor power are not used in this analysis due to the difficulty in modeling the time-dependent contribution from out-of-equilibrium fission products.

The predicted prompt spectrum is obtained from the expected reactor antineutrino flux at the TAO detector position with respect to the nearest Taishan core. The total antineutrino flux is calculated as
\begin{equation}\label{eq:flux_computation}
    \Phi(E_\nu) = \frac{W_{th}}{\sum_k f_{k} \cdot e_k} \cdot \frac{1}{4\pi L_c^2} \cdot S_{\rm tot}(E_\nu),
\end{equation}
where \(W_{th}\) is the reactor thermal power, \(f_k\) are the isotopic fission fractions for \(^{235}\mathrm{U}\), \(^{238}\mathrm{U}\), \(^{239}\mathrm{Pu}\), and \(^{241}\mathrm{Pu}\), \(e_k\) is the average energy release per fission of isotope \(k\), \(L_c=43.2\)~m is the distance to the reactor core, and \(S_{\rm tot}(E_\nu)\) is the total antineutrino flux spectrum based on the Huber--Mueller model~\cite{PhysRevC.84.024617,PhysRevC.83.054615}, weighted by the corresponding fission fractions.

Given their negligible residual contributions after the reactor-on/off subtraction, no additional correction for long-lived non-equilibrium isotopes or spent nuclear fuel is applied to the predicted spectrum. Contributions from the far Taishan core are negligible for the present analysis since it is off for the whole time of data analyzed. The input parameters in Eq.~\eqref{eq:flux_computation} are consistent with those used in the previous JUNO analysis~\cite{JUNO:2025gmd}.

The antineutrino flux in Eq.~\eqref{eq:flux_computation} is converted into a positron spectrum using the IBD kinematics and cross section~\cite{PhysRevD.60.053003,Strumia:2003zx}, together with the number of target protons in the CD. The positron spectrum is then forward folded into the prompt visible and reconstructed energy spectrum through the detector response inferred from simulations and calibration data. This response includes the conversion from positron energy to deposited energy, the convolution of energy leakage with the energy non-linearity, and the energy resolution discussed in Sections~\ref{sec:energyleakage}, \ref{sec:nonlinearity}, and \ref{sec:resolution}. Finally, the overall IBD selection efficiency described in Section~\ref{subsec:ibd_efficiency} is applied.

\begin{figure}[t]
    \centering
    \includegraphics[width=0.9\linewidth]{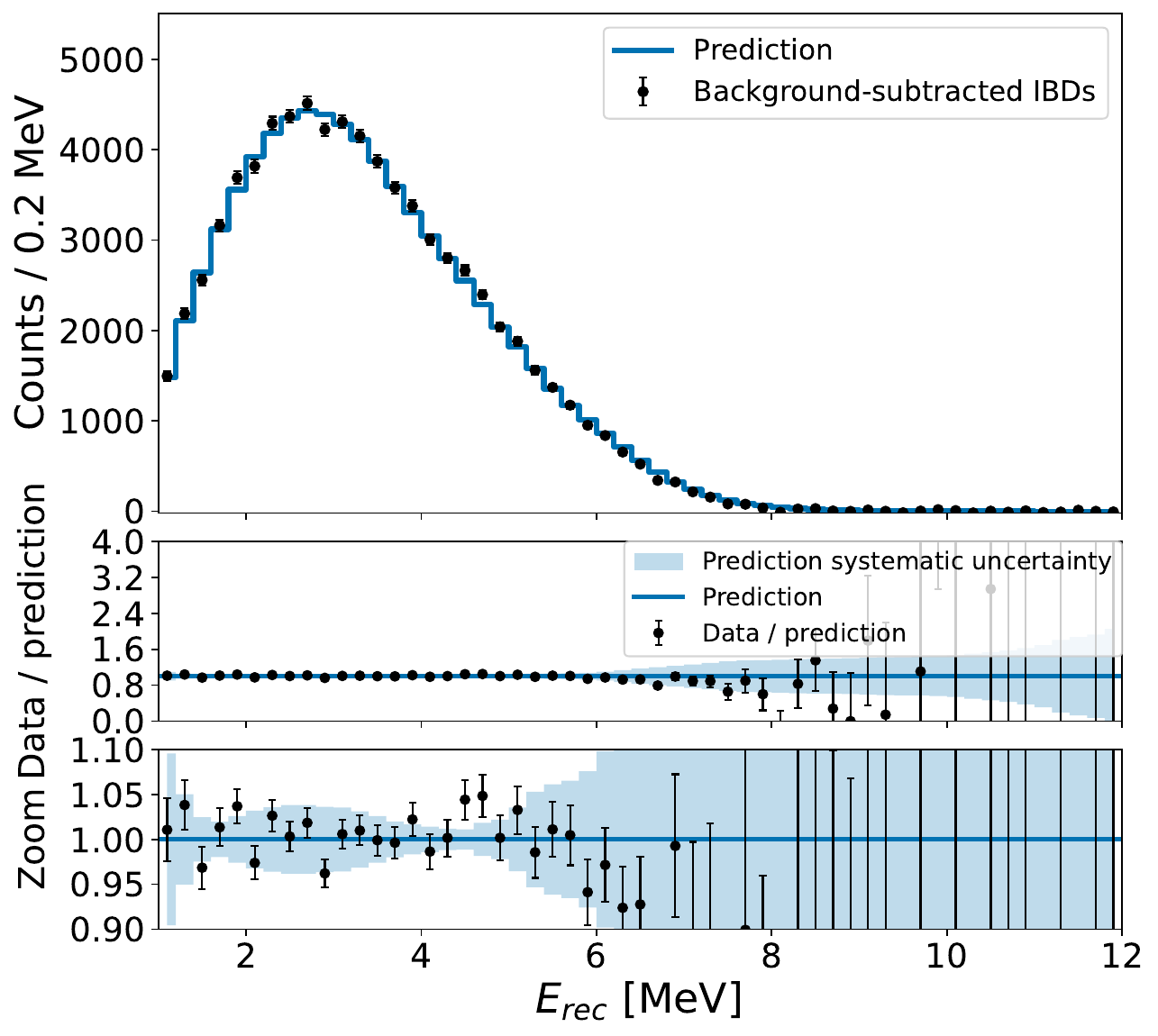}
    \caption{Measured IBD prompt-energy spectrum obtained from the reactor-on sample after subtraction of the DAQ-livetime-scaled reactor-off spectrum, compared with the Huber-Mueller prediction. The lower panel shows the ratio of data to prediction, and the shaded band indicates the total uncertainty of the prediction.} 
    \label{fig:ibd_prompt_spectrum_ihep}
\end{figure}

The predicted and observed prompt spectra are shown in Fig.~\ref{fig:ibd_prompt_spectrum_ihep}. The upper panel shows the reactor-on data after subtracting the livetime-scaled background together with the nominal IBD signal prediction. The lower panel shows the ratio of the measured spectrum to the total prediction with the total \(1\sigma\) uncertainty band propagated from the statistical and systematic uncertainties discussed in Section~\ref{subsec:ibd_uncertainty}. At the present stage, this comparison is primarily intended as a validation of the detector performance and analysis chain. A precision reactor-spectrum measurement will require larger statistics and further refinement of the detector response, background model, and systematic covariance.

\subsection{Event Selection Efficiency}
\label{subsec:ibd_efficiency}

The selection efficiency is evaluated individually for each step listed in Table~\ref{tab:ibd_efficiency_ihep} using simulation, data-driven estimates, or toy Monte Carlo studies, depending on the step. To reduce the dependence on
the detector simulation, data-driven methods are used whenever the corresponding
efficiency can be directly constrained from data. The simulation is used only
for efficiencies that cannot be obtained from data, such as the efficiency of
the energy-threshold requirement and the prompt--delayed time difference requirement. For
the remaining selection criteria, the central values of the efficiencies are
derived from data, while Monte Carlo samples are used only to estimate residual
corrections or systematic uncertainties where needed. 

The muon-veto efficiency for the delayed signal is obtained from the total livetime remaining after the veto is applied. Since the prompt signal occurs before the delayed signal, the muon-veto efficiency for the prompt side is evaluated by accounting for the possibility that the prompt event falls outside a live interval while the delayed event is still accepted. This effect is estimated using the recorded live-time fragments together with the measured prompt--delayed time distribution. The uncertainties on both muon-veto efficiencies are negligible.

\begin{table}[t]
    \centering
    \caption{Summary of IBD selection efficiencies and observed candidate rates. Uncertainties that are not quoted are negligible.}
    \begin{tabular}{lc}
        \hline
        Selection & Efficiency or rate \\
        \hline
        Muon veto for delayed signal & 95.1\% \\
        Muon veto for prompt signal & 97.4\% \\
        $E_{p},E_{d}\in[0.7,12]~\mathrm{MeV}$ & $90.0\pm1.5\%$ \\
        $R_{p},R_{d}<875~\mathrm{mm}$ & $97.8\pm0.8\%$ \\
        Multiplicity & 96.0\% \\
        $2<\Delta t_{pd}<100~\mu\mathrm{s}$ & $96.7\pm0.1\%$ \\
        $4<E_{d}<10~\mathrm{MeV}$ & $74.6\pm4.0\%$ \\
        $\Delta R_{pd}<0.5~\mathrm{m}$ & $96.4\pm2.6\%$ \\
        Fiducial volume, $R_{p}<800~\mathrm{mm}$ & $85.6\pm7.4\%$ \\
        \hline
        Total efficiency & $46.6\pm5.0\%$ \\
        Candidate rate, reactor on & $2900\pm10~\mathrm{cpd}$ \\
        Candidate rate, reactor off & $419\pm4~\mathrm{cpd}$ \\
        \hline
    \end{tabular}
    \label{tab:ibd_efficiency_ihep}
\end{table}

The pre-selection efficiency includes the requirement that both the prompt and delayed signals satisfy the loose reconstructed energy and radius cuts; the efficiencies of the energy cuts are evaluated using Monte Carlo simulation, whereas those of the radius cuts are determined from real data singles. The efficiency of the multiplicity requirement is evaluated semi-empirically as the probability of observing no uncorrelated prompt-like event within the multiplicity window, given the measured singles rate and the average prompt--delayed time interval derived from data. The prompt--delayed time-window efficiency is evaluated from simulation and checked against the measured time-difference distribution. The delayed-energy, prompt--delayed distance, and fiducial-volume efficiencies are obtained from Monte Carlo simulation, with conservative systematic uncertainties assigned to cover the modeling of the energy response, event vertex reconstruction, energy leakage, and data--Monte Carlo discrepancies. In particular, the uncertainty associated with the fiducial-volume requirement, dominated by the vertex reconstruction bias near the detector boundary, is the largest contribution to the total efficiency uncertainty.

The total IBD selection efficiency is estimated to be \(46.6\pm5.0\%\), while the observed candidate rates are \(2900\pm10\) counts per day (cpd) in the reactor-on sample and \(419\pm4\)~cpd in the reactor-off sample.

\subsection{Spectral Shape Uncertainty}
\label{subsec:ibd_uncertainty}

The shape uncertainty of the measured prompt-energy spectrum is evaluated using the fine non-uniform binning required for the combined fit to JUNO data in the oscillation analysis. The total statistical uncertainty is calculated directly from the observed event distribution after background subtraction. The background uncertainty is evaluated from the bin-by-bin background contributions normalized to the signal time window, with the spectral shape constrained by the reactor-off data.

The impact of the fiducial-volume uncertainty on the prompt spectrum is evaluated following the procedure described in Section~\ref{sec:leakage_uncertainty}. This treatment accounts for the combined effect of event-vertex bias, position smearing, and energy leakage near the detector boundary. The energy non-linearity uncertainty is evaluated by sampling positron non-linearity curves according to the fitted model uncertainties (see Fig.~\ref{fig:nl_curves}) and propagating them to reconstructed prompt spectra through the detector response model described in Section~\ref{subsec:ibd_prompt_spectrum}. The resulting ensemble is used to construct the non-linearity covariance matrix.

The energy-scale uncertainty is included by applying a common global energy shift to toy prompt spectra, with the shift sampled from a Gaussian distribution with a width of 0.6\%. Residual non-equilibrium and SNF contributions are estimated to be negligible after the reactor-on/off subtraction and are therefore not included in the uncertainty budget.

\begin{figure}[t]
    \centering
    \includegraphics[width=0.9\linewidth]{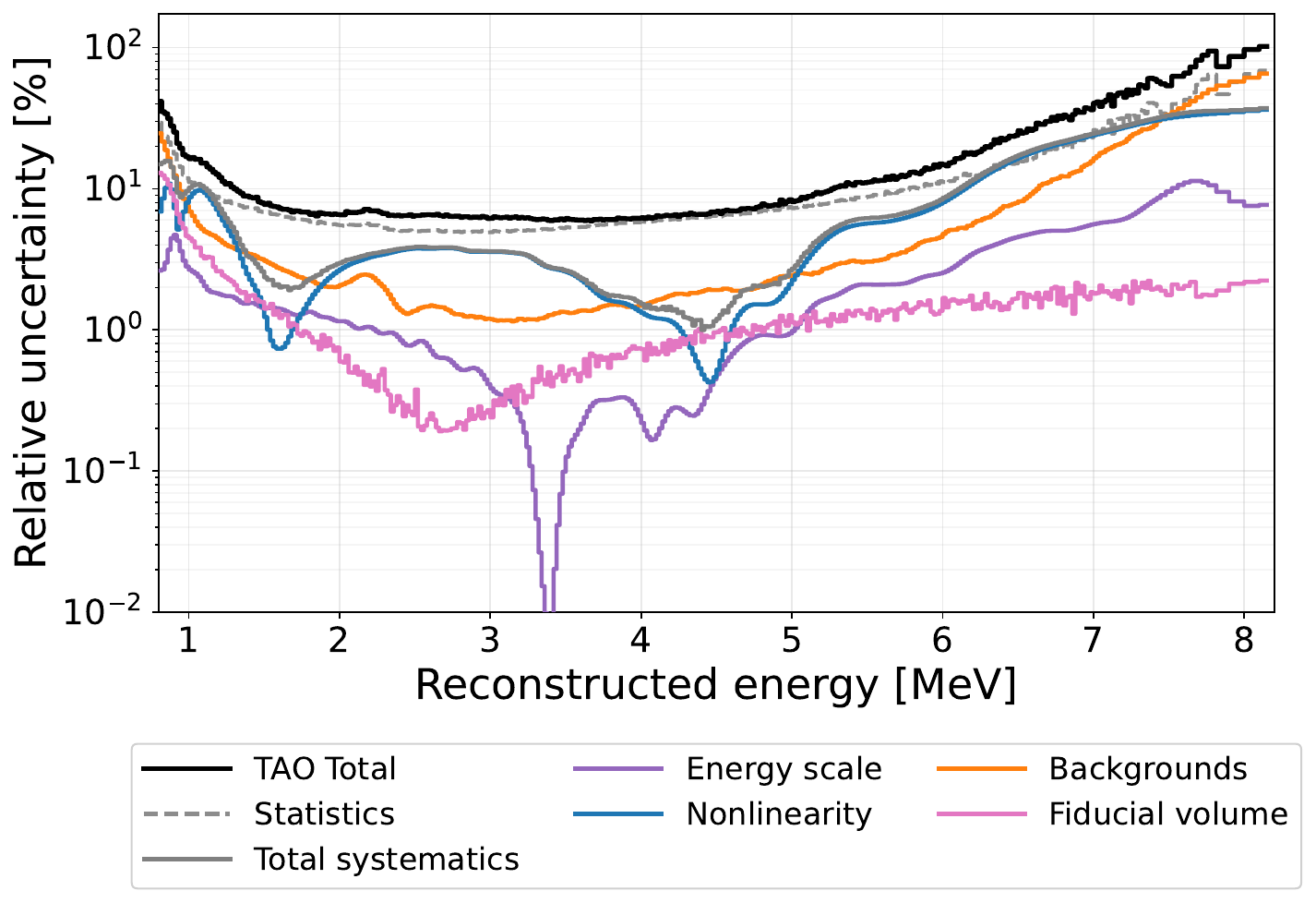}
    \caption{Relative uncertainty of the measured IBD prompt spectrum. Shown are the statistical, background, fiducial-volume, energy-scale, energy non-linearity and total uncertainty components.
    }
    \label{fig:ibd_uncertainty_ihep}
\end{figure}

The corresponding total fractional uncertainties are shown in Fig.~\ref{fig:ibd_uncertainty_ihep}. At the current exposure, statistical uncertainty dominates the measured prompt spectrum over most of the energy range.

%% file: sec-summary-v2.tex
The TAO detector has completed construction, commissioning, and the transition to physics data taking. Based on the commissioning data and the first physics data, the detector performance has been evaluated in terms of detector operation, calibration, event reconstruction, energy responses, and muon veto performance.

The central detector has achieved stable cryogenic operation. The temperature uniformity and long-term stability satisfy the design requirements. The SiPM operating voltages have been established on the basis of energy-resolution measurements from calibration data. After channel-quality selections, about 90\% of the 8,048 readout channels are retained for physics analysis. Most of the remaining unavailable channels are well understood and are expected to be addressed during the next detector maintenance period.

A channel-level calibration program has been developed for the SiPMs in the central detector. Run-by-run calibration constants have been obtained for the key SiPM parameters, including gain, DCR, time offset, optical crosstalk, and relative PDE. The detector light yield has been characterized using \(^{68}\)Ge calibration data at the detector center and neutron capture on hydrogen signals from IBD events, and is measured to be about 3614~PEs/MeV and 4000~PEs/MeV, respectively, after correction for dark noise and optical crosstalk. The current reconstruction chain provides a vertex bias controlled within \(\pm 20\)~mm in the central detector region and within \(\pm 30\)~mm near the boundary of the fiducial volume with radius 800~mm. The energy resolution is $\sim$2.8\% for \(^{68}\)Ge at the detector center and $\sim$3.7\% for \(^{137}\)Cs at \(R=193\)~mm. This energy-resolution performance is among the best reported for reactor-antineutrino detectors. Both the energy non-linearity model and the energy-resolution model for \(\gamma\), \(e^-\), and \(e^+\) have been established using calibration sources, radioactivity \(\gamma\)-ray backgrounds, and delayed signals from IBD events. The veto system, including the WT and TVT, has demonstrated stable operation, with a combined muon-tagging efficiency of about 97\%.

Several improvements are planned for future operation. New ADC cards have been produced and are scheduled for installation during the next detector maintenance period, which is expected to recover a significant fraction of the currently unavailable channels. Upgrades to the grounding, shielding, and filtering of the electronics system are also planned to reduce the noise level. In addition, the full functionality of the ACU is expected to be restored after maintenance of the combined-source deployment mechanism.

Finally, 31.3 days of reactor-on data have been used to obtain the first positron energy spectrum measured by TAO, together with 29.4 days of reactor-off data used to constrain the associated background. A complete event-selection procedure has been developed to isolate reactor antineutrino IBD interactions in TAO, resulting in an observed IBD candidate rate of \(2900 \pm 10\)~cpd with an estimated background rate of \(419 \pm 4\)~cpd. The selection efficiencies, the corresponding uncertainties, and their impact on the positron spectral shape have been studied individually. The final spectrum has been compared with the nominal prediction from theoretical models and is found to be consistent within the considered uncertainties. These results establish the current detector status and provide the basis for the first TAO positron-spectrum measurement. Further improvements are expected with continued detector optimization and the accumulation of additional physics data.